\documentclass[a4paper,11pt]{article}
\pdfoutput=1
\usepackage{jheppub}
\usepackage{hyperref}
\usepackage{bm}

\allowdisplaybreaks

\title{\boldmath Electroweak Precision Observables, New Physics and
  the Nature of a 126 GeV Higgs Boson}

\author[a]{Marco Ciuchini,}
\author[b]{Enrico Franco,}
\author[b,c]{Satoshi Mishima}
\author[b]{and Luca Silvestrini}

\affiliation[a]{INFN, Sezione di Roma Tre, Via della Vasca Navale 84, 
I-00146 Roma, Italy}
\affiliation[b]{INFN, Sezione di Roma, Piazzale A. Moro 2, 
I-00185 Roma, Italy}
\affiliation[c]{Dipartimento di Fisica, Universit\`a di Roma 
``La Sapienza'',\\ Piazzale A. Moro 2, I-00185 Roma, Italy}

\abstract{We perform the fit of electroweak precision observables
  within the Standard Model with a $126$ GeV Higgs boson, compare the
  results with the theoretical predictions and discuss the impact of
  recent experimental and theoretical improvements. We introduce New
  Physics contributions in a model-independent way and fit for the
  $S$, $T$ and $U$ parameters, for the $\epsilon_{1,2,3,b}$ ones, for
  modified $Zb\bar{b}$ couplings and for a modified Higgs coupling to
  vector bosons. We point out that composite Higgs models are very
  strongly constrained. Finally, we compute the bounds on
  dimension-six operators relevant for the electroweak fit.}
 
\begin{document}
\maketitle
\flushbottom

\section{Introduction}

Electroweak Precision Observables (EWPO) have played a key role in
constraining New Physics (NP) for the past twenty years
\cite{Amaldi:1987fu,Costa:1987qp,
  Langacker:1991an,Peskin:1991sw,Erler:1994fz,
  Altarelli:1990zd,Altarelli:1991fk,Altarelli:1993sz,Barbieri:2004qk,
  Grinstein:1991cd,Barbieri:1999tm}. The most striking examples of the
power of these indirect constraints are the prediction of the top and
Higgs masses. Concerning physics beyond the Standard Model (SM), the
$\epsilon_{1,2,3,b}$ parameterization
\cite{Altarelli:1990zd,Altarelli:1991fk,Altarelli:1993sz} allowed to
extract interesting information without knowing the Higgs mass,
although the constraining power of EWPO was somewhat diluted by the
missing information on the Higgs boson and by the approximations
necessary to write all LEP observables in terms of the
$\epsilon_{1,2,3,b}$ parameters.

The experimental situation improved dramatically in the past year,
with the precise measurement of the Higgs mass at the LHC
\cite{ATLAS:2013mma,Aad:2012tfa,CMS:yva,Chatrchyan:2013lba}. In addition, the
information on other key SM parameters such as the top and $W$ boson
masses has increased considerably, leading altogether to a sizable
progress in the electroweak (EW) fit. It is therefore
phenomenologically relevant to reassess the constraining power of the
EW fit in the light of these recent experimental improvements. To this
aim, we perform the EW fit in the SM and update the constraints on
oblique NP and on modified $Zb\bar b$ couplings. 
Although the direct measurement of the Higgs boson mass completes the SM
parameters relevant for the EW fit and thus makes the use of the
$\epsilon_{1,2,3,b}$ parameters unnecessary, for the sake of
comparison with previous analyses we will present also results for NP
in this parameterization.

On the theory side, the full two-loop fermionic EW contributions to
the $R_b^0$ observable have been recently numerically calculated in
ref.~\cite{Freitas:2012sy}. The implementation of this result in the
global fit has a large impact but represents a nontrivial problem, as we
illustrate in detail below.

A very interesting question that can be tackled with present data is
whether the Higgs boson is elementary or composite. Using a general
effective Lagrangian for Higgs boson interactions
\cite{Giudice:2007fh,Contino:2010mh,Azatov:2012bz,Contino:2013kra}, we
analyze the constraints on the Higgs coupling to vector bosons, and
find out that this coupling can be determined from the fit with an
uncertainty of $5\%$ at $95\%$ probability, while much larger
departures from the SM value are expected in generic composite Higgs
models. Thus, the EW fit points to an elementary Higgs or to composite
Higgs models in which additional contributions are present to restore
the agreement with EWPO.

Finally, we consider the most general effective Lagrangian relevant
for EWPO and compute the constraints on the coefficients of
dimension six operators, which can be translated into lower bounds on
the NP scale assuming a given value for the couplings.

To obtain our results, we perform a Bayesian analysis using 
the BAT library \cite{Caldwell:2008fw} and our own
implementation of the EWPO formul{\ae}. We have tested the agreement of
our code with the ZFITTER (v6.43) one
\cite{Bardin:1999yd,Bardin:1992jc,Arbuzov:2005ma,Akhundov:2013ons} 
and with outputs from the formul{\ae} in refs.~\cite{Cho:1999km,Cho:2011rk}. 

The paper is organized as follows. In Section \ref{sec:SM} we present
the ingredients of the SM fit, the fitting procedure and the SM
results. In Section \ref{sec:Oblique} we present the results for the
oblique parameters $S$, $T$ and $U$. In Section \ref{sec:Epsilon} we
discuss the results for $\epsilon_{1,2,3,b}$ parameters. In Section
\ref{sec:Zbb} we report the constraints on modified $Zb\bar b$
couplings. In Section \ref{sec:HVV} we present constraints on the
Higgs coupling to vector bosons. In Section \ref{sec:genNP} we discuss
the constraints on the effective Lagrangian relevant for EWPO and the
bounds on the NP scale.  Finally, in Section \ref{sec:concl} we
summarize our findings. Some technical details are presented in
Appendices \ref{app:STUformulae} and \ref{app:NonUnivCorr}, while more
information on the fit results is reported in Appendices \ref{app:corr},
\ref{app:obs_oldRb} and \ref{app:obs_newRb}.

\section{Standard Model fit}
\label{sec:SM}

The part of the SM Lagrangian relevant for the computation of EWPO can
be defined in terms of the following free parameters: the fine
structure constant $\alpha$, the muon decay constant $G_\mu$, the $Z$
boson mass $M_Z$, the strong coupling $\alpha_s(M_Z^2)$, the top
quark mass $m_t$ and the Higgs mass $m_h$. In addition, we introduce
the effective parameter $\Delta\alpha_{\rm had}^{(5)}(M_Z^2)$ to take into
account the hadronic contribution to the running of $\alpha$. In terms
of the seven parameters above, the SM prediction for all other EWPO
can be computed.\footnote{While they are negligible in most cases,
  we have kept all fermion masses whenever relevant. Furthermore, we
  have neglected fermion mixing.}

In the Bayesian approach we are following (see
ref.~\cite{Ciuchini:2000de} for details on the statistical treatment),
prior distributions for the parameters $\alpha$, $G_\mu$, $M_Z$, 
$\alpha_s(M_Z^2)$, $m_t$, $m_h$ and 
$\Delta\alpha_{\rm had}^{(5)}(M_Z^2)$ have to be specified. However, 
given the very accurate experimental measurements of these parameters 
(see below), the results are insensitive to the choice of any reasonable
prior.\footnote{In practice, any reasonable prior, convoluted with the
  experimental measurement, will coincide with the experimental
  likelihood. Thus, we can use directly as prior for the above
  parameters their experimental gaussian likelihood.} 
The numerical results presented in the following are derived
computing the region containing 68\% of a marginalized probability
distribution function (p.d.f.) starting from the mode and then
symmetrizing the error, {\it i.e.}, the central value corresponds to the
center of the 68\% probability region and not to the mode. Since all
p.d.f.'s obtained from the fit are almost gaussian, there is very
little dependence on the prescription adopted. 

\subsection{Experimental values of SM parameters}
\label{sec:expsmpar}

The recent measurements of $m_h$ by the ATLAS~\cite{ATLAS:2013mma} and
CMS~\cite{CMS:yva} experiments are given by 
\begin{equation}
m_h = 
\left\{
\begin{array}{ll}
125.5\pm 0.2\,({\rm stat})^{+ 0.5}_{-0.6}\,({\rm syst})\ {\rm GeV} &\ \ \ 
{\rm ATLAS},\\
125.7\pm 0.3\,({\rm stat})\pm 0.3\,({\rm syst})\ {\rm GeV} &\ \ \ 
{\rm CMS}. 
\end{array}
\right.
\end{equation}
We adopt the average $m_h = 125.6\pm 0.3$ GeV in the current
study.\footnote{This na{\"\i}ve average might underestimate the error
  neglecting possible correlations in the systematics, however even
  doubling the error would not affect any of the results in this paper.}

According to ref.~\cite{Beringer:1900zz,Bethke:2012jm}, the world average 
of $\alpha_s(M_Z^2)$ from the fit to various data, excluding the EW 
precision measurements, is given by $\alpha_s(M_Z^2) = 0.1184\pm 0.0006$.  

For the hadronic contribution to the running of the electromagnetic
coupling, we adopt the recent evaluation $\Delta\alpha_{\rm
  had}^{(5)}(M_Z^2) = 0.02750\pm 0.00033$ in
ref.~\cite{Burkhardt:2011ur}.  Note that other recent studies have
reported much smaller uncertainties, e.g., $\Delta\alpha_{\rm
  had}^{(5)}(M_Z^2) = 0.02757\pm 0.00010$~\cite{Davier:2010nc},
$0.027626\pm 0.000138$~\cite{Hagiwara:2011af} and $0.027498\pm
0.000135$~\cite{Jegerlehner:2011mw}, where the first result relies on
pQCD, and the last one has been derived with the Adler function
approach. The result of ref.~\cite{Hagiwara:2011af} differs from
ref.~\cite{Burkhardt:2011ur} mainly in the use of exclusive (instead
of inclusive) data in the range $1.2 - 2$ GeV. Since exclusive
determinations suffer from an unknown systematic uncertainty, we use
the conservative result of ref.~\cite{Burkhardt:2011ur}. We prefer not
to rely on the model-dependent results of refs. \cite{Davier:2010nc}
and \cite{Jegerlehner:2011mw}, although they are consistent with the
values we are using.

In the absence of a world average for the top pole mass, we adopt the
Tevatron average $m_t=173.18\pm 0.56\,({\rm stat})\pm 0.75\,({\rm
  syst})\ {\rm GeV} =173.2\pm 0.9$ GeV~\cite{Aaltonen:2012ra}, fully
compatible with the LHC result $m_t=173.3\pm 0.5\,({\rm stat})\pm
1.3\,({\rm syst})$ GeV~\cite{ATLAS:2012coa}. Since there might be
subtleties related to the precise definition of the pole mass measured
at Tevatron and LHC, we also use for comparison the determination of
the $\overline{{\rm MS}}$ mass 
$\overline{m}_t(\overline{m}_t)=163.3\pm 2.7$ GeV obtained from the
measurement of the $t\bar{t}$ production
cross-section~\cite{Alekhin:2012py}. This value corresponds to
$m_t=173.3\pm 2.8$ GeV. 

For completeness, the other quark masses are taken to be
$\overline{m}_u(2\, {\rm GeV})=0.0023$ GeV, $\overline{m}_d(2\, {\rm
  GeV})=0.0048$ GeV, $\overline{m}_s(2\, {\rm GeV})=0.095$ GeV,
$\overline{m}_c(\overline{m}_c)=1.275$ GeV and
$\overline{m}_b(\overline{m}_b)=4.18$ GeV~\cite{Beringer:1900zz}.

The renormalization group runnings of the strong coupling
constant and the fermion masses are taken into account up to
three-loop level~\cite{Chetyrkin:1997sg,Chetyrkin:1997dh,Chetyrkin:2000yt}.

The measurement of the $Z$ boson mass is taken from LEP: $M_Z =
91.1875\pm 0.0021$ GeV~\cite{LEPEWWG:WEB}. Finally, the 
parameters $G_\mu$ and $\alpha$ are fixed to be constants:
$G_\mu=1.1663787\times 10^{-5}$ GeV$^{-2}$ and 
$\alpha=1/137.035999074$, respectively~\cite{Beringer:1900zz}.

\subsection{Theoretical expressions for EWPO}
\label{sec:thewpo}

The SM contributions to the EWPO have been
calculated very precisely including higher-order radiative
corrections. 
We adopt the on-mass-shell renormalization
scheme~\cite{Sirlin:1980nh,Marciano:1980pb,Bardin:1980fe,Bardin:1981sv}, 
where the weak mixing angle is defined in terms of the physical masses
of the gauge bosons:    
\begin{equation}
s_W^2 \equiv \sin^2\theta_W = 1 - \frac{M_W^2}{M_Z^2}\,, 
\end{equation}
and $c_W^2=1-s_W^2$.

The Fermi constant $G_\mu$ in $\mu$ decay is taken as an input
quantity instead of the $W$-boson mass, since the latter has not been
measured very precisely compared to the former. The relation between
$G_\mu$ and $M_W$ is written as 
\begin{equation}
G_\mu = \frac{\pi\alpha}{\sqrt{2} s_W^2 M_W^2} (1+\Delta r)\,,
\label{eq:Gmu_Mw}
\end{equation}
where $\Delta r$ represents radiative corrections. 
From eq.~\eqref{eq:Gmu_Mw}, the $W$-boson mass is calculated as 
\begin{equation}
M_W^2 
= 
\frac{M_Z^2}{2}
\left( 1+\sqrt{1-\frac{4\pi\alpha}{\sqrt{2}G_\mu M_Z^2}\,(1+\Delta r)}\
\right). 
\label{eq:MW_Deltar}
\end{equation}
The radiative corrections to $\Delta r$ are known very precisely. In
the current study, we employ the approximate formula for $M_W$,
equivalently for $\Delta r$, in ref.~\cite{Awramik:2003rn}, which
includes the full one-loop EW corrections of
$O(\alpha)$~\cite{Sirlin:1980nh,Marciano:1980pb}, the full two-loop
QCD corrections of $O(\alpha\alpha_s)$~\cite{Djouadi:1987gn,
  Djouadi:1987di,Kniehl:1989yc,Halzen:1990je,Kniehl:1991gu,Kniehl:1992dx,Djouadi:1993ss}, 
three-loop QCD corrections of
$O\left(G_\mu\alpha_s^2m_t^2(1+M_Z^2/m_t^2
  +(M_Z^2/m_t^2)^2)\right)$~\cite{Avdeev:1994db,Chetyrkin:1995ix,Chetyrkin:1995js},
the full two-loop EW corrections of
$O(\alpha^2)$~\cite{Barbieri:1992nz,Barbieri:1992dq,
  Fleischer:1993ub,Fleischer:1994cb,Degrassi:1996mg,Degrassi:1996ps,Degrassi:1999jd,
  Freitas:2000gg,Freitas:2002ja,Awramik:2002wn,Onishchenko:2002ve,Awramik:2002vu,
  Awramik:2002wv,Awramik:2003ee,Awramik:2003rn}, and leading
three-loop corrections of $O(G_\mu^2\alpha_s m_t^4)$ and
$O(G_\mu^3m_t^6)$~\cite{vanderBij:2000cg,Faisst:2003px}.  Further
higher-order corrections are known to be negligibly
small~\cite{Weiglein:1998jz,Boughezal:2004ef,Boughezal:2005eb,Schroder:2005db,
  Chetyrkin:2006bj,Boughezal:2006xk}.  The remaining theoretical
uncertainty in $M_W$ coming from missing higher-order corrections is
estimated to be 4 MeV~\cite{Awramik:2003rn}. Since this residual
uncertainty is much smaller than the present experimental one, we do
not take it into account.\footnote{This uncertainty should however be
  added to the SM prediction quoted in table~\ref{tab:SMpred}.}
A comprehensive summary of the radiative corrections can be found in
ref.~\cite{Bardin:1999ak}. 

The interaction between the $Z$ boson and the neutral
current can be written in terms of the effective $Zf\bar{f}$ couplings
$g_{V}^f$ and $g_{A}^f$, of $g_{R}^f$ and $g_{L}^f$, 
or of $\rho_Z^f$ and $\kappa_Z^f$:  
\begin{eqnarray}
\label{eq:gvga}
\mathcal{L}
&=&
\frac{e}{2 s_W c_W}\,
Z_\mu \sum_f \bar{f}
\left( g_{V}^f\gamma_\mu - g_{A}^f \gamma_\mu\gamma_5 \right) f\,,
\\
&=&
\frac{e}{2s_W c_W}\,
Z_\mu \sum_f \bar{f}
\left[ g_{R}^f \gamma_\mu (1 + \gamma_5)
+ g_{L}^f \gamma_\mu (1 - \gamma_5) \right] f\,,
\\
&=&
\frac{e}{2 s_W c_W}\sqrt{\rho_Z^f}\,
Z_\mu \sum_f \bar{f}
\left[( I_3^f - 2Q_f\kappa_Z^f s_W^2)\gamma^\mu 
  - I_3^f\gamma^\mu\gamma_5\right]f\,,
\end{eqnarray}
where $e^2=4\pi\alpha$, $Q_f$ is the electric charge of the
fermion $f$ and $I_3^f$ is the third component of weak isospin. 
The effective mixing angle for a given fermion $f$ is defined through
the relation 
\begin{align}
\sin^2\theta_{\rm eff}^{f}
= {\rm Re}(\kappa_Z^f)s_W^2
= \frac{1}{4 |Q_f|}
\left[1 - {\rm Re}\left(\frac{g_{V}^f}{g_{A}^f}\right) \right].
\label{eq:EffectiveWeakAngle}
\end{align}
The radiative corrections to the effective couplings and the weak
mixing angle depend on the flavour of final-state fermions in general.
The corrections to $\sin^2\theta_{\rm eff}^{f}$ are given in the forms
of approximate
formul{\ae}~\cite{Awramik:2004ge,Awramik:2006uz,Awramik:2008gi},
including the full two-loop EW corrections of $O(\alpha^2)$ as well as
leading $O(G_\mu^2\alpha_s m_t^4)$ and $O(G_\mu^3m_t^6)$ corrections,
where the bosonic two-loop EW contribution is still missing only in
the $Z\to b\bar{b}$ channel.  The theoretical uncertainty from missing
higher-order corrections is estimated to be $4.7\times 10^{-5}$ for
the leptonic channels~\cite{Awramik:2004ge,Awramik:2006uz}, and we
neglect it in the following.  We use those formul{\ae} to calculate the
coupling Re($\kappa_Z^f$) through eq.~\eqref{eq:EffectiveWeakAngle},
while the imaginary part of $O(\alpha)$ is also included. 

The complete two-loop formul{\ae} for the coupling $\rho_Z^f$ are
currently missing. Recently, the complete fermionic two-loop EW
corrections have been calculated for $R_b^0 = \Gamma_b/\Gamma_h$ in
ref.~\cite{Freitas:2012sy}, where an approximate formula has been
presented. However, from this approximate formula alone we cannot
extract the values of $\rho_Z^f$ including fermionic two-loop
corrections, that are necessary to compute other $\rho_Z^f$-dependent
observables such as $R^0_\ell$, $R^0_c$, $\Gamma_Z$ and the hadronic
cross section (see below for their definitions). 
The authors of ref.~\cite{Freitas:2012sy} have kindly
provided us with the approximate formul{\ae} for $\Gamma_u/\Gamma_b$
and $\Gamma_d/\Gamma_b$ \cite{freitasprivate}, which allow us to use
the experimental information on one more observable in addition to
$R_b^0$. To illustrate the impact of these two-loop corrections, we
present our results for the SM fit in two scenarios. First, we use
only the previously known leading and (where available)
next-to-leading two-loop EW contributions of $O(G_\mu^2 m_t^4)$ and
$O(G_\mu^2 m_t^2 M_Z^2)$ in the large-$m_t$ expansion, 
together with the leading three-loop
corrections of $O(G_\mu^2\alpha_s m_t^4)$ and $O(G_\mu^3m_t^6)$.
Second, we use the approximate formul{\ae} for $\Gamma_u/\Gamma_b$ and
$\Gamma_d/\Gamma_b$ adding three free parameters to the fit, which
represent the unknown corrections to $\rho^\nu_Z$, $\rho^\ell_Z$ and
$\rho^b_Z$. The corrections to $\rho^{u,d}_Z$ can then be determined
using the formul{\ae} for $\Gamma_u/\Gamma_b$ and $\Gamma_d/\Gamma_b$.
This is the optimal use we can make of the presently available
theoretical information. It will be interesting to compare the fitted
values of $\delta \rho^\nu_Z$, $\delta \rho^\ell_Z$ and $\delta
\rho^b_Z$ with the theoretical expressions, once these will be
available. As we shall see below, the corrections computed in
ref.~\cite{Freitas:2012sy,freitasprivate} are surprisingly large, so
that an independent check of the computation would be very useful.

In the following, we consider so-called pseudo observables at the $Z$
pole~\cite{ALEPH:2005ab,ALEPH:2010aa}, 
which are not directly measurable in experiments but can be extracted
from real observables by subtracting initial-state QED corrections and
a part of final-state QED/QCD corrections. 

The asymmetry parameter $\mathcal{A}_f$ for a channel $Z\to f\bar{f}$ is 
defined in terms of the effective couplings: 
\begin{equation}
\mathcal{A}_f = 
\frac{2\, {\rm Re}\left(g_{V}^f/g_{A}^f\right)}
{1+\left[{\rm Re}\left(g_{V}^f/g_{A}^f\right)\right]^2}\,.
\end{equation}
The left-right asymmetry, 
the forward--backward asymmetry and the longitudinal polarization of
the $\tau\bar{\tau}$ channel are written in terms of the asymmetry
parameters: 
\begin{eqnarray}
A_{\rm LR}^0 &=& \mathcal{A}_e\,,
\\
A_{\rm FB}^{0,f} &=& \frac{3}{4}\, \mathcal{A}_e\mathcal{A}_f\,,
\\
P_\tau^{\rm pol} &=& \mathcal{A}_\tau\,.
\end{eqnarray}
The partial width of $Z$ decaying into a charged-lepton pair
$\ell\bar{\ell}$, including contribution from final-state QED
interactions, is given in terms of the effective
couplings by~\cite{Bardin:1999ak,Bardin:1999yd}:  
\begin{align}
\Gamma_\ell &= 
\Gamma_0 \big|\rho_Z^f\big|
\sqrt{1-\frac{4m_\ell^2}{M_Z^2}}
\left[ \left(1+\frac{2m_\ell^2}{M_Z^2}\right) 
  \left(\left|\frac{g_{V}^\ell}{g_{A}^\ell}\right|^2 + 1 \right)
  - \frac{6m_\ell^2}{M_Z^2}
\right]
\left( 1 + \frac{3}{4}\frac{\alpha(M_Z^2)}{\pi}\, Q_\ell^2 \right),
\end{align}
where $\Gamma_0=G_\mu M_Z^3/(24\sqrt{2}\pi)$ and $m_\ell$ is the mass
of the final-state lepton. 
In the case of the $Z\to q\bar{q}$ channels, 
final-state QCD interactions have to be taken into account in addition
to the QED ones: 
\begin{align}
\Gamma_q &= 
N_c\,
\Gamma_0 \big|\rho_Z^q\big|
\left[ \left|\frac{g_{V}^q}{g_{A}^q}\right|^2 R_V^q(M_Z^2) 
  + R_A^q(M_Z^2)
\right]
+ \Delta_{\rm EW/QCD}\,,
\end{align}
where $N_c$ is the color factor, and $R_V^q(s)$ and $R_A^q(s)$ are the
so-called radiator factors for which we refer to
refs.~\cite{Chetyrkin:1994js,Bardin:1999ak,Bardin:1999yd}.  We add
recent results for $O(\alpha_s^4)$ corrections~\cite{Baikov:2012er} to
the radiator functions.  The last term $\Delta_{\rm EW/QCD}$ denotes
non-factorizable EW-QCD
corrections~\cite{Bardin:1999yd,Czarnecki:1996ei,Harlander:1997zb}:
$\Delta_{\rm EW/QCD} = -0.113$ MeV for $q=u,c$, $-0.160$ MeV for
$q=d,s$ and $-0.040$ MeV for $q=b$.\footnote{The non-factorizable
  EW-QCD corrections have been neglected in the results of
  ref.~\cite{Freitas:2012sy,freitasprivate}.}

The total decay width of the $Z$ boson, denoted by $\Gamma_Z$, is then
given by the sum of all possible channels: 
\begin{equation}
\Gamma_Z = 3\,\Gamma_\nu 
+ \Gamma_{e} + \Gamma_{\mu} + \Gamma_{\tau} + \Gamma_h\,,
\end{equation}
where we have defined the hadronic width $\Gamma_h=\sum_q \Gamma_q$. 
Moreover the ratios of the widths 
\begin{eqnarray}
R_\ell^0 = \frac{\Gamma_h}{\Gamma_\ell}\,,
\qquad
R_q^0 = \frac{\Gamma_q}{\Gamma_h}\,,
\end{eqnarray}
and the cross section for $e^+e^-\to Z\to \mathrm{hadrons}$ 
at the $Z$ pole
\begin{equation}
\sigma_h^0 =
\frac{12\pi}{M_Z^2}\frac{\Gamma_e\Gamma_h}{\Gamma_Z^2}
\end{equation}
are part of the EWPO. 

For the $W$-boson decay width $\Gamma_W$, 
we use the one-loop formula in 
refs.~\cite{Bardin:1999ak,Bardin:1986fi,Bardin:1981sv}.

\subsection{Experimental data for EWPO and fit results}
\label{sec:fit}

The latest Tevatron average of the $W$-boson mass is $M_W=80.385\pm
0.015$ GeV~\cite{Group:2012gb}. We use the results for $\Gamma_Z$,
$\sigma_{h}^{0}$, $P_\tau^{\rm pol}$, $\mathcal{A}_{f}$, $A_{\rm
  FB}^{0,f}$ and $R_f^0$ from
SLD/LEP-I~\cite{ALEPH:2005ab,LEPEWWG:WEB} and $\Gamma_W$ from
LEP-II/Tevatron~\cite{ALEPH:2010aa}. All experimental inputs are
summarized in the second column of table~\ref{tab:SMfit}, where we take
into account the correlations among the inputs that can be found in
ref.~\cite{ALEPH:2005ab}.

\begin{table}[tp]
\centering
\begin{tabular}{lcccc} 
\hline
& Data & Fit & Indirect & Pull \\
\hline
$\alpha_s(M_Z^2)$ &
  $0.1184\pm 0.0006$ & 
  $0.1184\pm 0.0006$ & 
  $0.078\pm 0.024$ & 
  $-1.9$ 
\\
$\Delta\alpha_{\rm had}^{(5)}(M_Z^2)$ &
  $0.02750\pm 0.00033$ & 
  $0.02742\pm 0.00026$ & 
  $0.02728\pm 0.00043$ & 
  $-0.4$ 
\\
$M_Z$ [GeV] &
  $91.1875\pm 0.0021$ & 
  $91.1878\pm 0.0021$ & 
  $91.204\pm 0.013$ & 
  $+1.2$ 
\\
$m_t$ [GeV] &
  $173.2\pm 0.9$ & 
  $173.5\pm 0.8$ & 
  $175.7\pm 2.6$ & 
  $+0.9$ 
\\
$m_h$ [GeV] &
  $125.6\pm 0.3$ & 
  $125.6\pm 0.3$ & 
  $98.5\pm 27.7$ & 
  $-0.8$ 
\\
$\delta\rho_Z^\nu$ &
  --- & 
  $-0.0052\pm 0.0031$ & 
  --- & 
  --- 
\\
$\delta\rho_Z^\ell$ &
  --- & 
  $-0.0002\pm 0.0010$ & 
  --- & 
  --- 
\\
$\delta\rho_Z^b$ &
  --- & 
  $-0.0021\pm 0.0011$ & 
  --- & 
  --- 
\\
\hline
$\delta\rho_Z^u$ &
  --- & 
  $0.0026\pm 0.0012$ & 
  --- & 
  --- 
\\
$\delta\rho_Z^d$ &
  --- & 
  $0.0023\pm 0.0012$ & 
  --- & 
  --- 
\\
\hline
$M_W$ [GeV] &
  $80.385\pm 0.015$ & 
  $80.366\pm 0.007$ & 
  $80.361\pm 0.007$ & 
  $-1.4$ 
\\
$\Gamma_W$ [GeV] &
  $2.085\pm 0.042$ & 
  $2.0890\pm 0.0006$ & 
  $2.0890\pm 0.0006$ & 
  $+0.1$ 
\\
$\Gamma_{Z}$ [GeV] &
  $2.4952\pm 0.0023$ & 
  $2.4952\pm 0.0023$ & 
  --- & 
  --- 
\\
$\sigma_{h}^{0}$ [nb] &
  $41.540\pm 0.037$ & 
  $41.539\pm 0.037$ & 
  --- & 
  --- 
\\
$\sin^2\theta_{\rm eff}^{\rm lept}(Q_{\rm FB}^{\rm had})$ &
  $0.2324\pm 0.0012$ & 
  $0.23145\pm 0.00009$ & 
  $0.23145\pm 0.00009$ & 
  $-0.8$ 
\\
$P_\tau^{\rm pol}$ &
  $0.1465\pm 0.0033$ & 
  $0.1476\pm 0.0007$ & 
  $0.1476\pm 0.0007$ & 
  $+0.3$ 
\\
$\mathcal{A}_\ell$ (SLD) &
  $0.1513\pm 0.0021$ & 
  $0.1476\pm 0.0007$ & 
  $0.1470\pm 0.0008$ & 
  $-1.9$ 
\\
$\mathcal{A}_{c}$ &
  $0.670\pm 0.027$ & 
  $0.6681\pm 0.0003$ & 
  $0.6681\pm 0.0003$ & 
  $-0.1$ 
\\
$\mathcal{A}_{b}$ &
  $0.923\pm 0.020$ & 
  $0.93466\pm 0.00006$ & 
  $0.93466\pm 0.00006$ & 
  $+0.6$ 
\\
$A_{\rm FB}^{0,\ell}$ &
  $0.0171\pm 0.0010$ & 
  $0.0163\pm 0.0002$ & 
  $0.0163\pm 0.0002$ & 
  $-0.8$ 
\\
$A_{\rm FB}^{0,c}$ &
  $0.0707\pm 0.0035$ & 
  $0.0739\pm 0.0004$ & 
  $0.0740\pm 0.0004$ & 
  $+0.9$ 
\\
$A_{\rm FB}^{0,b}$ &
  $0.0992\pm 0.0016$ & 
  $0.1034\pm 0.0005$ & 
  $0.1038\pm 0.0005$ & 
  $+2.7$ 
\\
$R^{0}_{\ell}$ &
  $20.767\pm 0.025$ & 
  $20.768\pm 0.025$ & 
  --- & 
  --- 
\\
$R^{0}_{c}$ &
  $0.1721\pm 0.0030$ & 
  $0.17247\pm 0.00002$ & 
  $0.17247\pm 0.00002$ & 
  $+0.1$ 
\\
$R^{0}_{b}$ &
  $0.21629\pm 0.00066$ & 
  $0.21492\pm 0.00003$ & 
  $0.21492\pm 0.00003$ & 
  $-2.1$ 
\\
\hline
\end{tabular}
\caption{Summary of experimental data and fit
  results in the SM, including the subleading two-loop fermionic EW
  corrections to $\rho_Z^f$ with the results of
  ref.~\cite{Freitas:2012sy,freitasprivate} and introducing the
  parameters $\delta\rho_Z^{\nu,\ell,b}$. 
  The values in the column ``Indirect'' are determined without using
  the corresponding experimental information. The last column shows
  the pulls in units of standard deviations evaluated from the
  p.d.f.'s of ``Data'' and ``Indirect'' as explained in
  ref.~\cite{Bona:2005vz}. For completeness we also report the fit
  result for $\delta \rho_Z^{u,d}$ computed from $\delta \rho_Z^{b}$
  using $\Gamma_{u,d}/\Gamma_b$.}  
\label{tab:SMfit}
\end{table}

In the third column of table~\ref{tab:SMfit} we present the results
of the SM fit obtained using the top pole mass and the
expressions for $\Gamma_u/\Gamma_b$ and
$\Gamma_d/\Gamma_b$ from refs.~\cite{Freitas:2012sy,freitasprivate}. 
As discussed above, in this case we do not have enough information to
compute $\Gamma_Z$, $\sigma^0_h$ and $R^0_\ell$ at the same level of
accuracy of $R^0_b$ and $R^0_c$. We therefore add three free
parameters to the fit, representing the fermionic two-loop corrections
$\delta \rho^\nu_Z$, $\delta \rho^\ell_Z$ and $\delta \rho^b_Z$. These
parameters affect only the observables $\Gamma_Z$, $\sigma^0_h$ and
$R^0_\ell$, since we have
\begin{equation}
  \label{eq:Freitasrb}
  R^0_b = \frac{\Gamma_b}{\Gamma_h} 
      = \frac{1}{\displaystyle 1+2 \left(
      \frac{\Gamma_u}{\Gamma_b} +
      \frac{\Gamma_d}{\Gamma_b}\right)}\,,\qquad
  R^0_c = \frac{\Gamma_c}{\Gamma_h} =
  \frac{\displaystyle \frac{\Gamma_u}{\Gamma_b}}
      {\displaystyle 1+2 \left( 
      \frac{\Gamma_u}{\Gamma_b} +
      \frac{\Gamma_d}{\Gamma_b}\right)}\,,
\end{equation}
where we have used the approximation $\Gamma_u =\Gamma_c$ and
$\Gamma_d = \Gamma_s$. In this way, while we cannot predict
$\Gamma_Z$, $\sigma^0_h$ and $R^0_\ell$, we obtain a posterior for the
parameters $\delta \rho^\nu_Z$, $\delta \rho^\ell_Z$ and $\delta
\rho^b_Z$, which can be compared to the theoretical expressions once
they become available. The other parameters $\delta\rho^u_Z$ and
$\delta\rho^d_Z$ are determined from $\delta\rho^b_Z$ through
$\Gamma_u/\Gamma_b$ and $\Gamma_d/\Gamma_b$, respectively. 
Notice that fits performed using the
formula for $R^0_b$ from ref.~\cite{Freitas:2012sy} and the
formul{\ae} for $\rho_Z^f$ from ref.~\cite{Degrassi:1999jd} are
inconsistent, since the change in $R^0_b$ implies a change in
$R^0_{c,\ell}$, $\Gamma_Z$ and $\sigma^0_h$. Furthermore, the results of
ref.~\cite{Freitas:2012sy} imply much larger two-loop fermionic
corrections than expected from the expansion in
ref.~\cite{Degrassi:1999jd}. In fact, we can estimate the size of the
unknown two-loop corrections as follows:
\begin{align}
\delta\rho_Z^q - \delta\rho_Z^b
&\approx 
\frac{\Gamma_q/\Gamma_b - \Gamma'_q/\Gamma'_b}{\Gamma'_q/\Gamma'_b}
= \left\{
\begin{array}{ll}
4.8\times 10^{-3} & {\rm for}\ q=u\,,\\[1mm]
4.4\times 10^{-3} & {\rm for}\ q=d\,,
\end{array}
\right.
\end{align}
where $\Gamma_f$ ($\Gamma_f'$) denotes a partial width including
(omitting) the contribution from $\delta\rho_Z^f$, and the
approximation $\rho_Z^f\approx 1$ has been used. 
Since these
corrections are comparable in size to one-loop contributions, it would
be desirable to have an independent confirmation of the calculation of
ref.~\cite{Freitas:2012sy}.

\begin{figure}[tp]
  \centering
  \includegraphics[width=.35\textwidth]{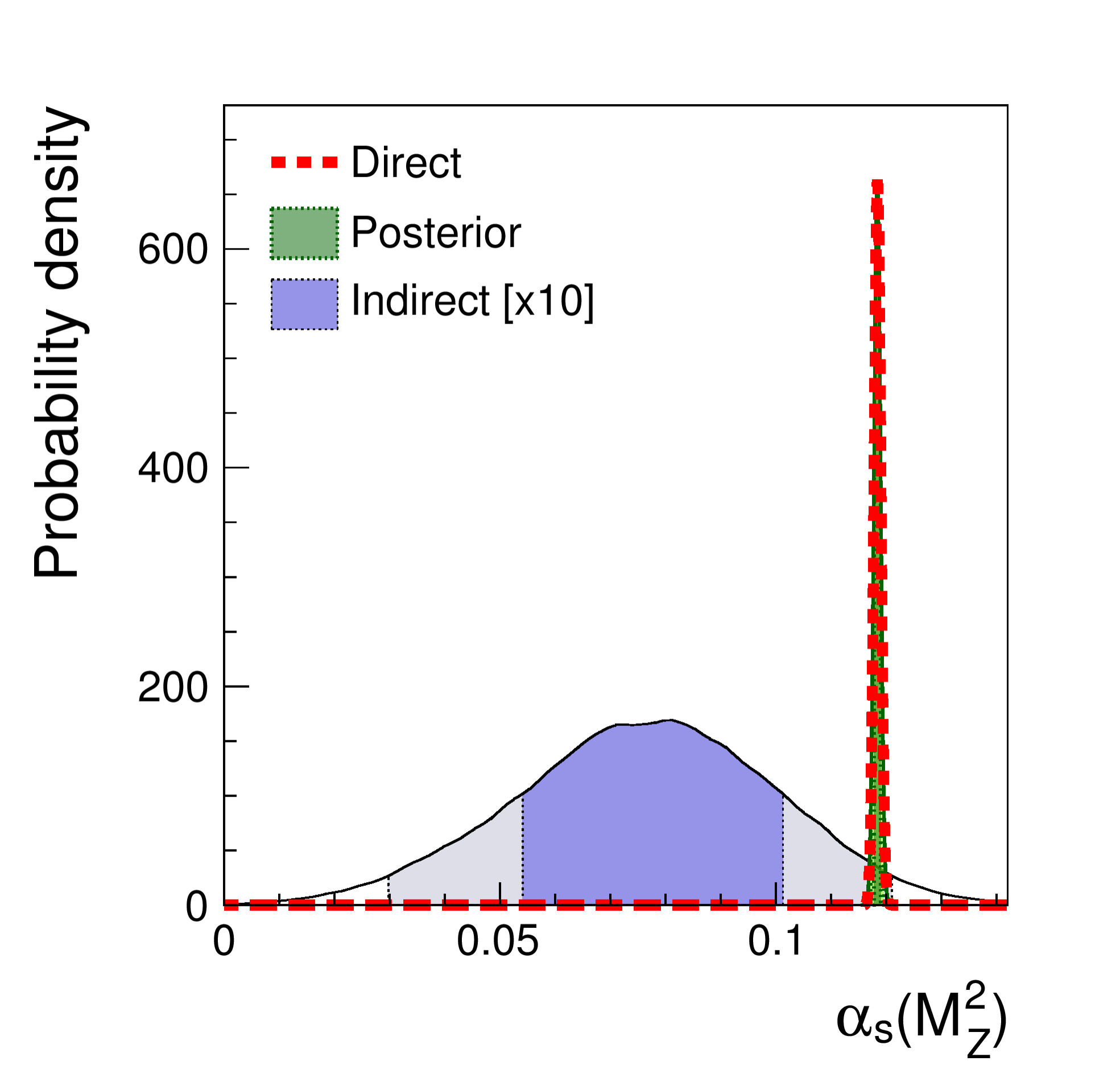}
  \hspace{-7mm}
  \includegraphics[width=.35\textwidth]{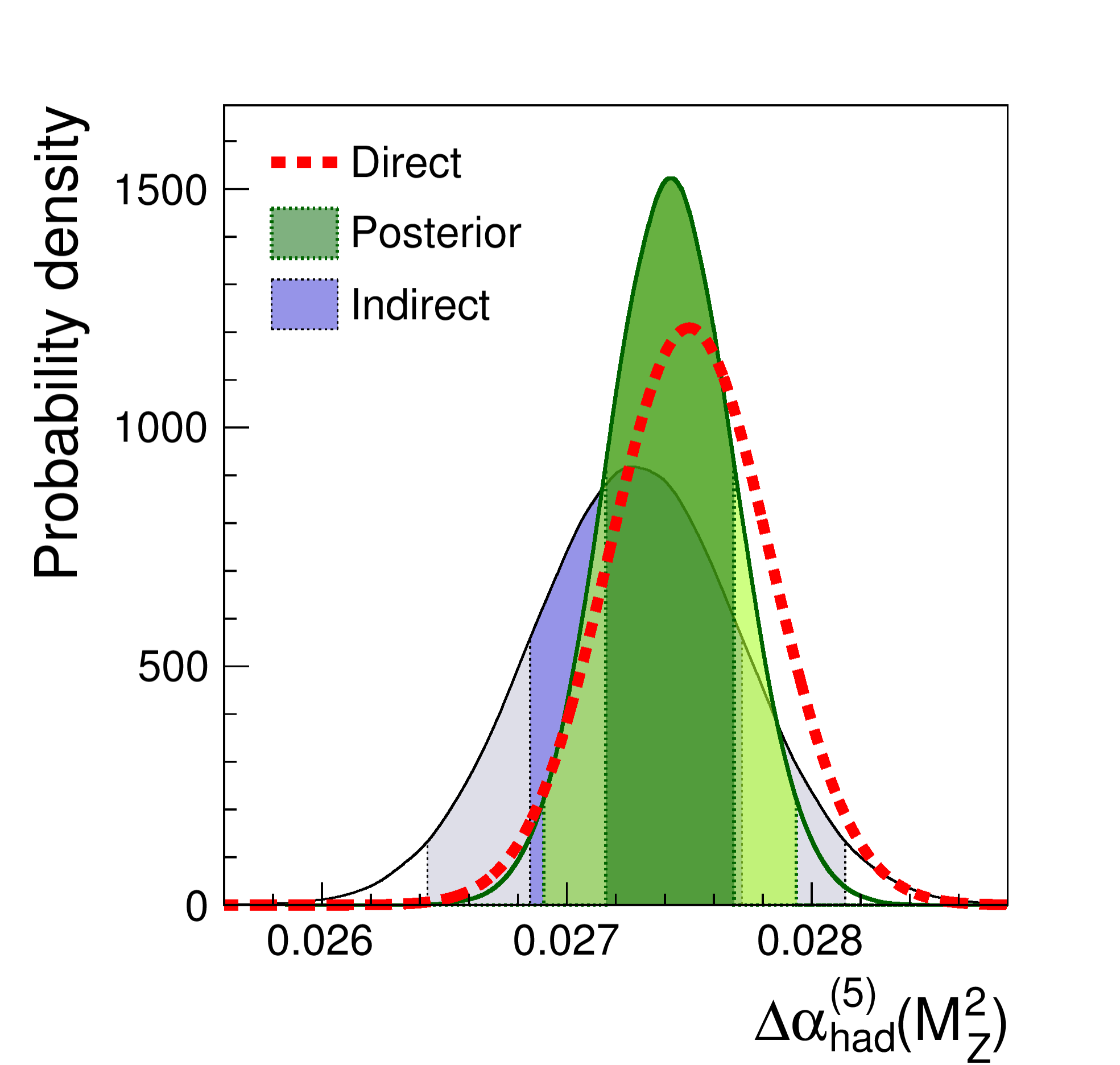}
  \hspace{-7mm}
  \includegraphics[width=.35\textwidth]{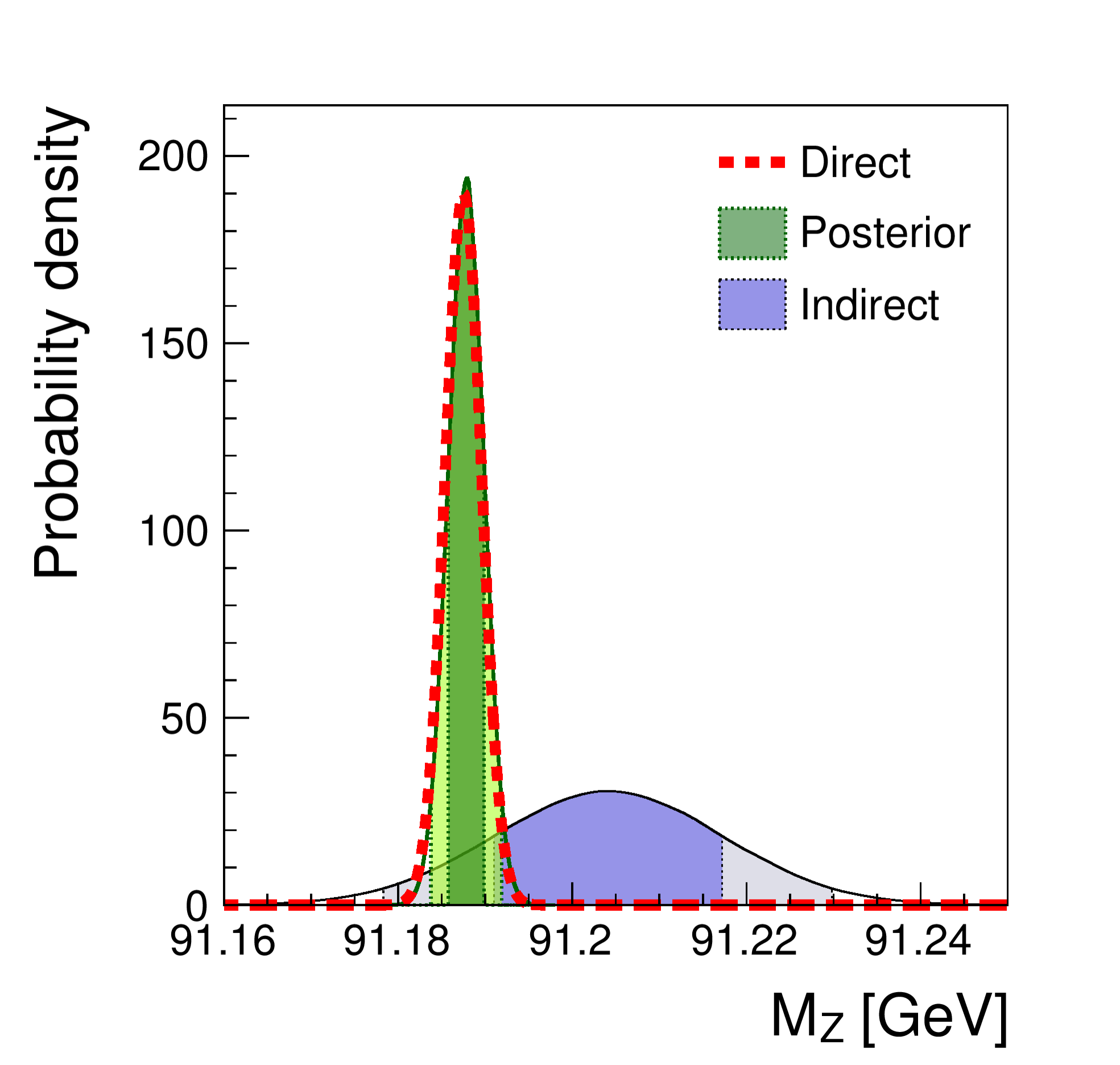}
  \\
  \includegraphics[width=.35\textwidth]{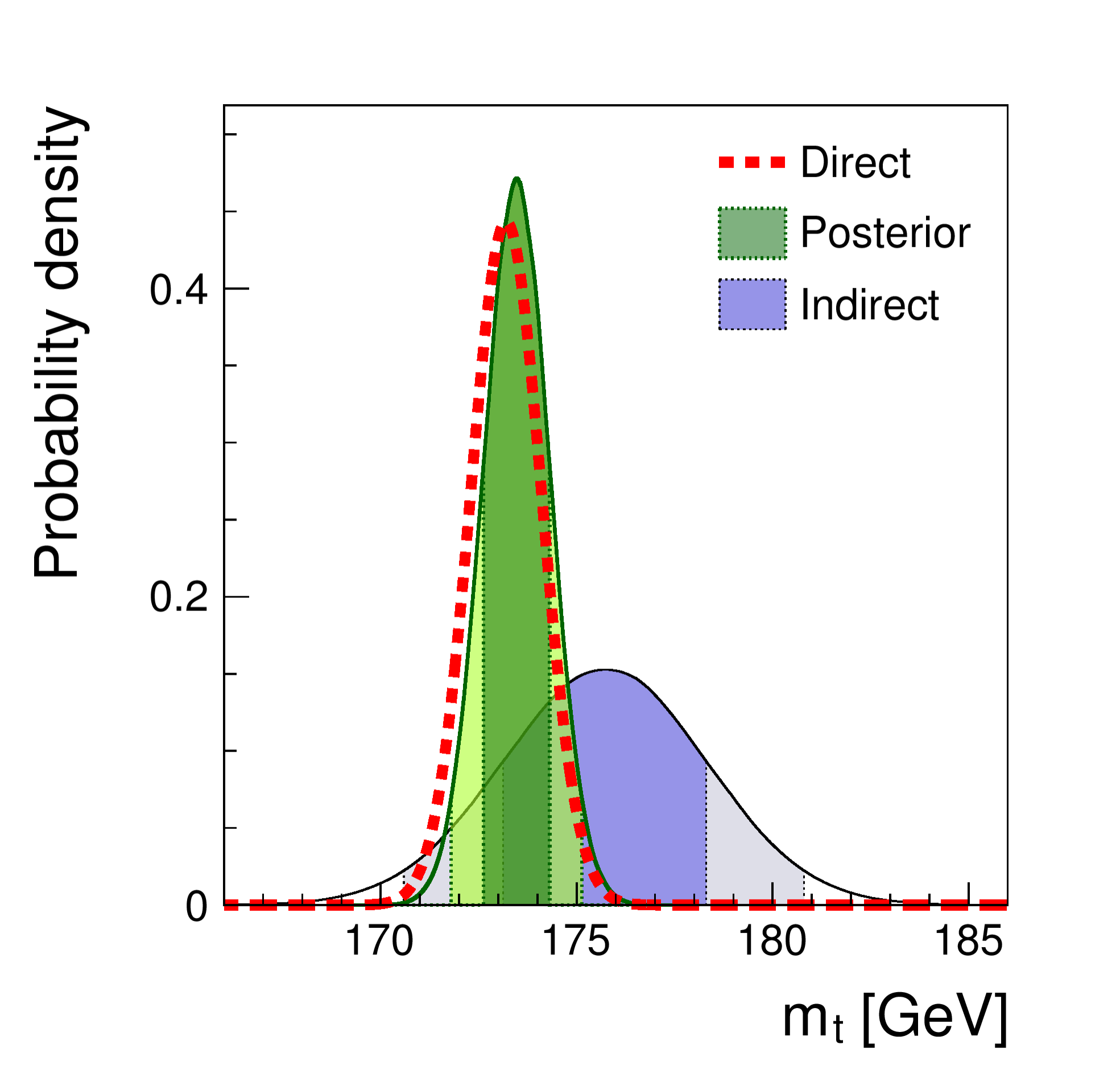}
  \hspace{-7mm}
  \includegraphics[width=.35\textwidth]{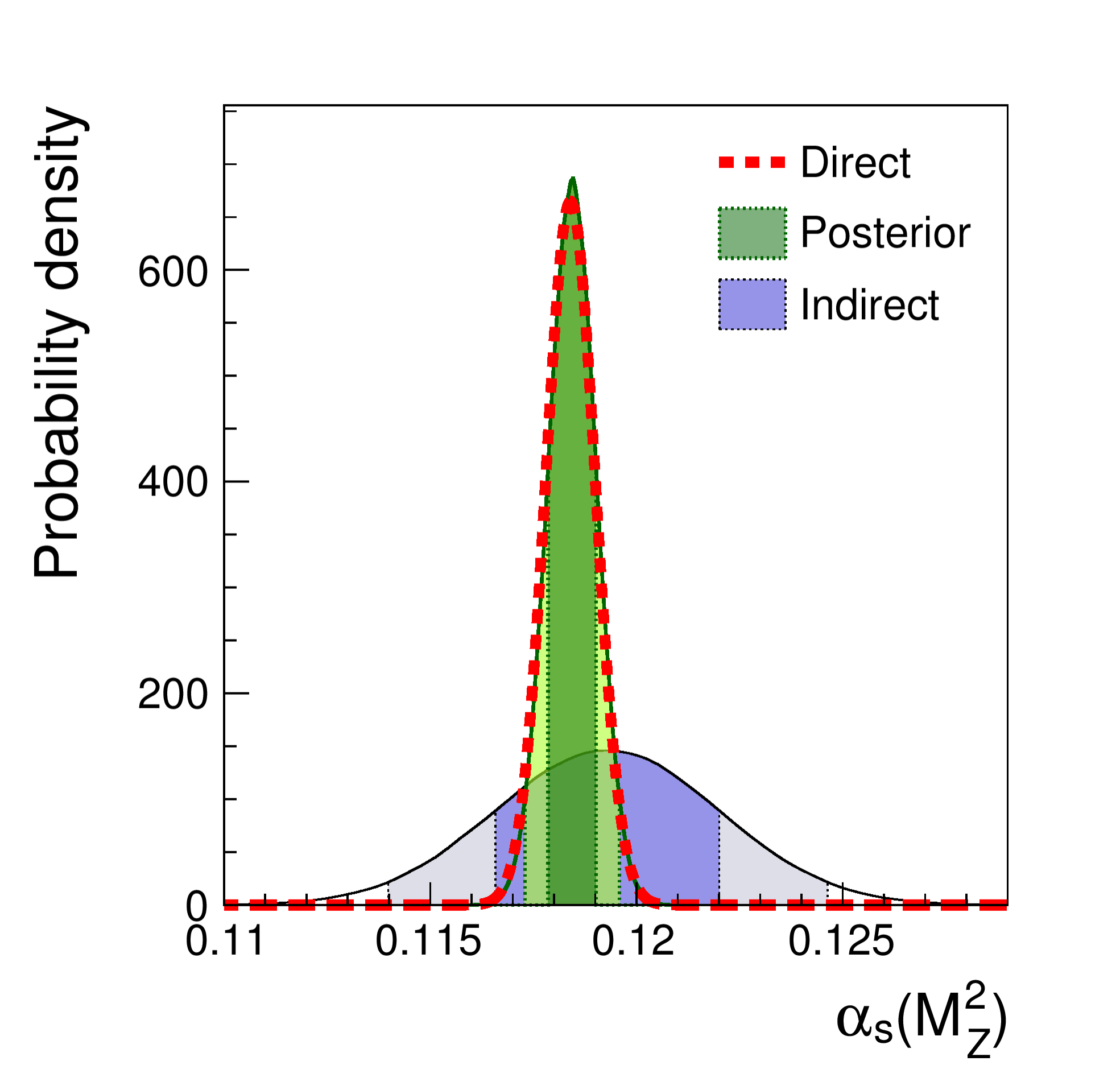}
  \hspace{-3mm}
  \includegraphics[width=.32\textwidth]{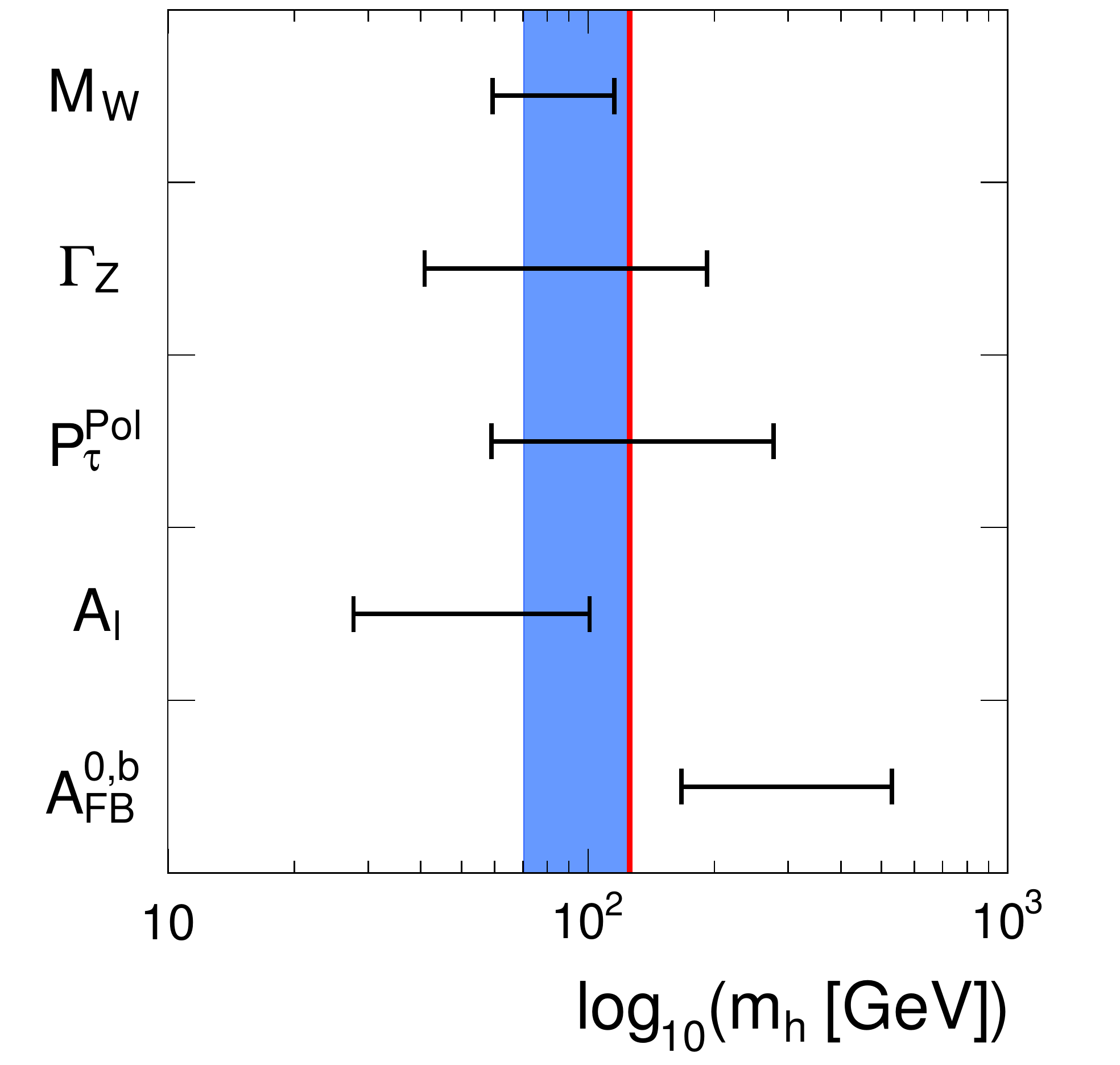}
  \caption{Comparisons between the direct measurement and the
    posterior probability distributions for the input parameters in
    the SM fit, together with their indirect determinations from the
    EWPO measurements, obtained by assuming a flat prior for the
    single parameter under consideration. Using the results of
    ref.~\cite{Freitas:2012sy,freitasprivate} and introducing the
    parameters $\delta \rho_Z^{\nu,\ell,b}$, the subleading two-loop
    fermionic EW corrections to $\rho_Z^f$ have been taken into
    account in the plots, except for the bottom-centre and
    bottom-right plots, in which the corrections have been
    omitted. Here and in the following, the dark (light) regions
    correspond to $68\%$ ($95\%$) probability. In the bottom-right
    plot, we report the indirect determinations of the Higgs mass
    excluding the observables $M_W$, $\Gamma_Z$, $P_\tau^{\rm pol}$,
    $\mathcal{A}_l$ and $A_{\rm FB}^{0,b}$, except for the one specified in
    each row. The vertical blue (red) band represents the one obtained
    from the the fit with all the observables (from the direct
    measurement). We assume a flat prior for the Higgs mass ranging
    from 10 MeV to 1 TeV.}
  \label{fig:SMinputs}
\end{figure}

From the fit we also obtain posteriors for the SM parameters 
$\alpha_s(M_Z^2)$, $\Delta\alpha_{\rm had}^{(5)}(M_Z^2)$, $M_Z$, $m_t$ 
and $m_h$ (see table~\ref{tab:SMfit}). As can be seen in
figure~\ref{fig:SMinputs}, while the posteriors are dominated by the
experimental input (as desirable for fit input parameters), the fit
would provide an indirect determination with a compatible result and a
remarkable accuracy (with the well-known exception of the Higgs mass
which is poorly indirectly determined).\footnote{Actually the indirect
determination of $\alpha_s(M_Z^2)$ is not very precise when we use
the results of ref.~\cite{Freitas:2012sy}, due to the uncertainty related
to $\delta \rho^\nu_Z$, $\delta \rho^\ell_Z$ and $\delta
\rho^b_Z$. This can be seen by comparing the first and the next-to-last
plots in figure~\ref{fig:SMinputs}.}
The correlation matrix for the
posteriors is given in table~\ref{tab:SMCorr_newRb}.

To show the impact on the fit of the new calculation of
$R_b^0$~\cite{Freitas:2012sy}, we present in
table~\ref{tab:SMfit_oldRb} the results obtained using instead
refs.~\cite{Barbieri:1992nz,Barbieri:1992dq,Fleischer:1993ub,Fleischer:1994cb,Degrassi:1996mg,Degrassi:1996ps,Degrassi:1999jd}
for the leading and next-to-leading 
terms in the large-$m_t$ expansion for two-loop
fermionic EW corrections to $\rho_Z^f$. The corresponding 
correlation matrix for the posteriors is given in
table~\ref{tab:SMCorr_oldRb}. As can be seen by comparing with
the full results, the tension in $R^0_b$ is reduced. The predictions
for EWPO are reported in table~\ref{tab:SMpred}. Notice that the
indirect determination of $\alpha_s(M_Z^2)$ 
is much more precise in this case since we are not considering large
unknown fermionic corrections to $\rho_Z^f$ (see the bottom center plot
in figure~\ref{fig:SMinputs}).

\begin{table}[tp]
\centering
\begin{tabular}{lcccc} 
\hline
& Data & Fit & Indirect & Pull \\
\hline
$\alpha_s(M_Z^2)$ &
  $0.1184\pm 0.0006$ & 
  $0.1184\pm 0.0006$ & 
  $0.1193\pm 0.0027$ & 
  $+0.3$ 
\\
$\Delta\alpha_{\rm had}^{(5)}(M_Z^2)$ &
  $0.02750\pm 0.00033$ & 
  $0.02740\pm 0.00026$ & 
  $0.02725\pm 0.00042$ & 
  $-0.5$ 
\\
$M_Z$ [GeV] &
  $91.1875\pm 0.0021$ & 
  $91.1878\pm 0.0021$ & 
  $91.197\pm 0.012$ & 
  $+0.8$ 
\\
$m_t$ [GeV] &
  $173.2\pm 0.9$ & 
  $173.5\pm 0.8$ & 
  $176.3\pm 2.5$ & 
  $+1.1$ 
\\
$m_h$ [GeV] &
  $125.6\pm 0.3$ & 
  $125.6\pm 0.3$ & 
  $97.3\pm 26.9$ & 
  $-0.9$ 
\\
\hline
$M_W$ [GeV] &
  $80.385\pm 0.015$ & 
  $80.367\pm 0.007$ & 
  $80.362\pm 0.007$ & 
  $-1.4$ 
\\
$\Gamma_W$ [GeV] &
  $2.085\pm 0.042$ & 
  $2.0891\pm 0.0006$ & 
  $2.0891\pm 0.0006$ & 
  $+0.1$ 
\\
$\Gamma_{Z}$ [GeV] &
  $2.4952\pm 0.0023$ & 
  $2.4953\pm 0.0004$ & 
  $2.4953\pm 0.0004$ & 
  $+0.0$ 
\\
$\sigma_{h}^{0}$ [nb] &
  $41.540\pm 0.037$ & 
  $41.484\pm 0.004$ & 
  $41.484\pm 0.004$ & 
  $-1.5$ 
\\
$\sin^2\theta_{\rm eff}^{\rm lept}(Q_{\rm FB}^{\rm had})$ &
  $0.2324\pm 0.0012$ & 
  $0.23145\pm 0.00009$ & 
  $0.23144\pm 0.00009$ & 
  $-0.8$ 
\\
$P_\tau^{\rm pol}$ &
  $0.1465\pm 0.0033$ & 
  $0.1476\pm 0.0007$ & 
  $0.1477\pm 0.0007$ & 
  $+0.3$ 
\\
$\mathcal{A}_\ell$ (SLD) &
  $0.1513\pm 0.0021$ & 
  $0.1476\pm 0.0007$ & 
  $0.1471\pm 0.0008$ & 
  $-1.9$ 
\\
$\mathcal{A}_{c}$ &
  $0.670\pm 0.027$ & 
  $0.6682\pm 0.0003$ & 
  $0.6682\pm 0.0003$ & 
  $-0.1$ 
\\
$\mathcal{A}_{b}$ &
  $0.923\pm 0.020$ & 
  $0.93466\pm 0.00006$ & 
  $0.93466\pm 0.00006$ & 
  $+0.6$ 
\\
$A_{\rm FB}^{0,\ell}$ &
  $0.0171\pm 0.0010$ & 
  $0.0163\pm 0.0002$ & 
  $0.0163\pm 0.0002$ & 
  $-0.8$ 
\\
$A_{\rm FB}^{0,c}$ &
  $0.0707\pm 0.0035$ & 
  $0.0740\pm 0.0004$ & 
  $0.0740\pm 0.0004$ & 
  $+0.9$ 
\\
$A_{\rm FB}^{0,b}$ &
  $0.0992\pm 0.0016$ & 
  $0.1035\pm 0.0005$ & 
  $0.1039\pm 0.0005$ & 
  $+2.8$ 
\\
$R^{0}_{\ell}$ &
  $20.767\pm 0.025$ & 
  $20.735\pm 0.004$ & 
  $20.734\pm 0.004$ & 
  $-1.3$ 
\\
$R^{0}_{c}$ &
  $0.1721\pm 0.0030$ & 
  $0.17223\pm 0.00002$ & 
  $0.17223\pm 0.00002$ & 
  $+0.0$ 
\\
$R^{0}_{b}$ &
  $0.21629\pm 0.00066$ & 
  $0.21575\pm 0.00003$ & 
  $0.21575\pm 0.00003$ & 
  $-0.8$ 
\\
\hline
\end{tabular}
\caption{Same as table~\ref{tab:SMfit}, but using the large-$m_t$
  expansion for the two-loop fermionic EW corrections to $\rho_Z^f$.}
\label{tab:SMfit_oldRb}
\end{table}

Using as SM input $m_t=173.3\pm 2.8$ GeV obtained from the
$\overline{\mathrm{MS}}$ mass instead of the Tevatron pole mass
average, one obtains the posterior $m_t = 174.6 \pm 1.9$ ($174.9\pm
1.9$) GeV using the results of
ref.~\cite{Freitas:2012sy,freitasprivate} (using the large-$m_t$
expansion), on the upper end of the Tevatron result. Concerning the
EWPO fit, the main observables affected by the change in $m_t$ are
$M_W$, $\Gamma_W$ and $R_b^0$, for which we obtain $M_W = 80.371\pm
0.011$ ($80.373\pm 0.010$) GeV, $\Gamma_W = 2.0894\pm 0.0009$
($2.0896\pm 0.0008$) GeV and $R_b^0=0.21488\pm 0.00007$ ($0.21570\pm
0.00007$).

\begin{table}[tp]
\centering
\begin{tabular}{lcllll} 
\hline
& Prediction &
\quad\ $\alpha_s$ & 
\ $\Delta\alpha_{\rm had}^{(5)}$ & 
\quad $M_Z$ & 
\quad $m_t$ 
\\
\hline
$M_W$ [GeV] & $80.362\pm 0.008$
& $\pm 0.000 $ 
& $\pm 0.006 $ 
& $\pm 0.003 $ 
& $\pm 0.005 $ 
\\
$\Gamma_W$ [GeV] & $2.0888\pm 0.0007$
& $\pm 0.0002 $ 
& $\pm 0.0005 $ 
& $\pm 0.0002 $ 
& $\pm 0.0004 $ 
\\
$\Gamma_{Z}$ [GeV] & $2.4951\pm 0.0005$
& $\pm 0.0003 $ 
& $\pm 0.0003 $ 
& $\pm 0.0002 $ 
& $\pm 0.0002 $ 
\\
$\sigma_{h}^{0}$ [nb] & $41.484\pm 0.004$
& $\pm 0.003 $ 
& $\pm 0.000 $ 
& $\pm 0.002 $ 
& $\pm 0.001 $ 
\\
$\sin^2\theta_{\rm eff}^{\rm lept}(Q_{\rm FB}^{\rm had})$ & $0.23149\pm 0.00012$
& $\pm 0.00000 $ 
& $\pm 0.00012 $ 
& $\pm 0.00001 $ 
& $\pm 0.00003 $ 
\\
$P_\tau^{\rm pol}=\mathcal{A}_\ell$ & $0.1472\pm 0.0009$
& $\pm 0.0000 $ 
& $\pm 0.0009 $ 
& $\pm 0.0001 $ 
& $\pm 0.0002 $ 
\\
$\mathcal{A}_{c}$ & $0.6680\pm 0.0004$
& $\pm 0.0000 $ 
& $\pm 0.0004 $ 
& $\pm 0.0001 $ 
& $\pm 0.0001 $ 
\\
$\mathcal{A}_{b}$ & $0.93464\pm 0.00008$
& $\pm 0.00000 $ 
& $\pm 0.00007 $ 
& $\pm 0.00001 $ 
& $\pm 0.00001 $ 
\\
$A_{\rm FB}^{0,\ell}$ & $0.0163\pm 0.0002$
& $\pm 0.0000 $ 
& $\pm 0.0002 $ 
& $\pm 0.0000 $ 
& $\pm 0.0000 $ 
\\
$A_{\rm FB}^{0,c}$ & $0.0738\pm 0.0005$
& $\pm 0.0000 $ 
& $\pm 0.0005 $ 
& $\pm 0.0001 $ 
& $\pm 0.0001 $ 
\\
$A_{\rm FB}^{0,b}$ & $0.1032\pm 0.0007$
& $\pm 0.0000 $ 
& $\pm 0.0006 $ 
& $\pm 0.0001 $ 
& $\pm 0.0002 $ 
\\
$R^{0}_{\ell}$ & $20.734\pm 0.004$
& $\pm 0.004 $ 
& $\pm 0.002 $ 
& $\pm 0.000 $ 
& $\pm 0.000 $ 
\\
$R^{0}_{c}$ & $0.17222\pm 0.00002$
& $\pm 0.00001 $ 
& $\pm 0.00001 $ 
& $\pm 0.00000 $ 
& $\pm 0.00001 $ 
\\
$R^{0}_{b}$ & $0.21576\pm 0.00003$
& $\pm 0.00000 $ 
& $\pm 0.00000 $ 
& $\pm 0.00001 $ 
& $\pm 0.00003 $ 
\\
\hline
$R^{0}_{c}$ & $0.17247\pm 0.00002$
& $\pm 0.00001 $ 
& $\pm 0.00001 $ 
& $\pm 0.00000 $ 
& $\pm 0.00001 $ 
\\
$R^{0}_{b}$ & $0.21493\pm 0.00004$
& $\pm 0.00001 $ 
& $\pm 0.00000 $ 
& $\pm 0.00000 $ 
& $\pm 0.00003 $ 
\\
\hline
\end{tabular}
\caption{SM predictions computed using the theoretical expressions for
  EWPO without the experimental constraints on the observables, and 
  individual uncertainties associated with each
  input parameter: $\alpha_s(M_Z^2)=0.1184\pm 0.0006$, 
  $\Delta\alpha_{\rm had}^{(5)}(M_Z^2) = 0.02750\pm 0.00033$, 
  $M_Z = 91.1875\pm 0.0021$ GeV and $m_t = 173.2\pm 0.9$ GeV, where
  the uncertainty associated to $m_h = 125.6\pm 0.3$ GeV is always
  negligible. The predictions are computed with the large-$m_t$
  expansion for the two-loop fermionic EW corrections to $\rho_Z^f$,
  except for $R^{0}_{c}$ and $R^{0}_{b}$ in the last two rows, which
  are computed with the results of
  ref.~\cite{Freitas:2012sy,freitasprivate}.} 
\label{tab:SMpred}
\end{table}

Let us now discuss the compatibility of the SM prediction with
experimental data. To this aim, we use the compatibility plots
introduced in ref.~\cite{Bona:2005vz}, where the
difference in standard deviations between the fit prediction and the
experimental result is given by the color coding. 

\begin{figure}[tp]
  \centering
  \hspace{-5mm}
  \includegraphics[width=.36\textwidth]{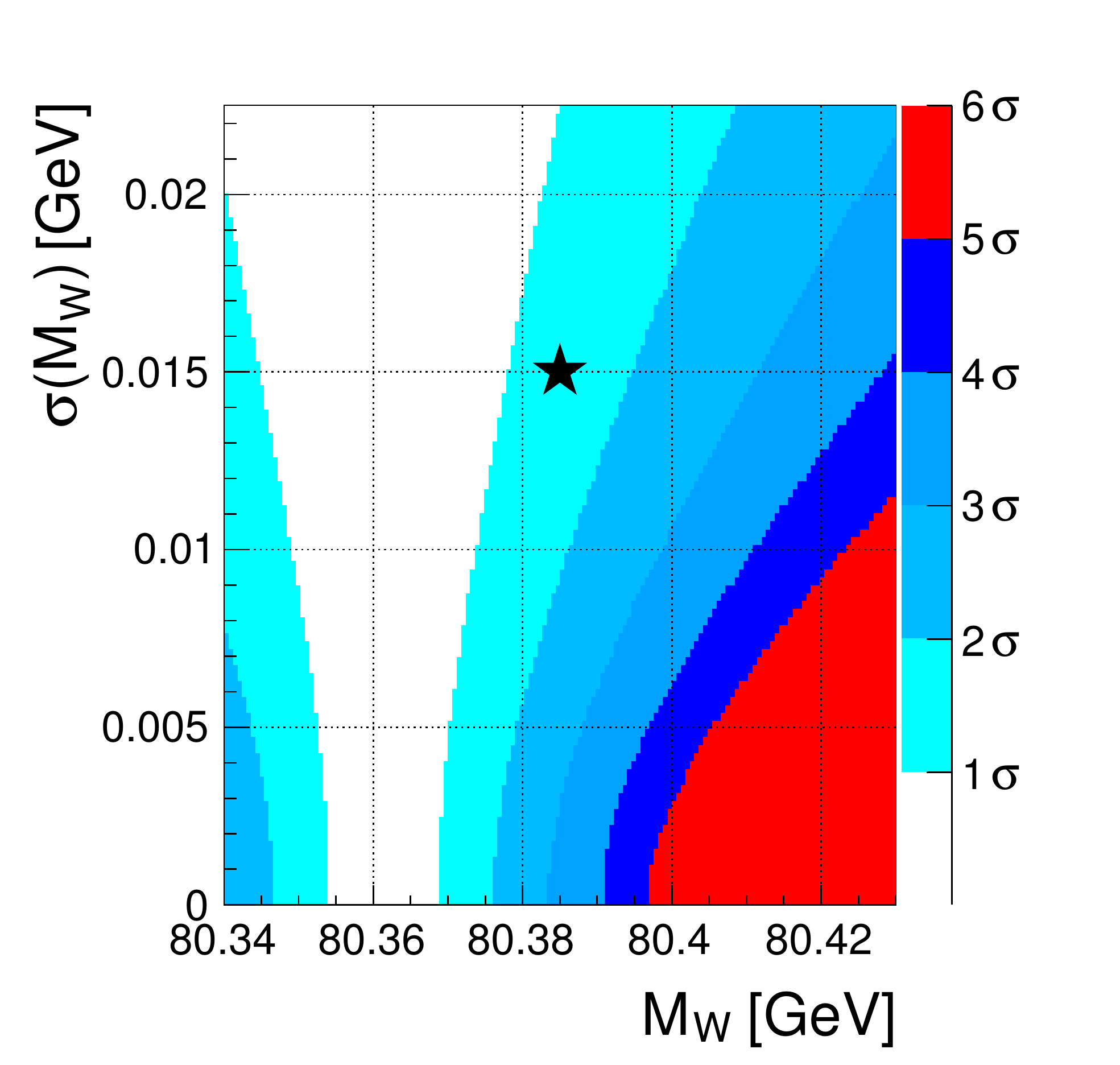}
  \hspace{-6mm}
  \includegraphics[width=.36\textwidth]{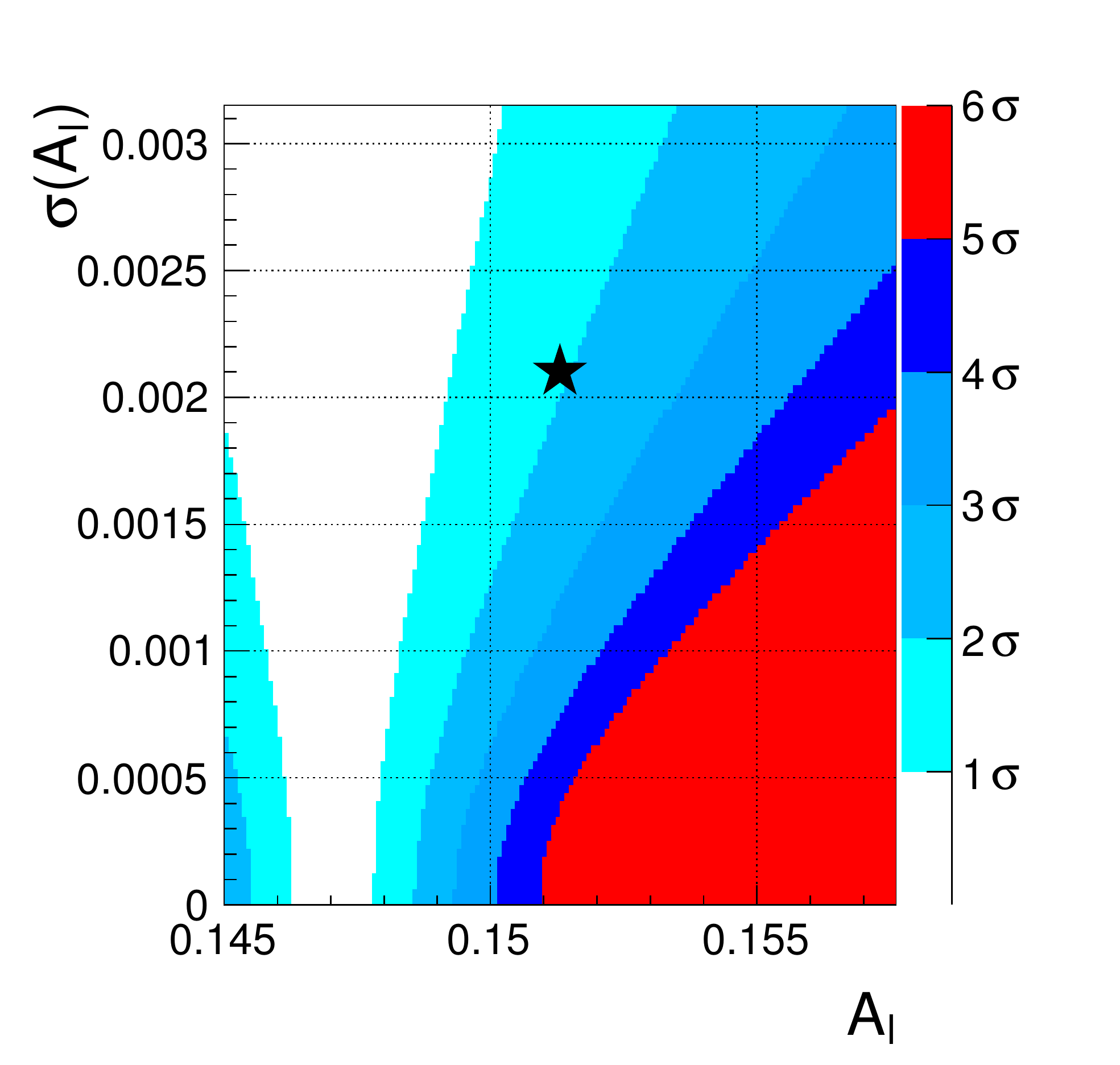}
  \hspace{-6mm}
  \includegraphics[width=.36\textwidth]{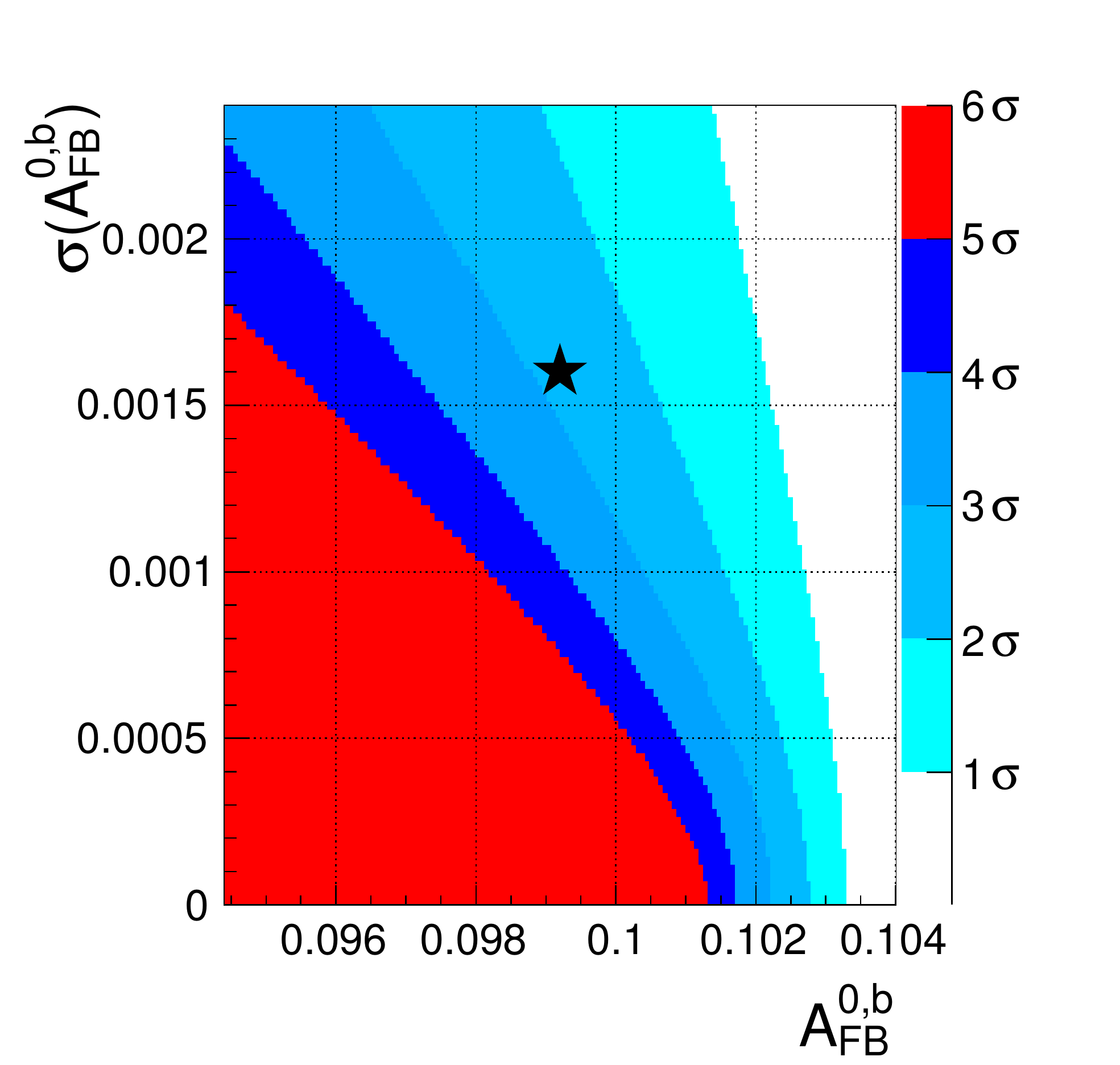}
  \hspace{-9mm}
  \caption{Compatibility plots of $M_W$, $\mathcal{A}_\ell$ and $A_{\rm
      FB}^{0,b}$. Any direct measurement corresponds to a point in the
    (central value, experimental error) plane, and its compatibility
    with the indirect determination is given in numbers of standard
    deviations by the color coding. The present experimental result is
    indicated by a star.}
  \label{fig:SMcompat}
\end{figure}

\begin{figure}[tp]
  \centering
  \includegraphics[width=.44\textwidth]{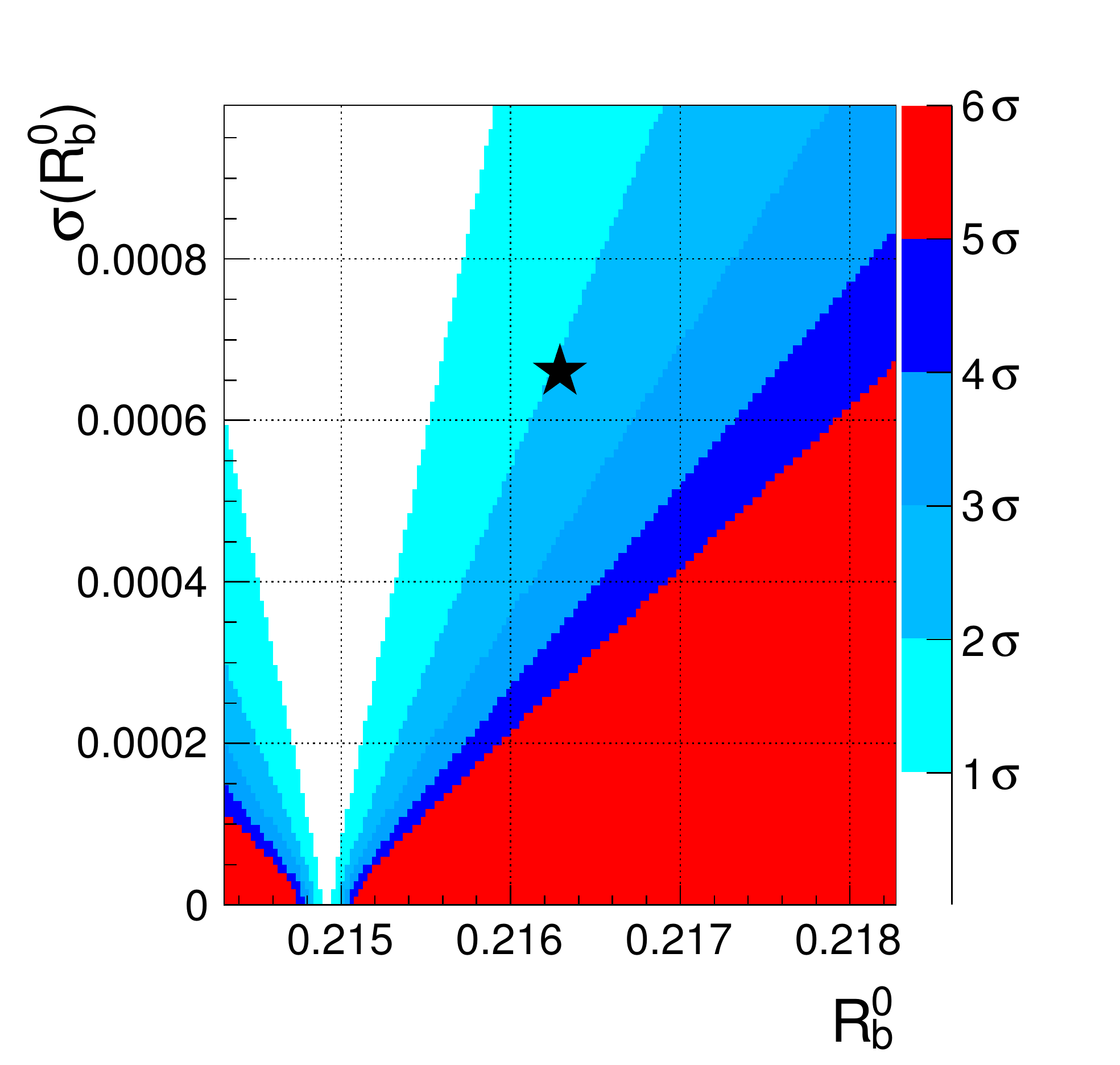}
  \qquad
  \includegraphics[width=.44\textwidth]{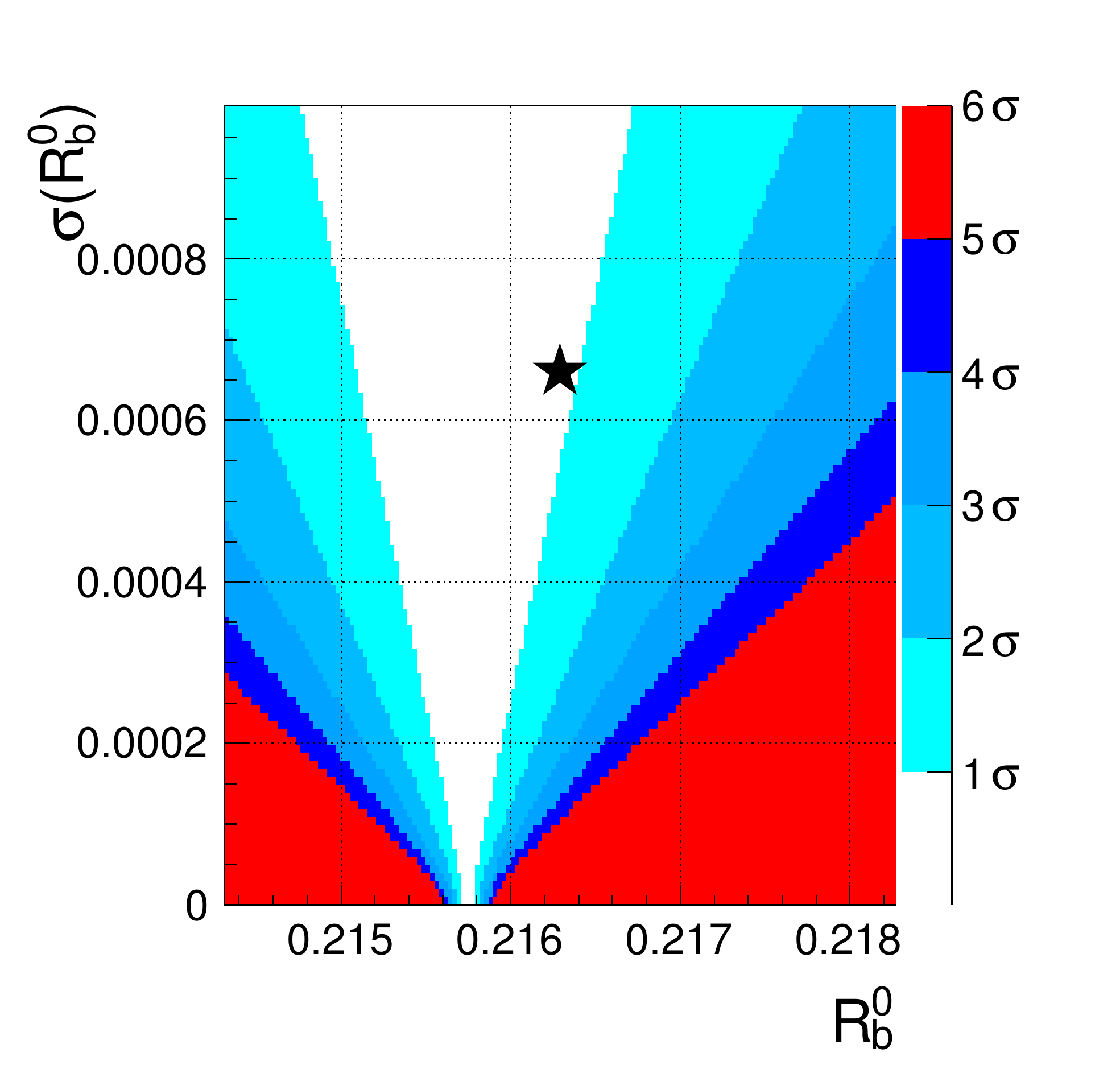}
  \caption{Compatibility plot of $R_b^0$ computed using the results of
    ref.~\cite{Freitas:2012sy} (left) or the large $m_t$ expansion for
    the two-loop fermionic EW corrections to $\rho_Z^f$ (right).}
  \label{fig:SMcompatRb}
\end{figure}

The compatibility of $M_W$, $\mathcal{A}_\ell$ and $A_{\rm FB}^{0,b}$ is
shown in figure~\ref{fig:SMcompat}. While these results are stable
against the inclusion of the recently calculated two-loop fermionic
corrections to $R^0_b$, the compatibility of $R_b^0$ is worsened by
the inclusion of the results in ref.~\cite{Freitas:2012sy}, as can be
seen by comparing the plots in 
figure~\ref{fig:SMcompatRb}.

In the bottom-right plot in figure~\ref{fig:SMinputs} we report the
indirect determinations of the Higgs mass obtained considering the
constraints from $M_W$, $\Gamma_Z$, $P_\tau^{\rm pol}$, $A_l^0$ and
$A_{\rm FB}^{0,b}$ one at a time, as well as the full fit result and
the direct measurement, omitting the results of
ref.~\cite{Freitas:2012sy}.

Our numerical results agree with those obtained using the ZFITTER
package
\cite{Bardin:1999yd,Bardin:1992jc,Arbuzov:2005ma,Akhundov:2013ons}. Our
fit results are compatible with the ones obtained by the LEP
Electroweak Working Group~\cite{ALEPH:2010aa} and also with the ones
in refs.~\cite{Erler:2012wz,Eberhardt:2012gv}. A comparison with the recent
Gfitter group fits~\cite{Baak:2012kk,Baak:2013ppa} is not
straightforward since the result for $R^0_b$ of
ref.~\cite{Freitas:2012sy} has been used without correspondingly
modifying other $\Gamma_q$-related observables and without accounting
for other possibly large fermionic two-loop corrections.

\section{Constraints on New Physics}
\label{sec:NP}

Let us now discuss the EW fit beyond the SM, using several widely
adopted model-independent parameterizations of NP
contributions. Before dwelling into the details of the different
analyses, a discussion on the inclusion of the results of
ref.~\cite{Freitas:2012sy,freitasprivate} is mandatory. In our SM fit
(see Section~\ref{sec:SM}), we parameterized the unknown two-loop
fermionic EW corrections to $\rho_Z^f$ with three free parameters. The
fit result selects values of these corrections that are as large as
the ones computed by Freitas and Huang, and much larger than naively
expected from the large-$m_t$ expansion. Waiting for a complete
calculation of these corrections, we cannot use consistently the
results of ref.~\cite{Freitas:2012sy,freitasprivate} in NP fits where
the use of $R^0_\ell$, $\Gamma_Z$ and $\sigma^0_h$ is
necessary to constrain NP contributions. Thus, in these cases we only
present results obtained using the large-$m_t$ expansion for the 
two-loop fermionic EW corrections to $\rho_Z^f$, while in
other cases we present results using both the large-$m_t$ expansion
and the expressions in ref.~\cite{Freitas:2012sy,freitasprivate},
leaving the choice of the preferred option to the reader. In the
latter case, we do not use the observables $\Gamma_Z$, $R^0_\ell$ and
$R^0_c$ in the fit. In all the NP fits reported below, the fit result
for SM parameters practically coincides with the input reported in
table~\ref{tab:SMfit}. 

\subsection{Constraints on the oblique parameters}
\label{sec:Oblique}

\begin{figure}[tp]
  \centering
  \hspace{-5mm}
  \includegraphics[width=.36\textwidth]{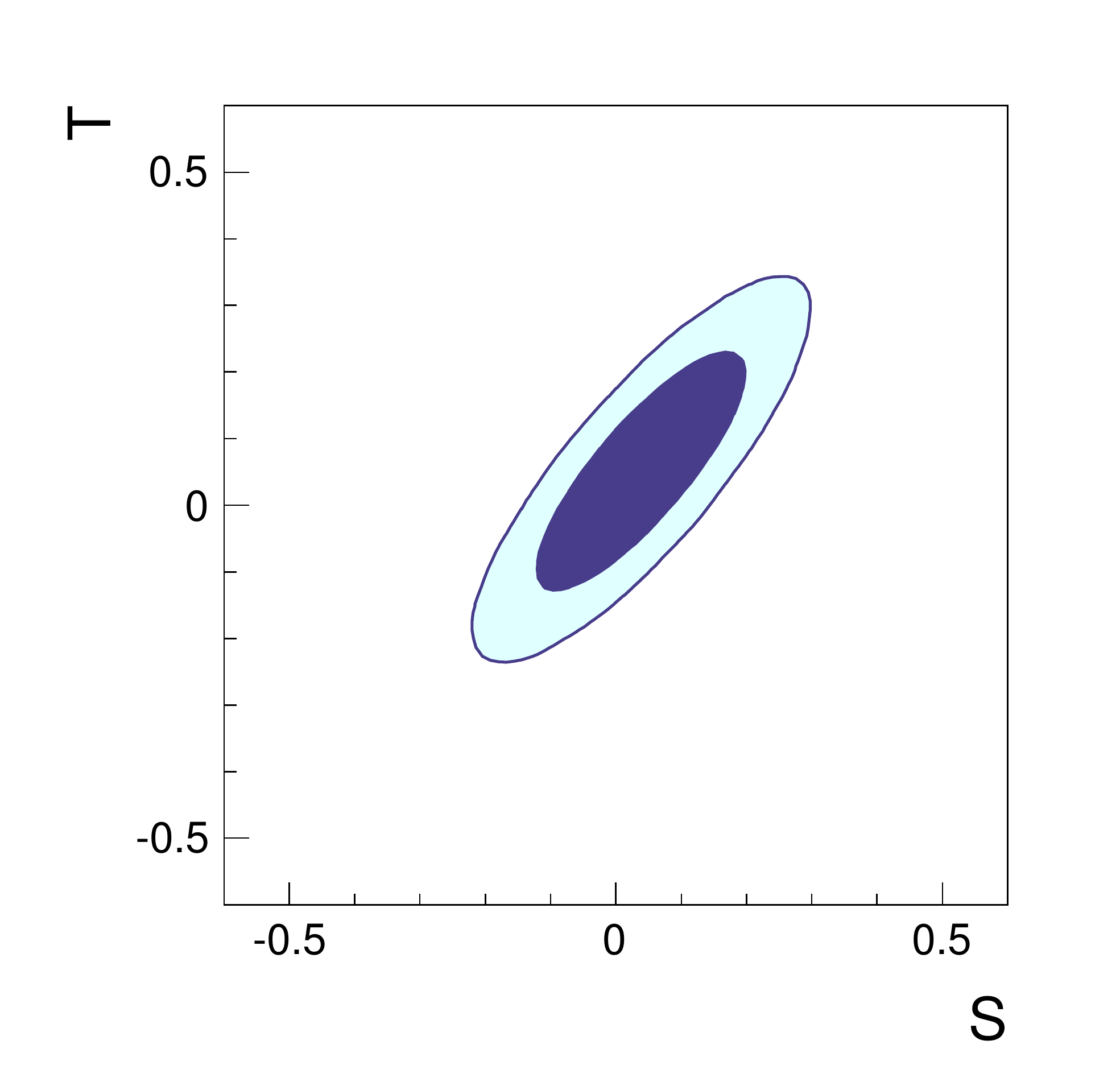}
  \hspace{-6mm}
  \includegraphics[width=.36\textwidth]{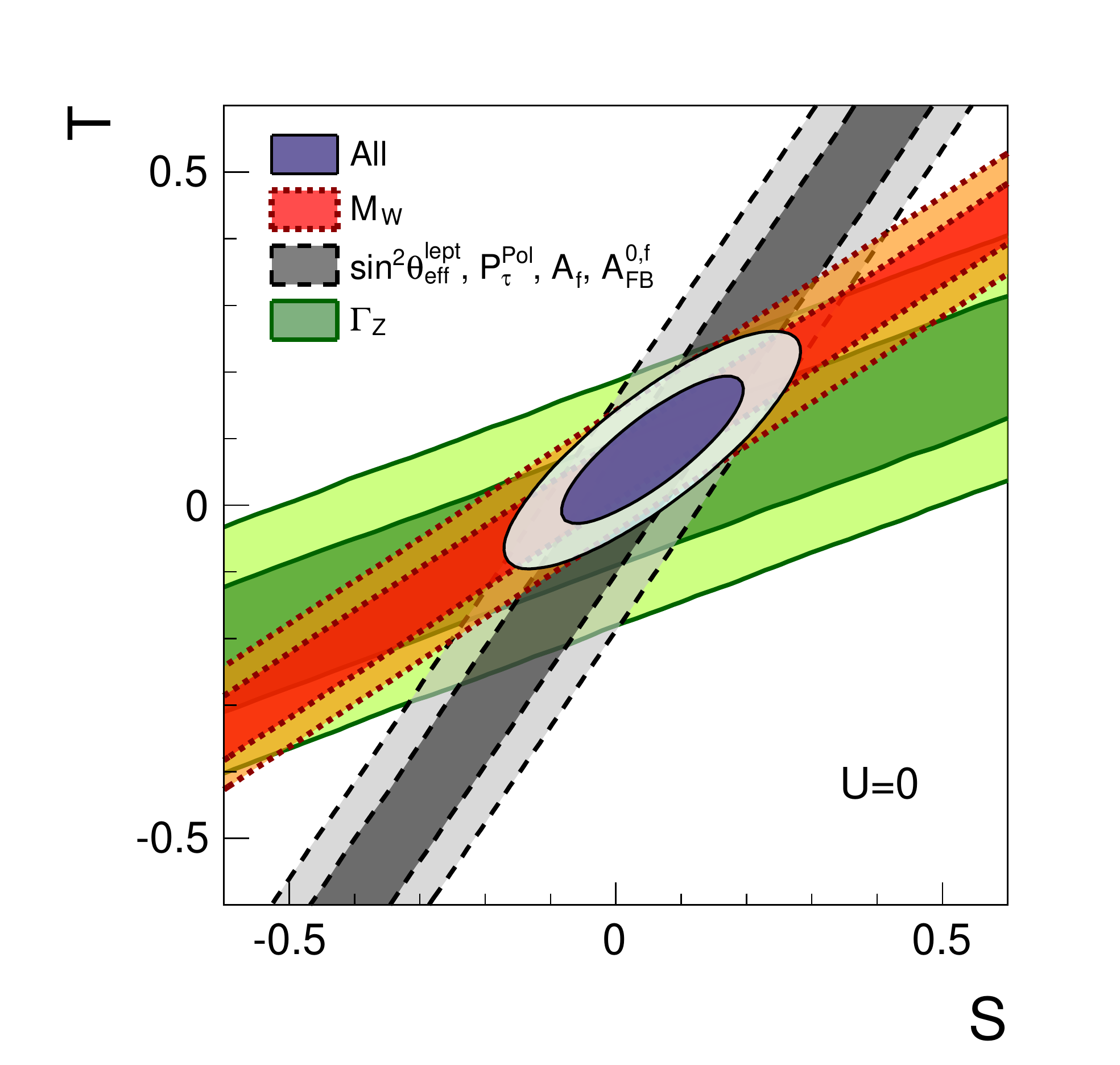}
  \hspace{-6mm}
  \includegraphics[width=.36\textwidth]{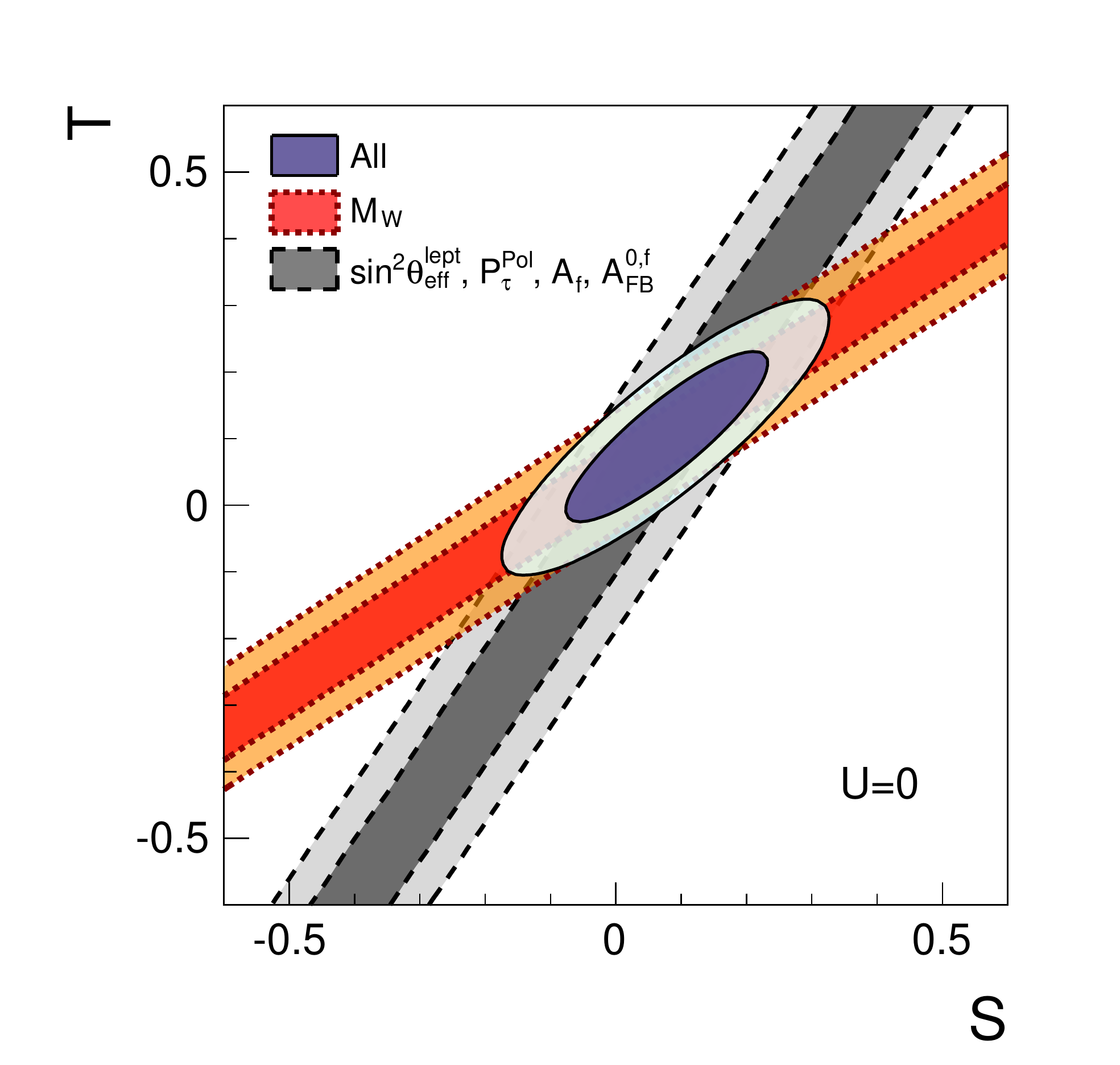}
  \hspace{-6mm}
  \caption{Left: Two-dimensional probability distribution for the
    oblique parameters $S$ and $T$ obtained from the fit with $S$,
    $T$, $U$ and the SM parameters, with the large-$m_t$ expansion for
    the two-loop fermionic EW corrections to $\rho_Z^f$. 
    Center: Two-dimensional probability distribution for the oblique
    parameters $S$ and $T$ obtained from the fit with $S$, $T$ and the
    SM parameters with $U=0$, with the large-$m_t$ expansion for
    the two-loop fermionic EW corrections to $\rho_Z^f$. 
    The individual constraints from $M_W$, the
    asymmetry parameters $\sin^2\theta_{\rm eff}^{\rm lept}$,
    $P_\tau^{\rm pol}$, $A_f$ and $A_{\rm FB}^{0,f}$ with
    $f=\ell,c,b$, and $\Gamma_Z$ are also presented, corresponding to
    the combinations of parameters $A$, $B$ and $C$ in
    eq.~(\ref{eq:abc}). Right: Same as center, but using 
    the results of ref.~\cite{Freitas:2012sy,freitasprivate}. 
    In this case, the constraint from $\Gamma_Z$ cannot be used.}
  \label{fig:STU}
\end{figure}

In several NP scenarios, the dominant NP effects appear in the
gauge-boson vacuum-polarization corrections, called oblique
corrections~\cite{Kennedy:1988sn,Kennedy:1988rt}. If the NP
scale is sufficiently higher than the weak scale, the oblique
corrections are effectively described by the three independent
parameters $S$, $T$ and $U$~\cite{Peskin:1990zt,Peskin:1991sw}:
\begin{eqnarray}
S &=& 
-16\pi \Pi^{\mathrm{NP}\prime}_{30}(0)
= 16\pi
\left[\Pi^{\mathrm{NP}\prime}_{33}(0) - \Pi^{\mathrm{NP}\prime}_{3Q}(0)
\right],
\\
T &=& \frac{4\pi}{s_W^2c_W^2 M_Z^2}
\left[\Pi^{\mathrm{NP}}_{11}(0) - \Pi^{\mathrm{NP}}_{33}(0)
\right],
\\
U &=& 16\pi
\left[\Pi^{\mathrm{NP}\prime}_{11}(0) - \Pi^{\mathrm{NP}\prime}_{33}(0)
\right],
\end{eqnarray}
where $\Pi^{\rm NP}_{XY}$ with $X,Y=0,1,3,Q$ denotes NP
contribution to the vacuum polarization amplitude of the gauge bosons
defined, e.g., in ref.~\cite{Peskin:1991sw}, 
$\Pi'_{XY}(q^2) = d\Pi_{XY}(q^2)/dq^2$, 
and $s_W^2$ and $c_W^2$ represent their SM values. 
NP contributions to an observable, parameterized by the above oblique
parameters, add up to the SM contribution:
\begin{equation}
\mathcal{O} = \mathcal{O}_{\rm SM} + \mathcal{O}_{NP}(S,T,U)\,,
\end{equation} 
where $S=T=U=0$ in the SM, and we linearize the NP contribution in
terms of the oblique
parameters~\cite{Peskin:1990zt,Peskin:1991sw,Maksymyk:1993zm,Burgess:1993mg,Burgess:1993vc}.
Explicit formul{\ae} for the observables are summarized in
Appendix~\ref{app:STUformulae}. Actually, all EWPO can be expressed in
terms of the following combinations of oblique parameters: 
\begin{eqnarray}
  \label{eq:abc}
  A &=& S - 2c_W^2\, T - \frac{(c_W^2-s_W^2)\,U}{2s_W^2}\,,
  \nonumber \\
  B &=& S - 4c_W^2 s_W^2\, T\,,\\
  C &=& -10(3-8s_W^2)\,S + (63-126s_W^2-40s_W^4)\,T\,. \nonumber
\end{eqnarray}
Note that the parameter $C$ describes the NP contribution to
$\Gamma_{Z}$, the parameter $A$ (the only one containing $U$)
describes the NP contribution to $M_W$ and $\Gamma_W$, and NP
contributions to all other EWPO are proportional to $B$. Clearly, for
$S$, $T$ and $U$ all different from zero, $\Gamma_{Z}$ is necessary to
obtain bounds on the NP parameters, so in this case we only use the
large-$m_t$ expansion. We fit the three oblique parameters together
with the SM parameters to the EW precision data in
table~\ref{tab:SMfit}. The fit results are summarized in the second 
column of table~\ref{tab:STU}, and the correlation matrix is given
in table~\ref{tab:STUcorr_oldRb}.  The two-dimensional probability
distribution for $S$ and $T$ is shown in the left plot of
figure~\ref{fig:STU}.  

\begin{table}[tp]
\centering
\begin{tabular}{c|cc|c} 
\hline
& \multicolumn{2}{c|}{Large-$m_t$ expansion}
&  Using
ref.~\cite{Freitas:2012sy,freitasprivate} 
\\
Parameter & 
$STU$ fit & $ST$ fit with $U=0$ &
$ST$ fit with $U=0$ 
\\
\hline
$S$ &
$0.04\pm 0.10$ & $0.06\pm 0.09$ &
$0.08\pm 0.10$ 
\\
$T$ &
$0.05\pm 0.12$ & $0.08\pm 0.07$ &
$0.10\pm 0.08$ 
\\
$U$ &
$0.03\pm 0.09$ & --- & 
---
\\
\hline
\end{tabular}
\caption{Fit results for the oblique parameters with floating $U$ or
  fixing $U=0$, using the large-$m_t$ expansion or with the results of 
  ref.~\cite{Freitas:2012sy,freitasprivate} for the two-loop
  fermionic EW corrections to $\rho_Z^f$. In the latter case, we do
  not consider constraints from $\Gamma_Z$, $\sigma^0_h$ and $R_\ell^0$.
}
\label{tab:STU}
\end{table}

If one fixes $U=0$, which is the case in many NP models where $U \ll
S,T$, the fit yields the results in the third (fourth) column of
table~\ref{tab:STU}, with correlation matrices given in
table~\ref{tab:STcorr_oldRb} (\ref{tab:STcorr_newRb}) omitting (using) the
formul{\ae} of ref.~\cite{Freitas:2012sy,freitasprivate}. The corresponding
two-dimensional distribution is given in the center and right plots in
figure~\ref{fig:STU}.  As expected, the results in the case $U=0$ do
not depend sizably on the choice made for the two-loop fermionic EW 
corrections.

\subsection{Constraints on the $\epsilon$ parameters}
\label{sec:Epsilon}

Aiming at a fully model-independent analysis of EWPO in the absence of
experimental information on the Higgs sector, Altarelli and Barbieri
introduced the parameters $\epsilon_1$, $\epsilon_2$ and
$\epsilon_3$~\cite{Altarelli:1990zd,Altarelli:1991fk}:
\begin{eqnarray}
\epsilon_1
&=&
\Delta\rho', 
\label{eq:eps1}
\\
\epsilon_2
&=&
c_0^2  \Delta\rho'
+ \frac{s_0^2}{c_0^2 - s_0^2} \Delta r_W
- 2 s_0^2 \Delta\kappa',
\label{eq:eps2}
\\
\epsilon_3
&=&
c_0^2\Delta\rho' 
+ (c_0^2-s_0^2)\Delta\kappa',
\label{eq:eps3}
\end{eqnarray}
where $\Delta r_W$, $\Delta\rho'$ and $\Delta\kappa'$ are defined
through the relations  
\begin{eqnarray}
s_W^2c_W^2 
&=&  
\frac{\pi\alpha(M_Z^2)}{\sqrt{2}\,G_\mu M_Z^2 (1-\Delta r_W)}\,,
\label{eq:Delta_r_W}
\\
\sqrt{{\rm Re}\,\rho_Z^e} 
&=&
1 + \frac{\Delta\rho'}{2}\,,
\label{eq:Delta_rho_Prime}
\\
\sin^2\theta_{\rm eff}^{e}
&=& 
(1+\Delta\kappa')\, s_0^2
\label{eq:Delta_kappa_Prime}
\end{eqnarray}
with 
\begin{equation}
s_0^2\, c_0^2 = \frac{\pi\alpha(M_Z^2)}{\sqrt{2}G_\mu M_Z^2}\,,
\end{equation}
and $c_0^2= 1 - s_0^2$. 
Unlike the oblique parameters $S$, $T$ and $U$ discussed in Section
\ref{sec:Oblique}, the $\epsilon$ parameters include the SM
contribution in addition to possible NP contributions. Moreover, they
involve not only oblique corrections, but also vertex corrections.
The $\epsilon$ parameters are defined in such a way that the
logarithmic corrections are separated from the large quadratic
corrections proportional to the top-quark mass.  The quadratic
corrections are then parameterized by $\epsilon_1$, while the other
corrections are included in $\epsilon_2$ and $\epsilon_3$.

In the SM, the $Z\to b\bar{b}$ vertex receives large corrections from
the top-quark loop, which can be parametrized by an additional
parameter $\epsilon_b$~\cite{Altarelli:1993sz}. However, given the
present experimental accuracy on EWPO, the flavour non-universal vertex
corrections in the SM have to be taken into account in all
channels. We define
\begin{eqnarray}
  \label{eq:rhofnonuniv}
  \rho_Z^f &=& \rho_Z^e + \Delta\rho_Z^f\,,\\
  \kappa_Z^f&=& \kappa_Z^e + \Delta\kappa_Z^f\nonumber
\end{eqnarray}
for $f\neq b$, and 
\begin{eqnarray}
  \label{eq:rhobnonuniv}
  \rho_Z^b&=& \left(\rho_Z^e + \Delta\rho_Z^b\right) (1 +
  \epsilon_b)^2\,, \\
  \kappa_Z^b&=& \frac{\kappa_Z^e + \Delta\kappa_Z^b}{1 +
  \epsilon_b}\,, \nonumber
\end{eqnarray}
where the non-universal corrections $\Delta\rho_Z^f$ and
$\Delta\kappa_Z^f$ are defined in Appendix~\ref{app:NonUnivCorr}. In
refs.~\cite{Altarelli:1994iz,Altarelli:1997et}, the relations between
the observables and the $\epsilon$ parameters are linearized.
However, in the case of the $W$-boson mass, the difference between the
values derived with and without the linearization is comparable in
size to the current experimental uncertainty. Therefore, we do not
employ any linearization in our analysis.

\begin{figure}[tp]
  \centering
  \includegraphics[width=.49\textwidth]{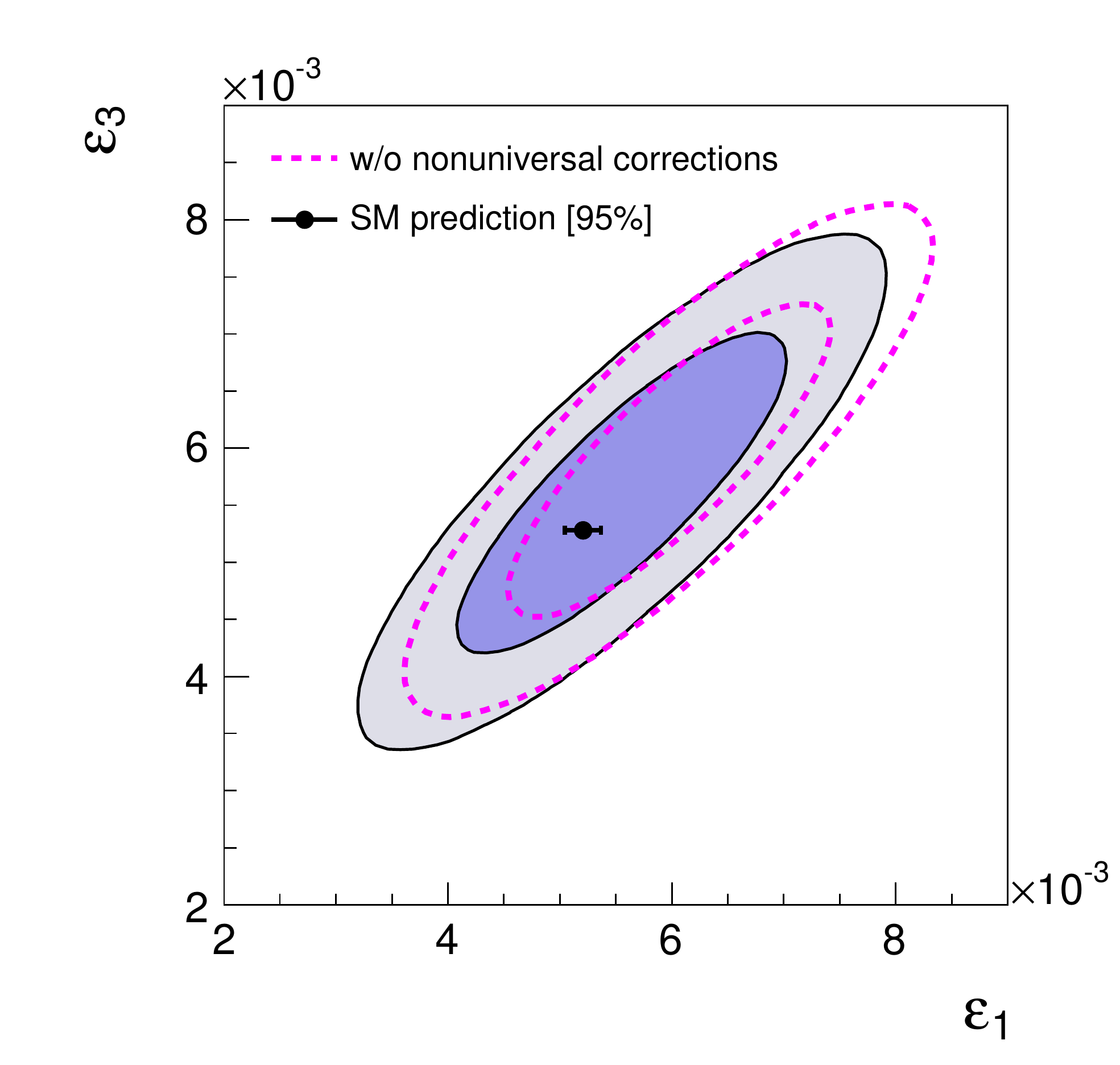}
  \hfill
  \includegraphics[width=.49\textwidth]{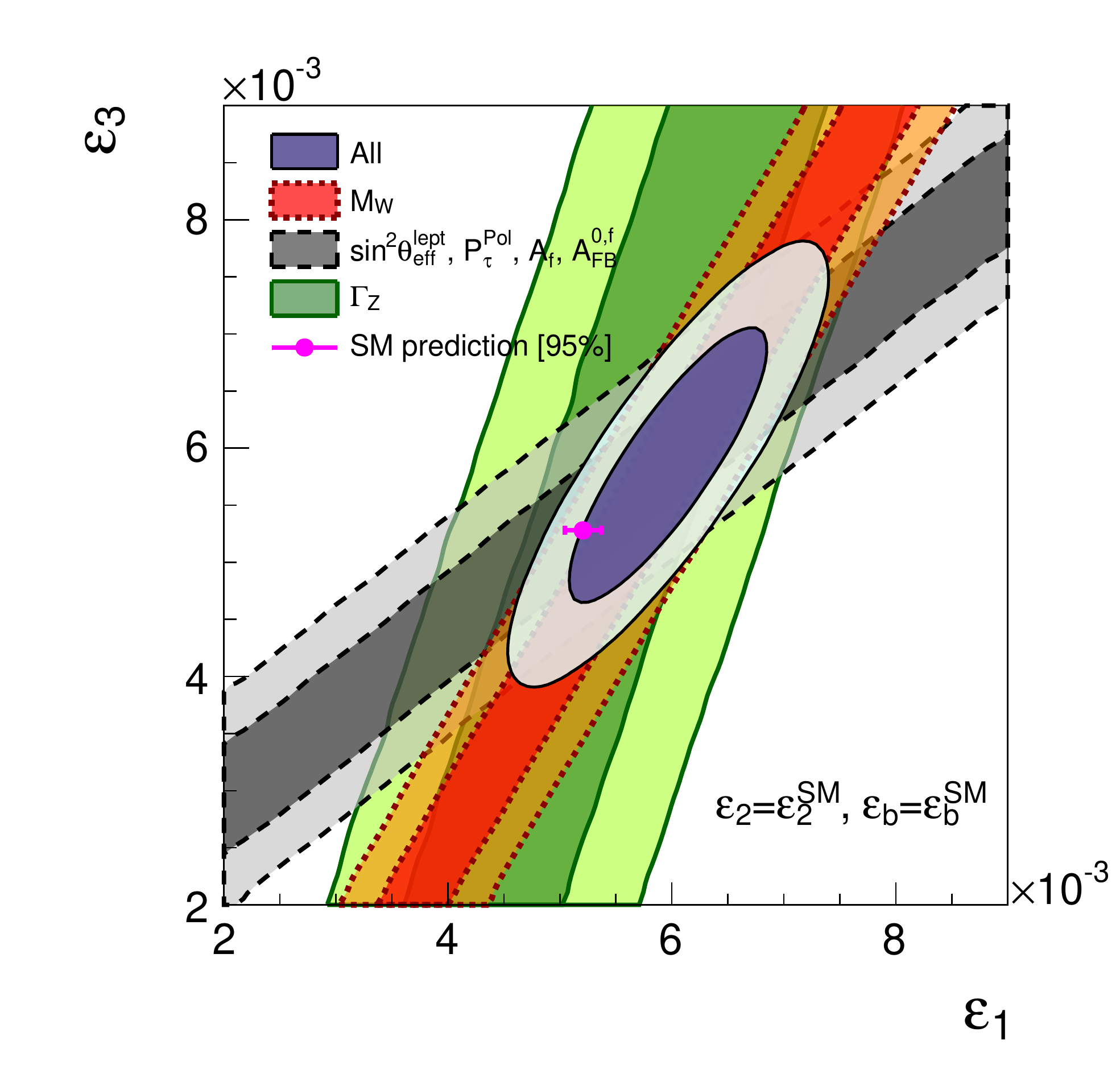}
  \caption{Two-dimensional probability distributions for $\epsilon_1$
    and $\epsilon_3$ in the fit, with floating $\epsilon_{1,2,3,b}$
    (left), or with assuming $\epsilon_2=\epsilon_2^{\rm SM}$ and
    $\epsilon_b=\epsilon_b^{\rm SM}$ (right). In the left plot, the effect of
    non-universal vertex corrections is presented. In the
    right plot, we also show the impact of different
    constraints. The SM prediction at 95\% is denoted by a point with
    an error bar.}
  \label{fig:EpsilonoldRb}
\end{figure}

We fit the four $\epsilon$ parameters together with the SM parameters
to the precision observables listed in table~\ref{tab:SMfit}, except
for $\Gamma_W$, which is not directly related to $\epsilon$'s. The fit
results are given in the second column of table~\ref{tab:Epsilon},  
and the corresponding correlation matrix is summarized in
table~\ref{tab:4epsCorr_oldRb}. 
Fixing $\epsilon_2=\epsilon_2^{\rm SM}$ and
$\epsilon_b=\epsilon_b^{\rm SM}$ in the fit, we obtain the results in
the third column of table~\ref{tab:Epsilon} 
with the correlation matrix in table~\ref{tab:2epsCorr_oldRb}.  The
two-dimensional probability distributions for $\epsilon_1$ and
$\epsilon_3$ in both fits are shown in
figure~\ref{fig:EpsilonoldRb}, where in the case of
$\epsilon_2=\epsilon_2^\mathrm{SM}$ and
$\epsilon_b=\epsilon_b^\mathrm{SM}$ we also plot the individual
constraints. To show the impact of including
non-universal vertex corrections, we also report in
figure~\ref{fig:EpsilonoldRb} the probability regions obtained
omitting these terms.

\begin{table}[tp]
\centering
\begin{tabular}{c|cc} 
\hline
& \multicolumn{2}{c}{Large-$m_t$ expansion}
\\
Parameter & 
$\epsilon_{1,2,3,b}$ fit &
$\epsilon_{1,3}$ fit 
\\
\hline
$\epsilon_1$ [$10^{-3}$] &
$\phantom{-}5.6\pm 1.0$ &
$6.0\pm 0.6$ 
\\ 
$\epsilon_2$ [$10^{-3}$] &
$-7.8\pm 0.9$ &
--- 
\\
$\epsilon_3$ [$10^{-3}$] & 
$\phantom{-}5.6\pm 0.9$ &
$5.9\pm 0.8$ 
\\
$\epsilon_b$ [$10^{-3}$] & 
$-5.8\pm 1.3$ &
--- 
\\
\hline
\end{tabular}
\caption{Fit results for the $\epsilon$ parameters, with floating
  $\epsilon_{1,2,3,b}$, or with assuming
  $\epsilon_2=\epsilon_2^{\rm SM}$ and 
  $\epsilon_b=\epsilon_b^{\rm SM}$. 
  The non-universal vertex corrections and the SM values for the
  $\epsilon$ parameters are computed with the large-$m_t$
  expansion for the two-loop fermionic EW corrections to $\rho_Z^f$. }
\label{tab:Epsilon}
\end{table}

The corresponding SM predictions for the $\epsilon$ parameters with 
the large-$m_t$ expansion for the two-loop fermionic EW corrections to
$\rho_Z^f$ are given by: 
\begin{eqnarray}
\epsilon_1^{\rm SM} &=& (5.21\pm 0.08)\, 10^{-3} 
\quad ([5.04,5.37]\, 10^{-3}\ @ 95\%\ \mathrm{prob.})\,,
\nonumber\\
\epsilon_2^{\rm SM} &=&-(7.37\pm 0.03)\, 10^{-3} 
\quad ([-7.43,-7.32]\, 10^{-3}\ @ 95\%\ \mathrm{prob.})\,,
\nonumber\\
\epsilon_3^{\rm SM} &=& (5.279\pm 0.004)\, 10^{-3} 
\quad ([5.271,5.288]\, 10^{-3}\ @ 95\%\ \mathrm{prob.})\,,
\nonumber\\
\epsilon_b^{\rm SM} &=&-(6.94\pm 0.15)\, 10^{-3} 
\quad ([-7.24,-6.64]\, 10^{-3}\ @ 95\%\ \mathrm{prob.})\,,
\label{eq:SMepsilons_oldRb}
\end{eqnarray}
where the uncertainties are dominated by the top-quark mass, and the
quadratic dependence in $\epsilon_1^{\rm SM}$ and $\epsilon_b^{\rm
  SM}$ results in the larger uncertainties.  The 95\% ranges of
$\epsilon_1^{\rm SM}$ and $\epsilon_b^{\rm SM}$ become $[4.71, 5.72]\,
10^{-3}$ and $[-7.49,-6.41]\, 10^{-3}$, respectively, if adopting
$m_t=173.3\pm 2.8$ GeV instead of $m_t=173.2\pm 0.9$ GeV.  Notice that
one can define $\epsilon_b^{\rm SM}$ either from the first or from the
second of eq.~(\ref{eq:rhobnonuniv}). We choose to define it from
$\kappa_Z^b$, so that the prediction is insensitive to the inclusion
of two-loop fermionic contributions to $\rho_Z^b$ (this is possible
within the approximations inherent in the $\epsilon$
parameterization). In figure~\ref{fig:EpsilonoldRb} we report the
one-dimensional 95\% probability range of the SM predictions for
$\epsilon_1$ and $\epsilon_3$, where the latter is invisible due to
the tiny error band.

\subsection{Constraints on the $Zb\bar{b}$ couplings}
\label{sec:Zbb}

Motivated phenomenologically by the long-standing pull in
$A_\mathrm{FB}^{0,b}$ and by the more recent pull in $R_b^0$, and
theoretically by the larger coupling to NP in the third generation
realized in many explicit models, the possibility of modified
$Zb\bar{b}$ couplings has been extensively studied (see for example
refs.~\cite{Bamert:1996px,Haber:1999zh,Choudhury:2001hs,
Agashe:2006at,Djouadi:2006rk,Kumar:2010vx,delAguila:2010mx,
DaRold:2010as,Alvarez:2010js,Dermisek:2011xu,Djouadi:2011aj,
Dermisek:2012qx,Batell:2012ca,Guadagnoli:2013mru}). 

\begin{figure}[tp]
  \centering
  \includegraphics[width=.49\textwidth]{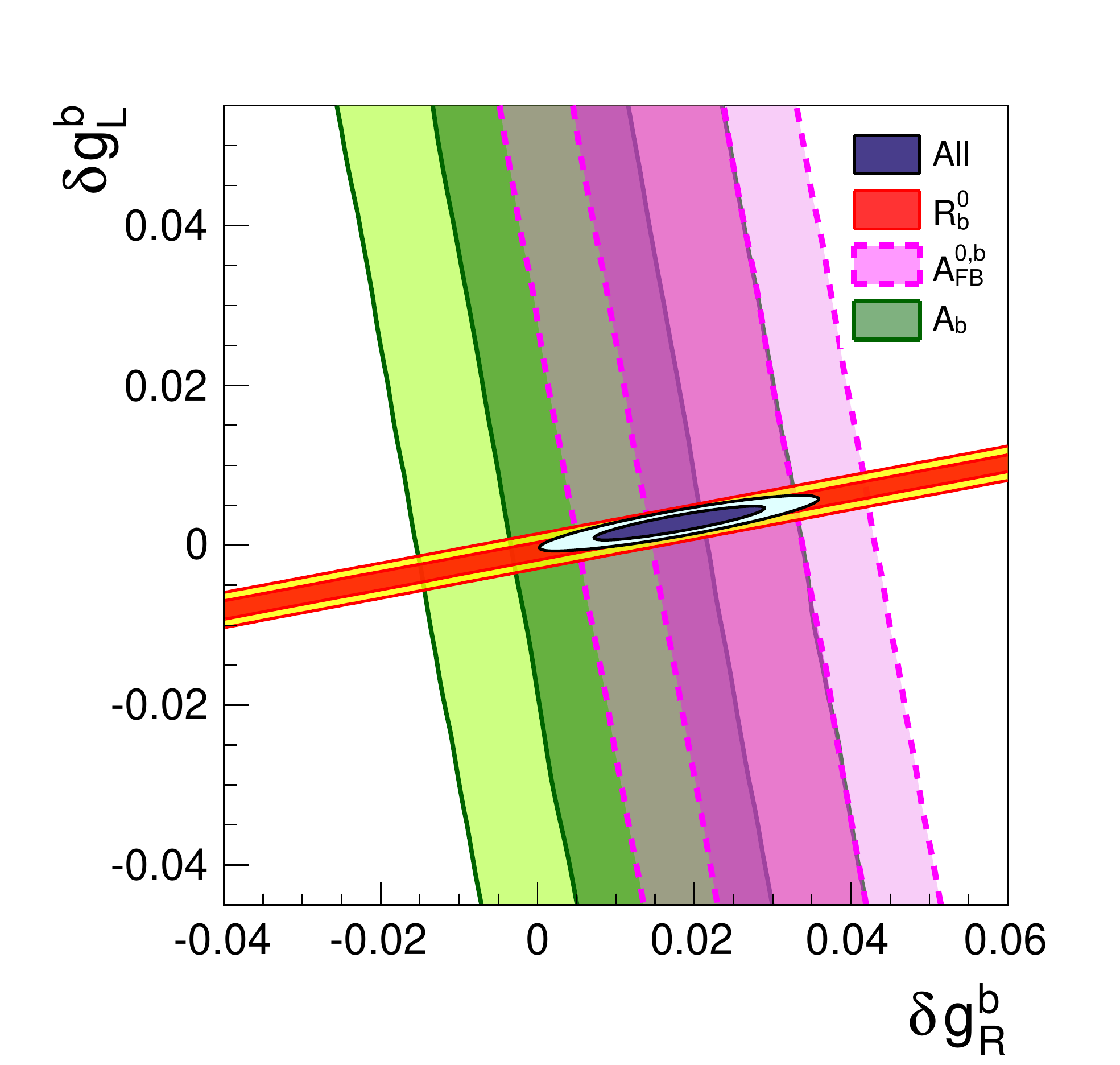}
  \hfill
  \includegraphics[width=.49\textwidth]{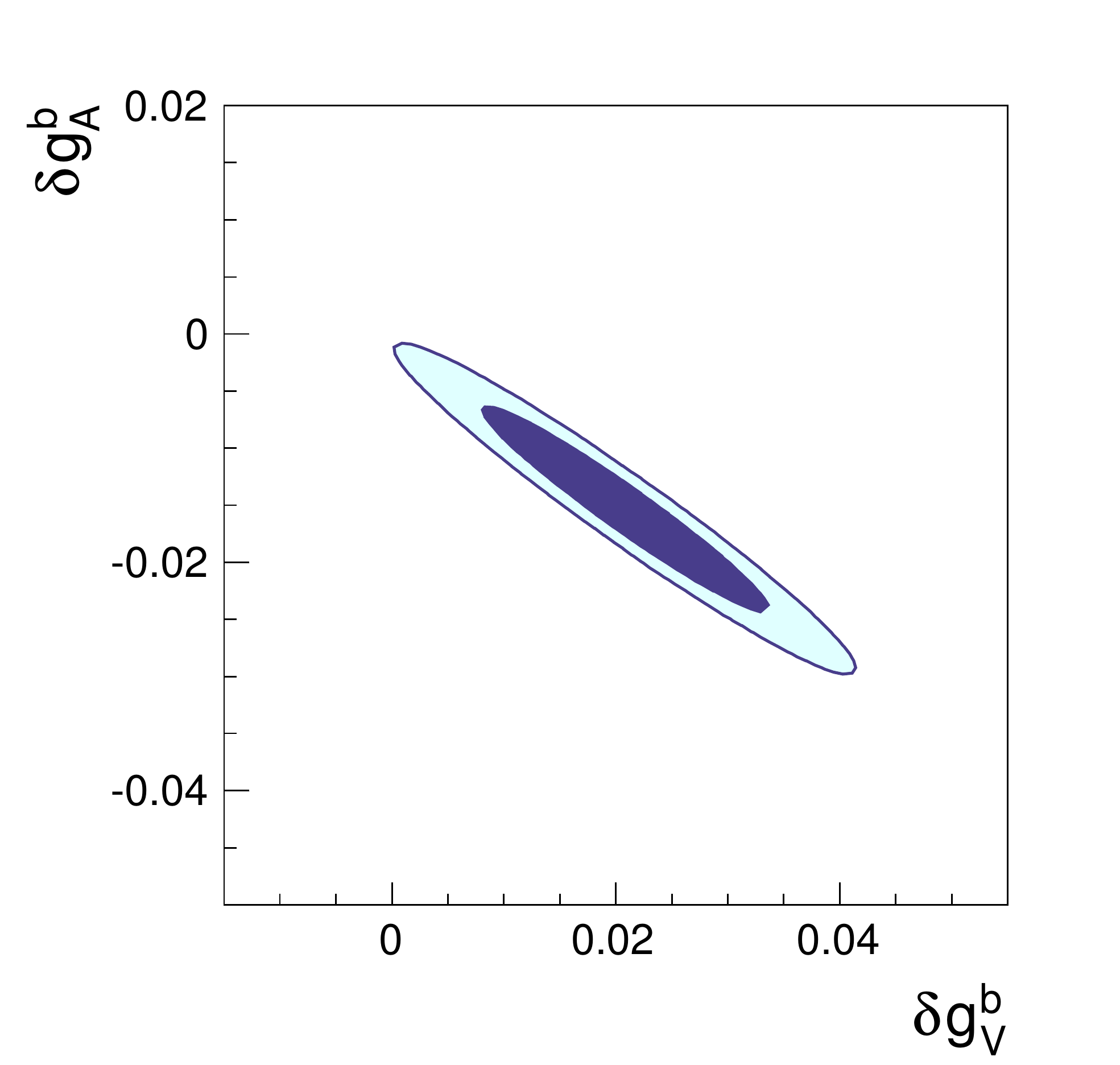}
  \\
  \includegraphics[width=.49\textwidth]{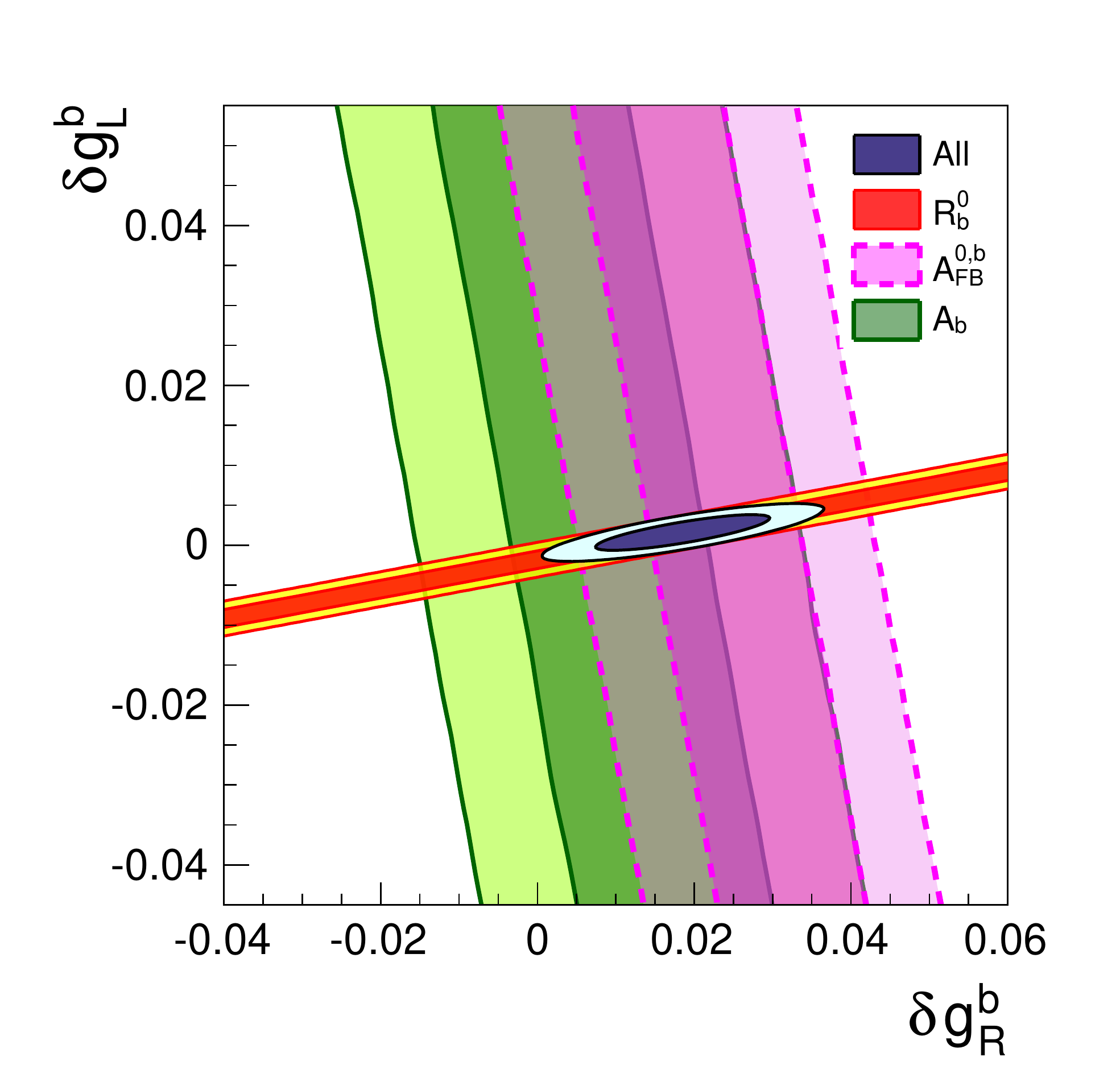}
  \hfill
  \includegraphics[width=.49\textwidth]{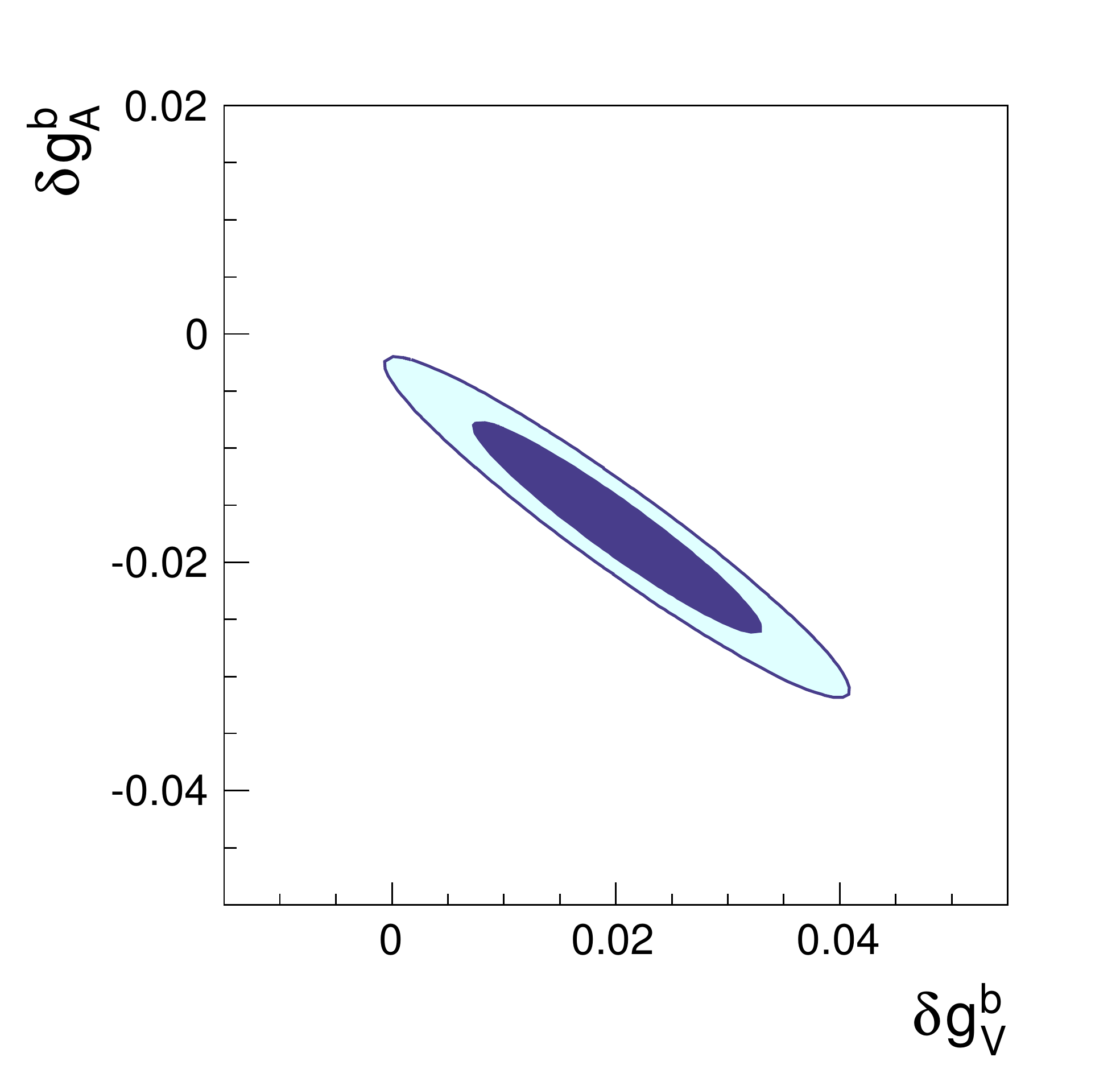}
  \caption{Two-dimensional probability distributions for the
    parameters $\delta g_R^b$ and $\delta g_L^b$ (left), or 
    $\delta g_V^b$ and $\delta g_A^b$ (right), using the large-$m_t$
    expansion (top) or the results of
    ref.~\cite{Freitas:2012sy,freitasprivate} (bottom) for the two-loop 
    fermionic EW corrections to $\rho_Z^f$. The individual
    constraints in the left plots are computed by omitting 
    $\mathcal{A}_b$, $A_{\rm FB}^{0,b}$, $\Gamma_Z$,
    $\sigma_h^0$, $R_\ell^0$, $R_c^0$ and $R_b^0$ except for the one
    specified in the legend.}
  \label{fig:Zbb}
\end{figure}

We parameterize NP contributions to the $Zb\bar{b}$ vertex by
modifying the couplings in eq.~(\ref{eq:gvga}) in the
following way:
\begin{equation}
  \label{eq:zbb}
  g_V^b =
  (g_V^b)_\mathrm{SM} + \delta g_V^b \,,\qquad
  g_A^b =
  (g_A^b)_\mathrm{SM} + \delta g_A^b\,,
\end{equation}
or equivalently by introducing $\delta g_R^b = (\delta g_V^b - \delta
g_A^b)/2$ and $\delta g_L^b = (\delta g_V^b + \delta g_A^b)/2$.
We may assume flat priors either for $\delta g_V^b$ and $\delta
g_A^b$ or for $\delta g_R^b$ and $\delta g_L^b$, but both choices 
yield almost identical results. Here we perform a 
fit with flat priors for $\delta g_R^b$ and $\delta g_L^b$. 
The results are summarized in table~\ref{tab:Zbb}, 
where the correlation matrices for the posteriors are 
given in tables~\ref{tab:ZbbCorr_oldRb} and \ref{tab:ZbbCorr_newRb}. 
There is also a second region in the fit (not
shown in table~\ref{tab:Zbb} nor in figure~\ref{fig:Zbb}) where $g_R$
flips its sign.\footnote{The other two allowed regions from the EWPO
  fit are disfavored by the off $Z$-pole data~\cite{Choudhury:2001hs}.}

As shown in the left plots in figure~\ref{fig:Zbb}, 
the asymmetries $\mathcal{A}_b$ and $A_{\rm FB}^{0,b}$ are mainly
sensitive to $\delta g_R^b$, since their shifts  
are given in terms of the combination 
$(g_L^b)_\mathrm{SM}\delta g_R^b - (g_R^b)_\mathrm{SM}\delta g_L^b$ 
with $|(g_R^b)_\mathrm{SM}|\ll |(g_L^b)_\mathrm{SM}|$. On the other
hand, $R_b^0$ is associated with 
$(g_R^b)_\mathrm{SM}\delta g_R^b + (g_L^b)_\mathrm{SM}\delta g_L^b$, 
and mainly constrains $\delta g_L^b$.

\begin{table}[tp]
\centering
\begin{tabular}{c|c|c} 
\hline
Parameter 
& Large-$m_t$ expansion
&  Using
ref.~\cite{Freitas:2012sy,freitasprivate} 
\\
\hline
$\delta g_R^b$ &
$0.018\pm 0.007$ &
$0.019\pm 0.007$ 
\\
$\delta g_L^b$ &
$0.0028\pm 0.0014$ &
$0.0016\pm 0.0015$ 
\\
\hline
$\delta g_V^b$ &
$\phantom{-}0.021\pm 0.008$ &
$\phantom{-}0.020\pm 0.008$ 
\\
$\delta g_A^b$ &
$-0.015\pm 0.006$ &
$-0.017\pm 0.006$ 
\\
\hline
\end{tabular}
\caption{Fit results for the shifts in the 
  $Zb\bar b$ couplings, using the large-$m_t$ expansion or the results
  in ref.~\cite{Freitas:2012sy,freitasprivate} for the two-loop
  fermionic two-loop EW corrections. In the latter case, we do not
  consider constraints from $\Gamma_Z$, $\sigma^0_h$ and $R_\ell^0$. 
}
\label{tab:Zbb}
\end{table}

\subsection{Constraints on a non-standard Higgs coupling}
\label{sec:HVV}

A key question to understand the mechanism of EWSB is whether the
underlying dynamics is weak or strong. As we shall see below, EWPO
strongly constrain the Higgs coupling to vector bosons, and this hints
either at a weakly interacting Higgs or at a non-trivial strongly
interacting sector in which additional contributions to EWPO are
present and restore the agreement with experimental data.

To investigate the question above, it is useful to consider a general
Lagrangian for a light Higgs-like scalar field
$h$~\cite{Giudice:2007fh,Contino:2010mh,Azatov:2012bz,Contino:2013kra}. Under
the assumption of an approximate custodial symmetry, the longitudinal
$W$ and $Z$ polarizations can be described by the two-by-two matrix
$\Sigma(x) = \exp(i\tau^a\chi^a(x)/v)$, with $\tau^a$ the Pauli
matrices and $v^2=1/(\sqrt{2} G_\mu)$. Then, assuming that there are
no other light states and no new sources of flavour violation, the
most general Lagrangian for $h$ can be written
as~\cite{Contino:2010mh,Azatov:2012bz}:
\begin{eqnarray}
\mathcal{L} &=& 
\frac{1}{2}(\partial_\mu h)^2 - V(h)
+ \frac{v^2}{4}\,{\rm Tr}\big(D_\mu\Sigma^\dagger D^\mu\Sigma\big)
  \left( 1 + 2a\,\frac{h}{v} + b\, \frac{h^2}{v^2} + \cdots \right)
\nonumber\\
&&
- m_{u,i}
  (\bar{u}_{L,i}, \bar{d}_{L,i})\, \Sigma\, 
  \bigg(\! \begin{array}{c} u_{R,i} \\ 0 \end{array} \!\bigg)
  \left( 1 + c_u\,\frac{h}{v} + c_{2u}\,\frac{h^2}{v^2} + \cdots \right)
+ {\rm h.c.}
\nonumber\\
&&
- m_{d,i}
  (\bar{u}_{L,i},\bar{d}_{L,i})\, \Sigma\, 
  \bigg(\! \begin{array}{c} 0 \\ d_{R,i} \end{array} \!\bigg)
  \left( 1 + c_d\,\frac{h}{v} + c_{2d}\,\frac{h^2}{v^2} +\cdots \right)
+ {\rm h.c.}
\nonumber\\
&&
- m_{\ell,i}
  (\bar{\nu}_{L,i}, \bar{\ell}_{L,i})\, \Sigma\, 
  \bigg(\! \begin{array}{c} 0 \\ \ell_{R,i} \end{array} \!\bigg)
  \left( 1 + c_\ell\,\frac{h}{v} + c_{2\ell}\,\frac{h^2}{v^2} +\cdots \right)
+ {\rm h.c.},
\end{eqnarray}
where $V(h)$ is the potential of the scalar field
\begin{equation}
V(h) = \frac{m_h^2}{2}h^2 
+ \frac{d_3}{6}\left(\frac{3m_h^2}{v}\right)h^3
+ \frac{d_4}{24}\left(\frac{3m_h^2}{v^2}\right)h^4
+ \cdots.
\end{equation}

The SM corresponds to the choice $a=b=c_u=c_d=c_\ell=d_3=d_4=1$ and
$c_{2u}=c_{2d}=c_{2\ell}=0$. The dominant deviations from the SM in
EWPO are induced by the non-standard coupling $a\neq 1$. This
generates extra contributions to the $S$ and $T$
parameters~\cite{Barbieri:2007bh}:
\begin{eqnarray}
S &=& \frac{1}{12\pi} (1 - a^2)
       \ln\bigg(\frac{\Lambda^2}{m_h^2}\bigg)\,,\\
T &=& - \frac{3}{16\pi c_W^2} (1 - a^2)
      \ln\bigg(\frac{\Lambda^2}{m_h^2}\bigg)\,,
\end{eqnarray}
where $\Lambda = 4\pi v/\sqrt{|1-a^2|}$ is the cutoff of the light
Higgs effective Lagrangian. A sum rule for $1-a^2$ can be written in
terms of the total cross sections in different isospin channels of
longitudinal EW gauge boson
scattering~\cite{Falkowski:2012vh}, implying $a^2 \le 1$ unless
the $I=2$ channel dominates the cross section. Thus, we expect in
general a positive $S$ and a negative $T$.

\begin{figure}[tp]
\centering
\includegraphics[width=.35\textwidth]{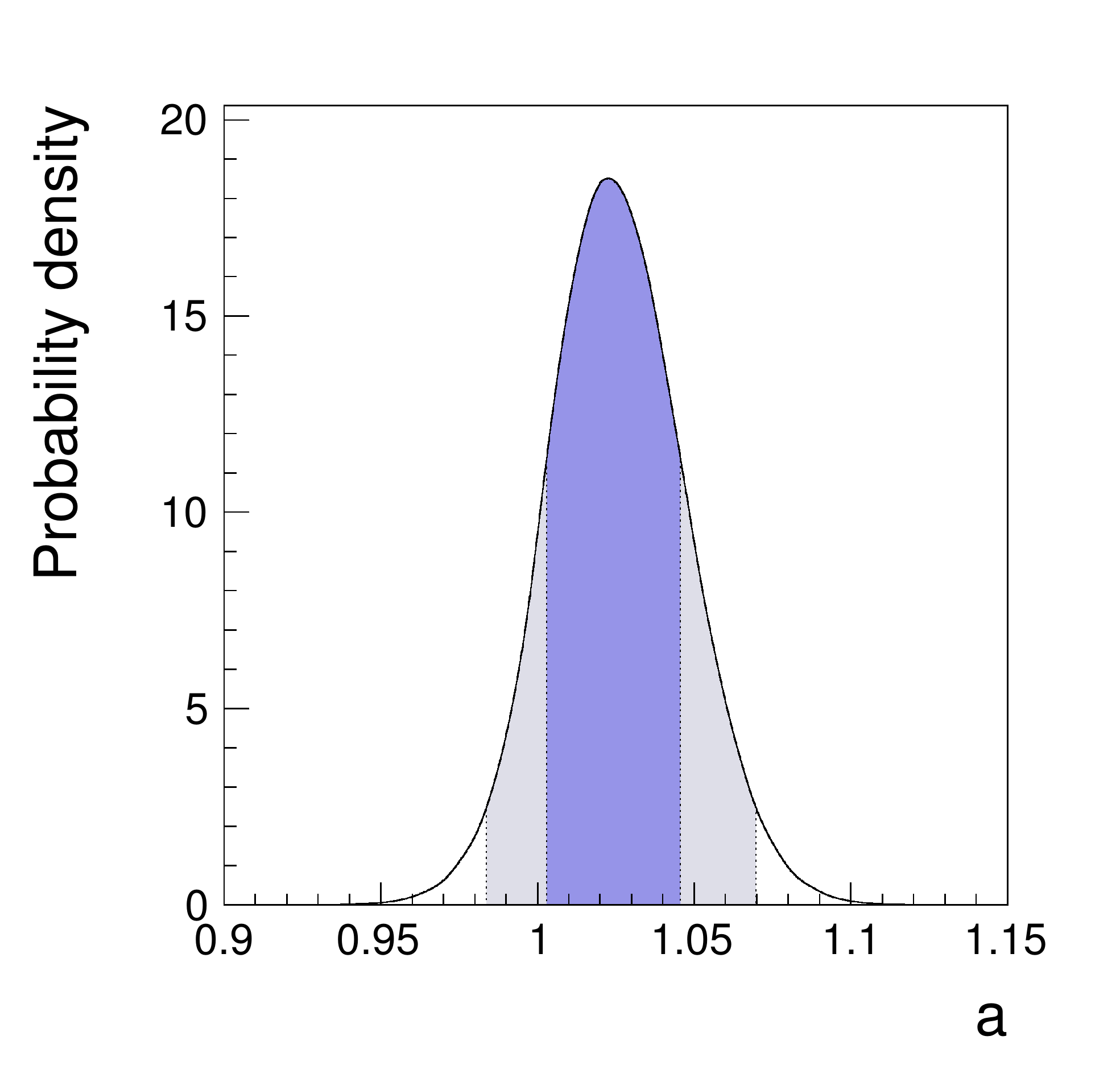}
\hspace{-4mm}
\includegraphics[width=.32\textwidth]{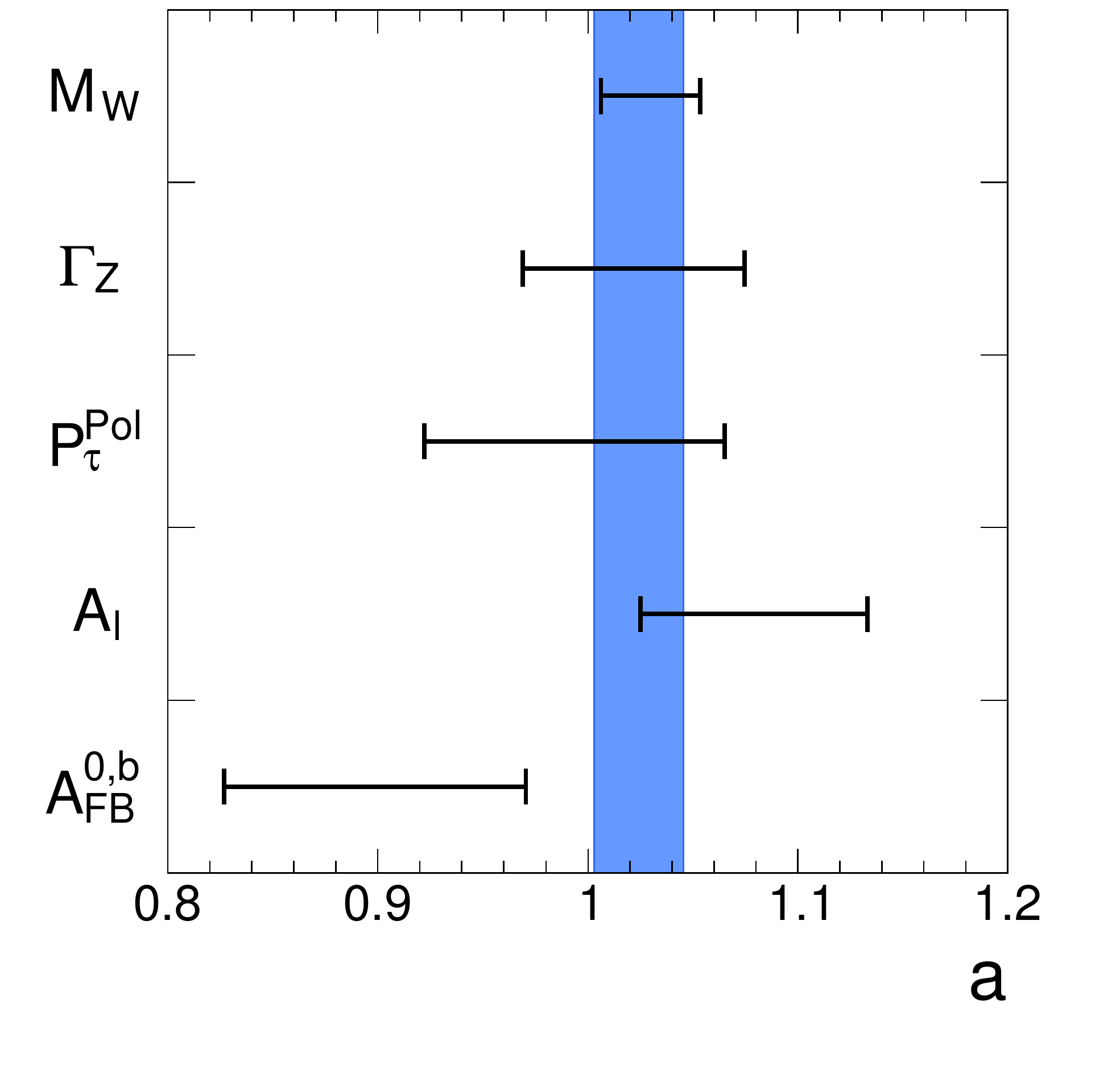}
\hspace{-4mm}
\includegraphics[width=.35\textwidth]{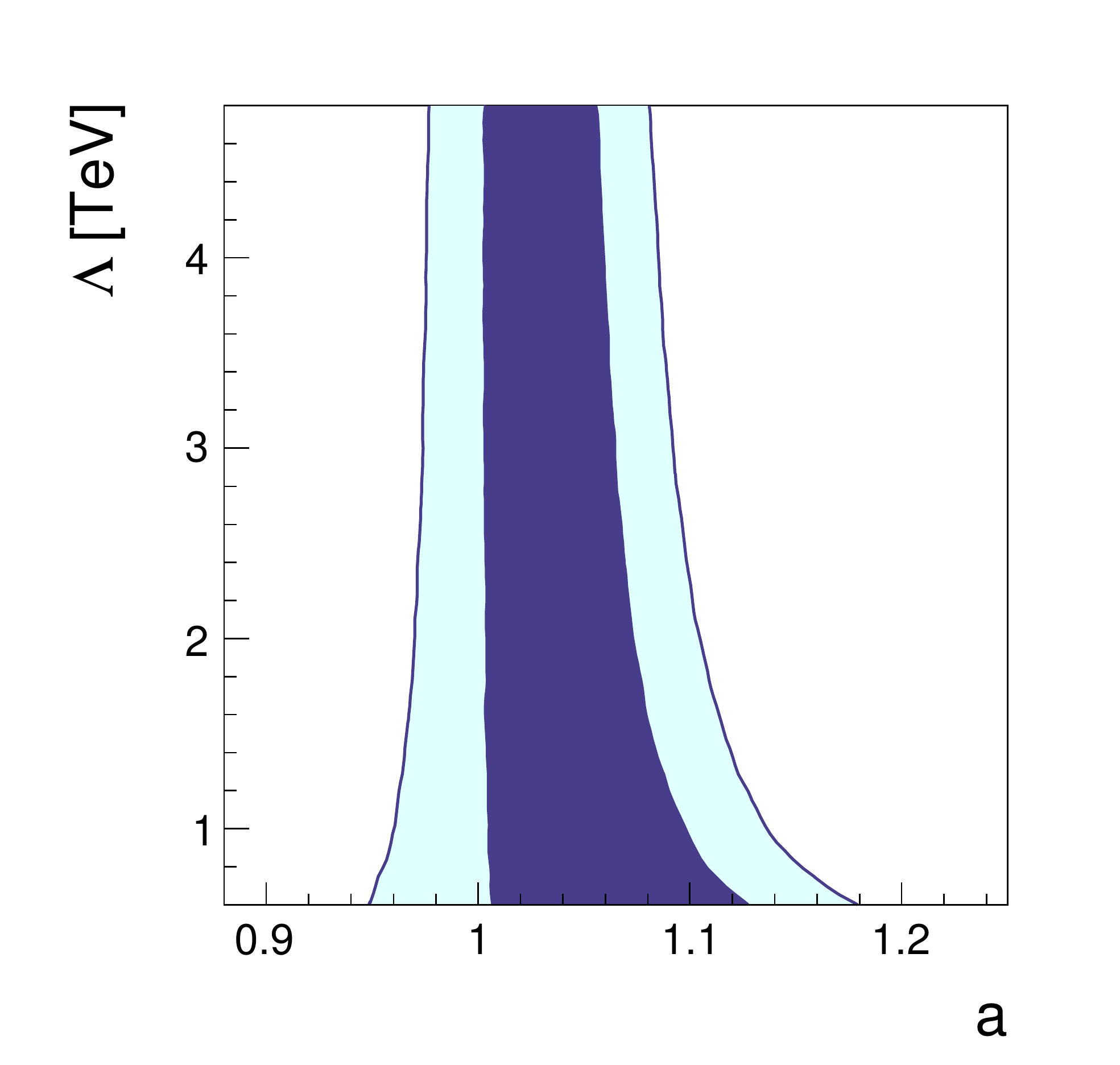}
\caption{Left: Probability distribution for the coupling $a$. Center:
  Indirect determinations of the coupling $a$, excluding the
  observables $M_W$, $\Gamma_Z$, $P_\tau^{\rm pol}$, $A_l^0$ and
  $A_{\rm FB}^{0,b}$, except for the one specified in each row.
  The vertical blue band represents the one obtained from
  the the fit with all the observables. Right: Probability regions
  in the $a$--$\Lambda$ plane.
  In all plots, the large-$m_t$
  expansion is adopted to the two-loop fermionic EW corrections to
  $\rho_Z^f$.}
\label{fig:Higgs}
\end{figure}

We fit the coupling $a$ together with the five SM parameters to the
precision observables using the large-$m_t$ expansion for the two-loop
fermionic EW corrections to $\rho_Z^f$, and obtain the results shown
in the left plot in figure~\ref{fig:Higgs} and reported in
table~\ref{tab:HVV}. The correlation matrices for the posteriors are
given in tables~\ref{tab:HVVCorr_oldRb} and \ref{tab:HVVCorr_newRb}
for the case of $m_t$ from Tevatron pole mass average. In
table~\ref{tab:HVV} we also present the result obtained using $m_t$
from the $\overline{\mathrm{MS}}$ mass and including the subleading
two-loop fermionic EW corrections to $\delta\rho_Z^f$ with the results
of ref.~\cite{Freitas:2012sy,freitasprivate}. As is evident
from the table, the results are stable against the treatment of
$\delta\rho_Z^f$, but the error is sensitive to the uncertainty in
$m_t$. This can be understood by looking at the impact of the
individual constraints on $a$ shown in the center plot in
figure~\ref{fig:Higgs}, from which it is evident that $M_W$ is giving
the strongest bound on the nonstandard Higgs coupling.  Our result is
compatible with the analysis of ref.~\cite{Falkowski:2013dza}.

\begin{table}[tp]
\centering
\begin{tabular}{c|c|c} 
  \hline
  $m_t$ [GeV] & large-$m_t$ expansion & Using
    ref.~\cite{Freitas:2012sy,freitasprivate} \\ \hline
  $173.2 \pm 0.9$ & $1.024\pm 0.021$ &  $1.024\pm 0.022$\\
  $173.3\pm 2.8$ & $1.025\pm 0.030$ & $1.027\pm 0.031$ \\
  \hline
\end{tabular}
\caption{Fit result for the $HVV$ coupling $a$, obtained with
  different choices for $m_t$ and for the two-loop fermionic
  EW corrections to $\rho_Z^f$. When using
  ref.~\cite{Freitas:2012sy,freitasprivate}, we do not impose
  constraints from $\Gamma_Z$, $\sigma_h^0$ and $R_\ell^0$.} 
\label{tab:HVV}
\end{table}

Since the fit prefers values of $a>1$, while the sum rule of
ref.~\cite{Falkowski:2012vh} gives in general $a<1$, additional
contributions to the EWPO, for example from additional light fermions
\cite{Barbieri:2007bh,Grojean:2013qca}, are required in order to restore the
agreement with experimental data in composite Higgs models.  If one
takes literally the model with no new particles below the cutoff and
assuming $a \le 1$, from the $95\%$ probability range $a \in [0.984,
1.070]$ ($[0.981, 1.071]$) one can derive a lower bound on $\Lambda$:
\begin{equation}
  \label{eq:lbound}
  \Lambda > 17\, (16)\, \mathrm{TeV}\, 
   @ 95\%\, \mathrm{probability}\,, 
\end{equation}
using the large-$m_t$ expansion 
(using the results of ref.~\cite{Freitas:2012sy,freitasprivate}). 
One can generalise the analysis allowing for $\Lambda < 4\pi v/\sqrt{|1-a^2|}$
and assuming that the dynamics at the cutoff does not contribute sizably
to $S$ and $T$. In this case one can determine regions in the $a$--$\Lambda$
plane as shown in right plot of figure~\ref{fig:Higgs}. Clearly the value of $a$ is tightly
constrained for values of $\Lambda$ compatible with direct searches.

\subsection{General bounds on the New Physics scale}
\label{sec:genNP}

Before concluding, let us take a more general approach and consider
the contributions to the EW fit of arbitrary dimension-six NP-induced
operators~\cite{Buchmuller:1985jz,Barbieri:1999tm,Contino:2013kra}:
\begin{equation}
  \label{eq:Leff}
  \mathcal{L}_\mathrm{eff} = \mathcal{L}_\mathrm{SM} + \sum_i
  \frac{C_i}{\Lambda^2} \mathcal{O}_i\,.
\end{equation}
For concreteness, let us use the same operator basis of
ref.~\cite{Barbieri:1999tm}: 
\begin{align}
\mathcal{O}_{WB} &= (H^\dagger\tau^a H) W^a_{\mu\nu}B^{\mu\nu},
&
\mathcal{O}_{H} &= | H^\dagger D_\mu H|^2\,,
\nonumber\\
\mathcal{O}_{LL} &= \frac{1}{2}(\overline{L}\gamma_\mu\tau^a L)^2\,,
&
\mathcal{O}^\prime_{HL} &= i(H^\dagger D_\mu\tau^a H)(\overline{L}\gamma^\mu\tau^a L)\,,
\nonumber\\
\mathcal{O}^\prime_{HQ} &= i(H^\dagger D_\mu\tau^a H)(\overline{Q}\gamma^\mu\tau^a Q)\,,
&
\mathcal{O}_{HL} &= i(H^\dagger D_\mu H)(\overline{L}\gamma^\mu L)\,,
\nonumber\\
\mathcal{O}_{HQ} &= i(H^\dagger D_\mu H)(\overline{Q}\gamma^\mu Q)\,,
&
\mathcal{O}_{HE} &= i(H^\dagger D_\mu H)(\overline{E}\gamma^\mu E)\,,
\nonumber\\
\mathcal{O}_{HU} &= i(H^\dagger D_\mu H)(\overline{U}\gamma^\mu U)\,,
&
\mathcal{O}_{HD} &= i(H^\dagger D_\mu H)(\overline{D}\gamma^\mu D)\,,
\end{align}
where we add the contribution of the Hermitian conjugate for operators
$\mathcal{O}^\prime_{HL}$ to $\mathcal{O}_{HD}$.  The Higgs field gets
a vev $\langle H\rangle = (0,v/\sqrt{2})^T$.  For fermions, we do not
consider generation mixing, and assume lepton-flavour universality:
$C_{HL}^\prime = C_{HL_i}^\prime$, $C_{HL} = C_{HL_i}$ and $C_{HE} =
C_{HE_i}$ for $i=1,2,3$.

The first two operators contribute to the oblique
parameters $S$ and $T$:
\begin{align}
S &=
\frac{4s_Wc_W\, C_{WB} }{\alpha(M_Z^2)}
\left(\frac{v}{\Lambda}\right)^2,
\\
T &= - \frac{C_H}{2\alpha(M_Z^2)}
\left(\frac{v}{\Lambda}\right)^2,
\end{align}
where $\mathcal{O}_H$ violates the custodial symmetry, since it
gives a correction to the mass of the $Z$ boson,
but not to that of the $W$ boson. 
The next two operators yield non-oblique corrections to the Fermi
constant: 
\begin{align}
G_\mu &= G_{\mu,\mathrm{SM}}
\left[ 1 - C_{LL}\left(\frac{v}{\Lambda}\right)^2
+ 2\,C^\prime_{HL}\left(\frac{v}{\Lambda}\right)^2\right],
\end{align}
where $G_{\mu,\mathrm{SM}}$ denotes the Fermi constant in the SM. 
The corrections to the Fermi constant affect 
the mass and width of the $W$ boson and the $Zf\bar{f}$
couplings as shown in Appendix~\ref{app:STUformulae}.

The width of the $W$ boson also receives the corrections from 
the operators $\mathcal{O}^\prime_{HL}$ and $\mathcal{O}^\prime_{HQ}$: 
\begin{align}
\Gamma_W
&= 
\Gamma_{W,\mathrm{SM}}
\left[
1 + \left( 3 C_{HL}^\prime + C_{HQ_1}^\prime + C_{HQ_2}^\prime\right)
\left(\frac{v}{\Lambda}\right)^2
\right]. 
\end{align}
Finally, the operators from
$\mathcal{O}^\prime_{HL}$ to $\mathcal{O}_{HD}$ contribute to the $Zf\bar{f}$ couplings:
\begin{align}
&\delta g_L^{\nu_i} = \frac{C_{HL_i}^\prime-C_{HL_i}}{2} \left(\frac{v}{\Lambda}\right)^2,
&&\delta g_L^{e_i} = -\frac{C_{HL_i}^\prime+C_{HL_i}}{2} \left(\frac{v}{\Lambda}\right)^2,
\\
&\delta g_L^{u_i} = \frac{C_{HQ_i}^\prime-C_{HQ_i}}{2} \left(\frac{v}{\Lambda}\right)^2,
&&\delta g_L^{d_i} = -\frac{C_{HQ_i}^\prime+C_{HQ_i}}{2} \left(\frac{v}{\Lambda}\right)^2,
\nonumber\\
&\delta g_R^{e_i} = -\frac{C_{HE_i}}{2} \left(\frac{v}{\Lambda}\right)^2,
&&\delta g_R^{u_i} = -\frac{C_{HU_i}}{2} \left(\frac{v}{\Lambda}\right)^2,
&&\delta g_R^{d_i} = -\frac{C_{HD_i}}{2}
\left(\frac{v}{\Lambda}\right)^2,
\nonumber
\end{align}
where the shifts in the vector and axial-vector couplings are given by 
$\delta g_V^f = \delta g_L^f + \delta g_R^f$ and
$\delta g_A^f = \delta g_L^f - \delta g_R^f$, respectively.

\begin{table}[tp]
\center
\begin{tabular}{c|c|rr|c|rr}
\hline
& \multicolumn{3}{c|}{Large-$m_t$ expansion}
& \multicolumn{3}{c}{Using ref.~\cite{Freitas:2012sy,freitasprivate}}
\\
\hline
& $C_i/\Lambda^2$ [TeV$^{-2}$]
& \multicolumn{2}{c|}{$\Lambda$ [TeV]}
& $C_i/\Lambda^2$ [TeV$^{-2}$]
& \multicolumn{2}{c}{$\Lambda$ [TeV]}
\\
Coefficient &  
at 95\% & $C_i=-1$ & $C_i=1$ &
at 95\% & $C_i=-1$ & $C_i=1$
\\
\hline
$C_{WB}$ &
$[-0.0096,\, 0.0042]$ & 
$10.2$\ \ \ \ & 
$15.4$\ \ \ & 
$[-0.0095,\, 0.0045]$ & 
$10.3$\ \ \ \ & 
$15.0$\ \ \ 
\\
$C_{H}$ &
$[-0.030,\, 0.007]$ & 
$5.8$\ \ \ \ & 
$12.1$\ \ \ & 
$[-0.031,\, 0.008]$ & 
$5.7$\ \ \ \ & 
$11.5$\ \ \ 
\\
$C_{LL}$ &
$[-0.011,\, 0.019]$ & 
$9.5$\ \ \ \ & 
$7.2$\ \ \ & 
$[-0.016,\, 0.023]$ & 
$8.0$\ \ \ \ & 
$6.6$\ \ \ 
\\
$C_{HL}^\prime$ &
$[-0.012,\, 0.005]$ & 
$9.2$\ \ \ \ & 
$14.1$\ \ \ & 
$[-0.017,\, 0.009]$ & 
$7.6$\ \ \ \ & 
$10.8$\ \ \ 
\\
$C_{HQ}^\prime$ &
$[-0.010,\, 0.015]$ & 
$10.2$\ \ \ \ & 
$8.2$\ \ \ & 
$[-0.40,\, 0.20]$ & 
$1.6$\ \ \ \ & 
$2.2$\ \ \ 
\\
$C_{HL}$ &
$[-0.007,\, 0.010]$ & 
$12.2$\ \ \ \ & 
$10.0$\ \ \ & 
$[-0.034,\, 0.022]$ & 
$5.5$\ \ \ \ & 
$6.7$\ \ \ 
\\
$C_{HQ}$ &
$[-0.023,\, 0.046]$ & 
$6.6$\ \ \ \ & 
$4.7$\ \ \ & 
$[-0.01,\, 0.11]$ & 
$11.7$\ \ \ \ & 
$3.0$\ \ \ 
\\
$C_{HE}$ &
$[-0.014,\, 0.008]$ & 
$8.4$\ \ \ \ & 
$11.1$\ \ \ & 
$[-0.029,\, 0.019]$ & 
$5.9$\ \ \ \ & 
$7.2$\ \ \ 
\\
$C_{HU}$ &
$[-0.061,\, 0.087]$ & 
$4.0$\ \ \ \ & 
$3.4$\ \ \ & 
$[-0.37,\, 0.08]$ & 
$1.6$\ \ \ \ & 
$3.5$\ \ \ 
\\
$C_{HD}$ &
$[-0.15,\, 0.05]$ & 
$2.6$\ \ \ \ & 
$4.6$\ \ \ & 
$[-1.1,\, -0.2]$ & 
$1.0$\ \ \ \ & 
---\ \ \ 
\\
\hline
\end{tabular}
\caption{Fit results for the coefficients of the dimension six
  operators at 95\% probability in units of $1/\Lambda^2$ TeV$^{-2}$,
  with quark-flavour universality in NP contribution. 
  The fit is performed switching on one operator at a time. 
  The corresponding lower bounds on the NP scale in TeV obtained by
  setting $C_i=\pm 1$ are also presented. When using the results from
  ref.~\cite{Freitas:2012sy,freitasprivate}, we do not consider
  constraints from $\Gamma_Z$, $\sigma^0_h$ and $R_\ell^0$.} 
\label{tab:NPscale-univ}
\end{table}

\begin{table}[tp]
\center
\begin{tabular}{c|c|rr|c|rr}
\hline
& \multicolumn{3}{c|}{Large-$m_t$ expansion}
& \multicolumn{3}{c}{Using ref.~\cite{Freitas:2012sy,freitasprivate}}
\\
\hline
& $C_i/\Lambda^2$ [TeV$^{-2}$]
& \multicolumn{2}{c|}{$\Lambda$ [TeV]}
& $C_i/\Lambda^2$ [TeV$^{-2}$]
& \multicolumn{2}{c}{$\Lambda$ [TeV]}
\\
Coefficient &  
at 95\% & $C_i=-1$ & $C_i=1$ &
at 95\% & $C_i=-1$ & $C_i=1$
\\
\hline
$C_{HQ_1}^\prime$ &
$[-0.026,\, 0.034]$ & 
$6.2$\ \ \ \ & 
$5.4$\ \ \ & 
$[-0.19,\, 0.01]$ & 
$2.3$\ \ \ \ & 
$11.9$\ \ \ 
\\
$C_{HQ_2}^\prime$ &
$[-0.026,\, 0.034]$ & 
$6.2$\ \ \ \ & 
$5.4$\ \ \ & 
$[-0.20,\, 0.01]$ & 
$2.3$\ \ \ \ & 
$10.8$\ \ \ 
\\
$C_{HQ_3}^\prime,C_{HQ_3}$ &
$[-0.025,\, 0.053]$ & 
$6.3$\ \ \ \ & 
$4.3$\ \ \ & 
$[0.00,\, 0.10]$ & 
$15.6$\ \ \ \ & 
$3.1$\ \ \ 
\\
$C_{HQ_1}$ &
$[-0.26,\, 0.34]$ & 
$2.0$\ \ \ \ & 
$1.7$\ \ \ & 
$[-1.9,\, 0.1]$ & 
$0.7$\ \ \ \ & 
$3.9$\ \ \ 
\\
$C_{HQ_2}$ &
$[-0.16,\, 0.18]$ & 
$2.5$\ \ \ \ & 
$2.4$\ \ \ & 
$[-0.25,\, 0.15]$ & 
$2.0$\ \ \ \ & 
$2.6$\ \ \ 
\\
$C_{HU_1}$ &
$[-0.13,\, 0.17]$ & 
$2.8$\ \ \ \ & 
$2.4$\ \ \ & 
$[-0.97,\, 0.03]$ & 
$1.0$\ \ \ \ & 
$5.6$\ \ \ 
\\
$C_{HU_2}$ &
$[-0.11,\, 0.17]$ & 
$3.0$\ \ \ \ & 
$2.4$\ \ \ & 
$[-0.39,\, 0.21]$ & 
$1.6$\ \ \ \ & 
$2.2$\ \ \ 
\\
$C_{HD_1},C_{HD_2}$ &
$[-0.34,\, 0.26]$ & 
$1.7$\ \ \ \ & 
$2.0$\ \ \ & 
$[-0.1,\, 1.9]$ & 
$3.8$\ \ \ \ & 
$0.7$\ \ \ 
\\
$C_{HD_3}$ &
$[-0.38,\, 0.03]$ & 
$1.6$\ \ \ \ & 
$6.3$\ \ \ & 
$[-0.66,\, -0.13]$ & 
$1.2$\ \ \ \ & 
---\ \ \ 
\\
\hline
\end{tabular}
\caption{Same as table~\ref{tab:NPscale-univ}, but without
  quark-flavour universality. 
  The operator $\mathcal{O}_{HU_3}$ does not contribute to the EWPO.}
\label{tab:NPscale-nonuniv}
\end{table}

Switching on one operator at a time (thus barring accidental
cancellations), one can constrain the coefficient of each of the above
operators using the EW fit. Clearly, as is the case for all indirect
constraints, one can either interpret this as a bound on the NP scale
fixing the coupling or as a bound on the coupling for fixed NP
scale. In tables~\ref{tab:NPscale-univ} and \ref{tab:NPscale-nonuniv}, 
we list for all the operators the 95\% probability regions of the
coefficients and the lower bound on the NP scale in TeV obtained by
setting $C_i=\pm 1$, with and without quark-flavour universality for
the operators. Comparing these results with the ones of
ref.~\cite{Barbieri:1999tm}, we see that the recent experimental
improvements strengthen the bounds on NP contributions, pushing the
lower bound on $\Lambda$ to scales as large as 15 TeV. 

\begin{table}[tp]
\center
\begin{tabular}{c|c|c}
\hline
& Large-$m_t$ expansion
& Using ref.~\cite{Freitas:2012sy,freitasprivate}
\\
\hline
Coefficient &  
$C_i/\Lambda^2$ [TeV$^{-2}$] at 95\% 
& $C_i/\Lambda^2$ [TeV$^{-2}$] at 95\% 
\\
\hline
$C_{WB}$ &
$[-0.009,\, 0.018]$ & 
$[-0.009,\, 0.021]$ 
\\
$C_{H}$ &
$[-0.058,\, 0.015]$ & 
$[-0.068,\, 0.016]$ 
\\
\hline
$C_{HL}^\prime$ &
$[-0.026,\, 0.008]$ & 
$[-0.029,\, 0.006]$ 
\\
$C_{HQ}^\prime$ &
$[-0.18,\, 0.00]$ & 
$[-0.34,\, 0.31]$ 
\\
$C_{HL}$ &
$[-0.013,\, 0.020]$ & 
--- 
\\
$C_{HQ}$ &
$[-0.11,\, 0.07]$ & 
$[-0.07,\, 0.12]$ 
\\
$C_{HE}$ &
$[-0.022,\, 0.018]$ & 
--- 
\\
$C_{HU}$ &
$[-0.22,\, 0.41]$ & 
$[-0.26,\, 0.49]$ 
\\
$C_{HD}$ &
$[-1.2,\, -0.2]$ & 
$[-1.2,\, -0.2]$ 
\\
$C[\mathcal{A}_{\ell}]$ &
--- & 
$[-0.0021,\, 0.0050]$ 
\\
\hline
$C_{HL}^\prime$ &
$[-0.026,\, 0.008]$ & 
$[-0.029,\, 0.006]$ 
\\
$C_{HL}$ &
$[-0.013,\, 0.020]$ & 
--- 
\\
$C_{HE}$ &
$[-0.022,\, 0.018]$ & 
--- 
\\
$C_{HU_2}$ &
$[-0.22,\, 0.45]$ & 
$[-0.32,\, 0.55]$ 
\\
$C_{HD_3}$ &
$[-1.2,\, -0.2]$ & 
$[-1.2,\, -0.2]$ 
\\
$C_{HQ_1}^\prime + C_{HQ_2}^\prime$ &
$[-0.59,\, 0.51]$ & 
$[-0.68,\, 0.61]$ 
\\
$C_{HQ_2}^\prime - C_{HQ_2}$ &
$[-0.30,\, 0.17]$ & 
$[-0.67,\, 0.55]$ 
\\
$C_{HQ_3}^\prime + C_{HQ_3}$ &
$[-0.22,\, -0.01]$ & 
$[-0.77,\, 0.63]$ 
\\
$C[\mathcal{A}_{\ell}]$ &
--- & 
$[-0.0021,\, 0.0050]$ 
\\
$C[\Gamma_{uds}]$ &
$[-0.039,\, 0.044]$ & 
$[-0.42,\, 0.43]$ 
\\
\hline
\end{tabular}
\caption{Fit results for the coefficients of the dimension six
  operators at 95\% probability in units of $1/\Lambda^2$ TeV$^{-2}$. 
  We perform three separate fits, for the oblique
  operators $\mathcal{O}_{WB}$ and $\mathcal{O}_{H}$, for the
  non-oblique operators, except for $\mathcal{O}_{LL}$, with
  quark-flavour universality, and for the non-oblique operators,
  except for $\mathcal{O}_{LL}$, without quark-flavour
  universality. When we use the results of
  ref.~\cite{Freitas:2012sy,freitasprivate}, we cannot determine
  individually the coefficients $C_{HL}$ and $C_{HE}$ but only the
  combination $C[\mathcal{A}_{\ell}]$, since the observables 
  $\Gamma_Z$, $\sigma_h^0$ and $R_\ell^0$ have been neglected.}
\label{tab:Dim6Fit}
\end{table}

Moreover, we also fit multiple coefficients simultaneously by dividing
the operators into three categories: the oblique operators
$\mathcal{O}_{WB}$ and $\mathcal{O}_{H}$, the four-fermion operator
$\mathcal{O}_{LL}$, and the operators with scalars and fermions.
Since one cannot determine all the operators simultaneously from the
EWPO alone, we fit a part of them turning on the operators in each
category. The fit results are summarized in table~\ref{tab:Dim6Fit},
with and without assuming quark-flavour universality (the results for
$\mathcal{O}_{LL}$ can be found in table~\ref{tab:NPscale-univ}). 
When we use the results of ref.~\cite{Freitas:2012sy,freitasprivate} 
dropping $\Gamma_Z$, $\sigma^0_h$ and $R_\ell^0$ from the fit, 
we cannot determine individually the coefficients $C_{HL}$ and
$C_{HE}$, but only the combination 
\begin{align}
C[\mathcal{A}_\ell] 
&=
\left(g_L^\ell\delta g_R^\ell - g_R^\ell\delta g_L^\ell\right)
\left(\frac{\Lambda}{v}\right)^2, 
\end{align}
which is associated with $\mathcal{A}_{\ell}$, can be constrained. 
For the fit without universality, we float the coefficients
$C_{HL}^\prime$, $C_{HQ_i}^\prime$, $C_{HL}$, $C_{HQ_i}$, $C_{HE}$,
$C_{HU_i}$, $C_{HD_i}$ for $i=1,2$ and $3$, except for $C_{HU_3}$,
together with the SM parameters, and obtain the posteriors listed 
in table~\ref{tab:Dim6Fit}. The combinations $C_{HQ_1}^\prime +
C_{HQ_2}^\prime$, $C_{HQ_2}^\prime -
C_{HQ_2}$, $C_{HQ_3}^\prime + C_{HQ_3}$ 
and $C[\Gamma_{uds}]$, are associated with
$\Gamma_W$, $g_L^c$, $g_L^b$ and the light-quark
contribution to $\Gamma_Z$ respectively, where the last 
combination is defined as
\begin{align}
C[\Gamma_{uds}] 
&=
\sum_{f=u,d,s}\left(g_L^f\delta g_L^f + g_R^f\delta g_R^f\right)
\left(\frac{\Lambda}{v}\right)^2.
\end{align}
The correlations of the fit results are summarized in
tables~\ref{tab:Dim6FitCorr1}-\ref{tab:Dim6FitCorr3newRb}. 

A similar analysis was recently performed in
ref.~\cite{Contino:2013kra}. The constraints on $C_{WB}/\Lambda^2$ and
$C_{H}/\Lambda^2$ correspond to those on $\tan\theta_W(\bar{c}_{W} +
\bar{c}_{B})/v^2$ and $-2 \bar{c}_T/v^2$ in
ref.~\cite{Contino:2013kra}, respectively, while the other
coefficients satisfy the relations
$C_{i}/\Lambda^2=\bar{c}_i/v^2$. Our results in table
\ref{tab:Dim6Fit} are generally similar to theirs, although one cannot
directly compare the results since we have floated a larger set of
operators simultaneously. Our fit results are also compatible with the
ones of ref.~\cite{delAguila:2011zs}, considering that in the latter
work $m_h$ was not yet available and that in the fit the other SM
parameters were not floated.

All the results presented here refer to coefficients computed at the
weak scale. While other choices of operator basis could be more
convenient to study running effects (see
refs.~\cite{Grojean:2013kd,Elias-Miro:2013gya,Jenkins:2013fya}) or additional
observables such as in ref.~\cite{Contino:2013kra}, for our purpose
the basis of ref.~\cite{Barbieri:1999tm} is perfectly adequate.

\section{Summary}
\label{sec:concl}

With the recent discovery of the Higgs boson and the persistent
absence of any direct signal of NP, indirect searches represent even
more than before the best strategy to probe physics beyond the SM. In
particular, EWPO offer a very powerful handle on the mechanism of EWSB
and allow us to strongly constrain any NP relevant to solve the
hierarchy problem. In this context, we have presented an updated fit
of EWPO in the SM and beyond, obtained using a new code tested against
the ZFITTER one. We have discussed in detail the impact of the
recently computed two-loop fermionic EW corrections to the $Zf\bar{f}$
vertices, stressing the need for an independent evaluation of these
corrections for individual fermions. Our results in the SM are
summarized in tables~\ref{tab:SMfit}, \ref{tab:SMfit_oldRb} and
\ref{tab:SMpred}. We have obtained bounds on oblique NP contributions
(see table \ref{tab:STU}) and on $\epsilon$ parameters (see table
\ref{tab:Epsilon}), as well as SM predictions for $\epsilon_i$ in
eq.~(\ref{eq:SMepsilons_oldRb}). We have derived constraints on
modified $Zb\bar b$ couplings, see table~\ref{tab:Zbb}. We have
studied the bounds from EWPO on the Higgs coupling to vector bosons,
obtaining the results in table \ref{tab:HVV}, hinting at an elementary
Higgs boson or at a nontrivial composite Higgs model. Finally, we have
updated the constraints on the NP-induced dimension-six operators
relevant for the EWPO, reported in tables~\ref{tab:NPscale-univ},
\ref{tab:NPscale-nonuniv} and \ref{tab:Dim6Fit}. 

A graphical summary of the result for each observable is presented in Appendices
\ref{app:obs_oldRb} and \ref{app:obs_newRb}.

While the results we obtained are consistent with the non-observation
of NP at the $7$ and $8$-TeV runs, the possibility of
weakly-interacting NP hiding behind the corner remains unscathed.

\acknowledgments

M.C. is associated to the Dipartimento di Matematica e Fisica,
Universit\`a di Roma
Tre, and E.F. and L.S. are associated to the Dipartimento di Fisica,
Universit\`a di Roma ``La Sapienza''. We are indebted to R.~Contino
for stimulating discussions and suggestions, and for his careful
comparison of our preliminary results with the literature. We are
grateful to A.~Freitas for providing us with details of his
computations and unpublished results.  We thank J.~de Blas, 
G.~Degrassi, G.~Grilli di Cortona and A.~Strumia for useful
discussions and comments. We acknowledge useful discussions with
T.~Riemann on the comparison of our results with the ZFITTER ones.
We thank the BAT authors, in particular F.~Beaujean and D.~Kollar,
for precious support on the usage of their toolkit.
M.C, E.F and L.S. acknowledge partial support from MIUR (Italy) under
the PRIN 2010-2011 program.
The research leading to these results has received funding from the
European Research Council under the European Union's Seventh Framework
Programme (FP/2007-2013) / ERC Grant Agreements n.~279972 and n.~267985.

\appendix

\section{NP contributions to the EW precision observables}
\label{app:STUformulae}

We express each observable as a linear function of the NP parameters as in 
refs.~\cite{Peskin:1990zt,Peskin:1991sw,Maksymyk:1993zm,Burgess:1993mg,Burgess:1993vc}. 
Here we use $s_W^2$, $c_W^2$, $g_{V}^f$ and $g_{A}^f$ for the
corresponding SM values, and write the shift to the Fermi constant 
as $G_\mu = G_{\mu,\mathrm{SM}}( 1 + \Delta G)$. The corrections to the mass and
width\footnote{Our formula for the $W$-boson width in
  eq.~\eqref{eq:GammaW_NP} differs from that in
  ref.~\cite{Maksymyk:1993zm,Burgess:1993mg}, since we have expressed 
  the $W$-boson mass appearing in the phase-space factor in terms of
  the NP parameters.}  
of the $W$ boson are then given by 
\begin{align}
M_W &=
M_{W,\mathrm{SM}}
\left[
1 - \frac{\alpha(M_Z^2)}{4(c_W^2-s_W^2)}
\left( S - 2c_W^2\,T - \frac{c_W^2-s_W^2}{2s_W^2}\,U \right)
- \frac{s_W^2}{2(c_W^2-s_W^2)}\,\Delta G 
\right],
\\
\Gamma_W &= 
\Gamma_{W,\mathrm{SM}}
\left[ 1 
- \frac{3\alpha(M_Z^2)}{4(c_W^2-s_W^2)}
\left( S - 2c_W^2\,T
- \frac{c_W^2-s_W^2}{2s_W^2}\,U \right) 
- \frac{1+c_W^2}{2(c_W^2-s_W^2)}\, \Delta G
\right], 
\label{eq:GammaW_NP}
\end{align}
where $\Gamma_{W,\mathrm{SM}}$ is given in terms of $G_\mu$,
$M_{W,\mathrm{SM}}$ and so forth. 
Moreover, the shifts in the $Zf\bar{f}$ couplings read 
\begin{align}
\delta g_V^f &=
\frac{g_{V}^f}{2}
\left[ \alpha(M_Z^2)\, T - \Delta G \right]
+ 
\frac{\big( g_{V}^f - g_{A}^f \big)
\left[
\alpha(M_Z^2)\left( S - 4\,c_W^2s_W^2\, T \right)
+ 4\,c_W^2s_W^2\, \Delta G
\right]}{4s_W^2\,(c_W^2-s_W^2)}\,,
\\
\delta g_A^f &=
\frac{g_{A}^f}{2}
\left[ \alpha(M_Z^2)\, T - \Delta G \right], 
\end{align}
where we neglect the imaginary
parts of the SM couplings in NP contributions below. 
Using these couplings and defining the following quantities 
\begin{align}
G_f \equiv (g_{V}^f)^2 + (g_{A}^f)^2,\ \ \ \ \
\delta G_f\equiv 
2(g_{V}^f\,\delta g_V^f + g_{A}^f\,\delta g_A^f)\,,
\end{align}
the $Z$-pole observables are written as  
\begin{align}
\sin^2\theta^{\rm lept}_{\mathrm{eff}}
&=
\sin^2\theta^{\rm lept}_{\mathrm{eff,SM}}
- 
\frac{g_{A}^e\, \delta g_{V}^e - g_{V}^e\, \delta g_{A}^e}{4(g_A^e)^2}\,,
\\
\mathcal{A}_f &= 
\mathcal{A}_{f,\mathrm{SM}}
- \frac{2\big[ (g_{V}^f)^2 - (g_{A}^f)^2\big]}{ G_f^2 }\,
\big(g_{A}^f\, \delta g_{V}^f - g_{V}^f\, \delta g_{A}^f\big)\,,
\\
A_{\rm FB}^{0,f} &=
A_{\rm FB,SM}^{0,f} 
- 
\left[
\frac{3\, g_{V}^f\, g_{A}^f 
  \big[(g_{V}^e)^2 - (g_{A}^e)^2\big]}{ G_f G_e^2 }
\big(g_{A}^e\, \delta g_{V}^e - g_{V}^e\, \delta g_{A}^e\big)
+ (e \leftrightarrow f)
\right],
\\
\Gamma_Z &= 
\Gamma_{Z,\mathrm{SM}} 
+ 
\frac{\alpha(M_Z^2)M_Z}{12s_W^2c_W^2}
\sum_f N_c^f\, \delta G_f\,,
\\
\sigma_h^0 &=
\sigma_{h,\mathrm{SM}}^0 
+ 
\frac{12\pi}{M_Z^2}
\frac{ G_e \big(\sum_q N_c^q\, G_q\big) }
{\big(\sum_f N_c^f\, G_f \big)^2}
\left(
\frac{\delta G_e}{G_e}
+ \frac{\sum_q \delta G_q}{\sum_q G_q}
- \frac{2\sum_f N_c^f\, \delta G_f}{\sum_f N_c^f\, G_f}
\right),
\\
R_\ell^0 &= 
R_{\ell,\mathrm{SM}}^0 
+ 
\frac{\sum_q N_c^q\, \delta G_q}{G_\ell}
- \frac{(\sum_q N_c^q\, G_q)\,\delta G_\ell}{G_\ell^2}\,,
\\
R_c^0 &= 
R_{c,\mathrm{SM}}^0 
+ \frac{\delta G_c}{\sum_q G_q}
- \frac{G_c \sum_q \delta G_q}{\big(\sum_q G_q\big)^2}\,,
\\
R_b^0 &= 
R_{b,\mathrm{SM}}^0 
+ \frac{\delta G_b}{\sum_q G_q}
- \frac{G_b \sum_q \delta G_q}{\big(\sum_q G_q\big)^2}\,,
\end{align}
where $N_c^f = 3$ for quarks and $N_c^f = 1$ for leptons. 

\section{Non-universal vertex corrections}
\label{app:NonUnivCorr}

As shown in eqs.~\eqref{eq:eps1}-\eqref{eq:Delta_kappa_Prime}, the
parameters $\epsilon_1$, $\epsilon_2$ and $\epsilon_3$ are defined
from the $Ze\bar{e}$ effective couplings. To apply the same parameters
to other decay channels, flavour non-universal vertex corrections have to
be taken into account. 
Below we summarize the formul{\ae} of the non-universal corrections at
one-loop level, which can be found in ref.~\cite{Bardin:1999ak} and
references therein. 

The non-universal corrections to the effective couplings $\rho_Z^f$ and
$\kappa_Z^f$ are given by 
\begin{eqnarray}
\Delta \rho_Z^f &=& \rho_Z^f - \rho_Z^e 
=
\frac{\alpha}{2\pi s_W^2}\left(u_f - u_e\right),
\nonumber\\
\Delta \kappa_Z^f &=& \kappa_Z^f - \kappa_Z^e
=
\frac{\alpha}{4\pi s_W^2}
\left( \frac{\delta_f^2-\delta_e^2}{4c_W^2}\,\mathcal{F}_Z(M_Z^2)
-u_f+u_e\right),
\label{eq:NonUnivRhoKappa}
\end{eqnarray}
respectively, where $u_f$ and $\delta_f$ are defined as 
\begin{eqnarray}
u_f 
&=&
\frac{3v_f^2+a_f^2}{4c_W^2}\mathcal{F}_Z(M_Z^2)
+ \mathcal{F}_W^f(M_Z^2)\,,
\nonumber\\
\delta_f
&=&
v_f - a_f
\end{eqnarray}
with the tree-level vector and axial-vector couplings 
$v_f = I_3^f - 2Q_f s_W^2$ and 
$a_f = I_3^f$. The so-called unified form factors $\mathcal{F}_Z$ and
$\mathcal{F}_W^f$, associated with the radiative corrections to the
$Zf\bar{f}$ vertices with a virtual $Z$ boson and with a virtual $W$
boson(s), respectively, are given as follows: 
\begin{eqnarray}
\mathcal{F}_Z(s)
&=&
\mathcal{F}_{Za}(s)\,,
\nonumber\\
\mathcal{F}_W^f(s)
&=&
c_W^2 \mathcal{F}_{Wn}(s) 
- \frac{1}{2}|\sigma_{f'}| \mathcal{F}_{Wa}(s) 
- \frac{1}{2}\overline{\mathcal{F}}_{Wa}(s)\,,
\end{eqnarray}
where $|\sigma_{f'}| = |v_{f'} + a_{f'}|$ with $f'$ being the partner
of $f$ in the $SU(2)_L$ doublet, and the subscripts ``\textit{a}'' and
``\textit{n}'' stand for contributions from abelian and non-abelian
diagrams, respectively.  In the limit of massless fermions, the form
factors are written with the loop functions $B_0$ and $C_0$:
\begin{eqnarray}
\mathcal{F}_{Va}(s) 
&=&
2(R_V + 1)^2 s\, C_0(s;0,(\widetilde{M}_V^2)^{1/2},0) 
- (2R_V+3) \ln\bigg(-\frac{\widetilde{M}_V^2}{s}\bigg)
- 2R_V - \frac{7}{2}\,,
\nonumber\\
\mathcal{F}_{Wn}(s) 
&=&
- 2 (R_W + 2) M_W^2 C_0(s;M_W,0,M_W)
+ 2R_W + \frac{9}{2} - \frac{11}{18R_W} + \frac{1}{18R_W^2}
\nonumber\\
&&
- \left( 2R_W + \frac{7}{3} - \frac{3}{2R_W} - \frac{1}{12R_W^2} \right)
  B_0(\mu; s; M_W, M_W)\,,
\nonumber\\
\overline{\mathcal{F}}_{Wa}(s) 
&=& 0\,,
\end{eqnarray}
where $\widetilde{M}_V^2 \equiv M_V^2 - i\, M_V\Gamma_V\approx
M_V^2-i\varepsilon$, and 
$R_V=M_V^2/s$ for $V=Z,W$. The scalar two-point function $B_0$ and the
scalar three-point function $C_0$ are defined by 
\begin{eqnarray}
&&B_0(\mu; p^2; m_0,m_1) 
= 
\frac{(2\pi\mu)^{4-d}}{i\pi^2}\int d^dk\, 
\frac{1}{(k^2-m_0^2+i\varepsilon)\left[(k+p)^2-m_1^2+i\varepsilon\right]}\,,
\\
&&C_0(p^2;\, m_0,m_1,m_0) 
\nonumber\\
&&\hspace{5mm}=
-\frac{1}{i\pi^2}\int d^4k\, 
\frac{1}{(k^2-m_1^2+i\varepsilon)\left[(k+p_1)^2-m_0^2+i\varepsilon\right]
\left[(k-p_2)^2-m_0^2+i\varepsilon\right]}\,,
\end{eqnarray}
where $\mu$ is the renormalization scale in the former, and 
$p_1^2=p_2^2=0$ and $p^2=(p_1+p_2)^2$ in the latter. 
Note that the contributions from $\mathcal{F}_{Wn}(s)$ cancel out in 
eq.~\eqref{eq:NonUnivRhoKappa}.

In the case of $f=b$, the additional non-universal corrections associated 
with the heavy top-quark loop are parameterized by $\epsilon_b$ as shown 
in eq.~\eqref{eq:rhobnonuniv}.

\section{Correlation matrices for fit results}
\label{app:corr}

\begin{table}[ht]
\centering
\begin{tabular}{crrrrcrcc} 
\hline
& $\alpha_s$\ \ & $\Delta\alpha_{\rm had}^{(5)}$\hspace{-2mm} & 
$M_Z$\ & $m_t$\ \ & $m_h$ &
$\delta\rho_Z^\nu$\ \ & $\delta\rho_Z^\ell$ & $\delta\rho_Z^b$ 
\\
\hline
$\alpha_s$ &
$1.00$ &  &  &  &  &  &  &  \\
$\Delta\alpha_{\rm had}^{(5)}$ &
$-0.01$ & $1.00$ &  &  &  &  &  &  \\
$M_Z$ &
$0.00$ & $0.08$ & $1.00$ &  &  &  &  &  \\
$m_t$ &
$0.01$ & $0.18$ & $-0.05$ & $1.00$ &  &  &  &  \\
$m_h$ &
$0.00$ & $-0.01$ & $0.00$ & $0.00$ & $1.00$ &  &  &  \\
$\delta\rho_Z^\nu$ &
$0.00$ & $-0.01$ & $-0.05$ & $-0.02$ & $0.00$ & $1.00$ &  &  \\
$\delta\rho_Z^\ell$ &
$0.00$ & $0.02$ & $-0.10$ & $-0.08$ & $0.00$ & $0.49$ & $1.00$ &  \\
$\delta\rho_Z^b$ &
$-0.18$ & $0.10$ & $-0.02$ & $-0.06$ & $0.00$ & $-0.28$ & $0.38$ & $1.00$ \\
\hline
\end{tabular}
\caption{Correlation matrix for the fit in table~\ref{tab:SMfit}.}
\label{tab:SMCorr_newRb}
\end{table}

\begin{table}[ht]
\centering
\begin{tabular}{crrrrc} 
\hline
& $\alpha_s$\ \ & $\Delta\alpha_{\rm had}^{(5)}$\hspace{-2mm} & 
$M_Z$\ & $m_t$\hspace{1mm} & $m_h$ 
\\
\hline
$\alpha_s$ &
$1.00$ &  &  &  &  \\
$\Delta\alpha_{\rm had}^{(5)}$ &
$0.01$ & $1.00$ &  &  &  \\
$M_Z$ &
$0.00$ & $0.09$ & $1.00$ &  &  \\
$m_t$ &
$0.01$ & $0.19$ & $-0.06$ & $1.00$ &  \\
$m_h$ &
$0.00$ & $0.00$ & $0.00$ & $0.00$ & $1.00$ \\
\hline
\end{tabular}
\caption{Correlation matrix for the fit in table~\ref{tab:SMfit_oldRb}.}
\label{tab:SMCorr_oldRb}
\end{table}

\begin{table}[ht]
\centering
\begin{tabular}{crrrccccc}
\hline
& $S$\ \, & $T$\ \, & $U$\ \,  & 
$\alpha_s$ & $\Delta\alpha_{\rm had}^{(5)}$ & 
$M_Z$ & $m_t$ & $m_h$ 
\\
\hline
$S$ &
$1.00$ &  &  &  &  &  &  &  \\
$T$ &
$0.85$ & $1.00$ &  &  &  &  &  &  \\
$U$ &
$-0.48$ & $-0.79$ & $1.00$ &  &  &  &  &  \\
$\alpha_s$ &
$-0.07$ & $-0.10$ & $0.10$ & $1.00$ &  &  &  &  \\
$\Delta\alpha_{\rm had}^{(5)}$ &
$-0.30$ & $0.00$ & $-0.09$ & $0.00$ & $1.00$ &  &  &  \\
$M_Z$ &
$-0.05$ & $-0.10$ & $0.03$ & $0.01$ & $0.00$ & $1.00$ &  &  \\
$m_t$ &
$0.02$ & $-0.07$ & $-0.04$ & $0.01$ & $0.00$ & $0.00$ & $1.00$ &  \\
$m_h$ &
$0.00$ & $0.00$ & $0.00$ & $0.00$ & $0.00$ & $0.00$ & $0.00$ & $1.00$ \\
\hline
\end{tabular}
\caption{Correlation matrix for the fit results in the second column
  of table~\ref{tab:STU}.}
\label{tab:STUcorr_oldRb}
\end{table}

\begin{table}[ht]
\centering
\begin{tabular}{crrccccc}
\hline
& $S$\ \, & $T$\ \, &
$\alpha_s$ & $\Delta\alpha_{\rm had}^{(5)}$ & 
$M_Z$ & $m_t$ & $m_h$ 
\\
\hline
$S$ &
$1.00$ &  &  &  &  &  &  \\
$T$ &
$0.86$ & $1.00$ &  &  &  &  &  \\
$\alpha_s$ &
$-0.03$ & $-0.03$ & $1.00$ &  &  &  &  \\
$\Delta\alpha_{\rm had}^{(5)}$ &
$-0.40$ & $-0.12$ & $0.01$ & $1.00$ &  &  &  \\
$M_Z$ &
$-0.04$ & $-0.12$ & $0.00$ & $0.00$ & $1.00$ &  &  \\
$m_t$ &
$0.00$ & $-0.17$ & $0.01$ & $0.00$ & $0.00$ & $1.00$ &  \\
$m_h$ &
$0.00$ & $0.00$ & $0.00$ & $0.00$ & $0.00$ & $0.00$ & $1.00$ \\
\hline
\end{tabular}
\caption{Correlation matrix for the fit results in the third column
  of table~\ref{tab:STU}.}
\label{tab:STcorr_oldRb}
\end{table}

\begin{table}[ht]
\centering
\begin{tabular}{crrccccc} 
\hline
& $S$\ \, & $T$\ \, & 
$\alpha_s$ & $\Delta\alpha_{\rm had}^{(5)}$\hspace{-2mm} & 
$M_Z$ & $m_t$ & $m_h$ 
\\
\hline
$S$ &
$1.00$ &  &  &  &  &  &  \\
$T$ &
$0.89$ & $1.00$ &  &  &  &  &  \\
$\alpha_s$ &
$0.00$ & $0.01$ & $1.00$ &  &  &  &  \\
$\Delta\alpha_{\rm had}^{(5)}$ &
$-0.40$ & $-0.14$ & $0.00$ & $1.00$ &  &  &  \\
$M_Z$ &
$-0.01$ & $-0.08$ & $0.00$ & $0.00$ & $1.00$ &  &  \\
$m_t$ &
$-0.01$ & $-0.16$ & $0.00$ & $0.00$ & $0.00$ & $1.00$ &  \\
$m_h$ &
$0.00$ & $0.00$ & $0.00$ & $0.00$ & $0.00$ & $0.00$ & $1.00$ \\
\hline
\end{tabular}
\caption{Correlation matrix for the fit results in the fourth column
  of table~\ref{tab:STU}.}
\label{tab:STcorr_newRb}
\end{table}

\begin{table}[ht]
\centering
\begin{tabular}{crrrrccccc} 
\hline
& 
$\epsilon_1$\ \ & $\epsilon_2$\ \ & $\epsilon_3$\ \ & 
$\epsilon_b$\ \, & $\alpha_s$ & $\Delta\alpha_{\rm had}^{(5)}$ & 
$M_Z$ & $m_t$ & $m_h$ 
\\
\hline
$\epsilon_1$ &
$1.00$ &  &  &  &  &  &  &  &  \\
$\epsilon_2$ &
$0.79$ & $1.00$ &  &  &  &  &  &  &  \\
$\epsilon_3$ &
$0.86$ & $0.50$ & $1.00$ &  &  &  &  &  &  \\
$\epsilon_b$ &
$-0.32$ & $-0.31$ & $-0.21$ & $1.00$ &  &  &  &  &  \\
$\alpha_s$ &
$-0.06$ & $-0.06$ & $-0.04$ & $-0.12$ & $1.00$ &  &  &  &  \\
$\Delta\alpha_{\rm had}^{(5)}$ &
$0.00$ & $0.07$ & $-0.30$ & $0.00$ & $0.00$ & $1.00$ &  &  &  \\
$M_Z$ &
$-0.10$ & $-0.03$ & $-0.05$ & $0.02$ & $0.00$ & $0.00$ & $1.00$ &  &  \\
$m_t$ &
$0.00$ & $0.00$ & $0.00$ & $0.00$ & $0.00$ & $0.00$ & $0.00$ & $1.00$ &  \\
$m_h$ &
$0.00$ & $0.00$ & $0.00$ & $0.00$ & $0.00$ & $0.00$ & $0.00$ & $0.00$ & $1.00$ \\
\hline
\end{tabular}
\caption{Correlation matrix for the fit results in the second column
  of table~\ref{tab:Epsilon}.}
\label{tab:4epsCorr_oldRb}
\end{table}

\begin{table}[ht]
\centering
\begin{tabular}{crrcrrcc} 
\hline
&
$\epsilon_1$\ \ & $\epsilon_3$\ \ & 
$\alpha_s$ & $\Delta\alpha_{\rm had}^{(5)}$\hspace{-2mm} & 
$M_Z$\ & $m_t$ & $m_h$ 
\\
\hline
$\epsilon_1$ &
$1.00$ &  &  &  &  &  &  \\
$\epsilon_3$ &
$0.87$ & $1.00$ &  &  &  &  &  \\
$\alpha_s$ &
$-0.04$ & $-0.03$ & $1.00$ &  &  &  &  \\
$\Delta\alpha_{\rm had}^{(5)}$ &
$-0.10$ & $-0.39$ & $0.01$ & $1.00$ &  &  &  \\
$M_Z$ &
$-0.12$ & $-0.04$ & $0.02$ & $0.00$ & $1.00$ &  &  \\
$m_t$ &
$-0.03$ & $-0.01$ & $0.01$ & $-0.01$ & $-0.01$ & $1.00$ &  \\
$m_h$ &
$0.00$ & $0.00$ & $0.00$ & $0.00$ & $0.00$ & $0.00$ & $1.00$ \\
\hline
\end{tabular}
\caption{Correlation matrix for the fit results in the third column
  of table~\ref{tab:Epsilon}.}
\label{tab:2epsCorr_oldRb}
\end{table}

\begin{table}[ht]
\centering
\begin{tabular}{crrcrrcc} 
\hline
& 
$\delta g_R^b$\ \ & $\delta g_L^b$\ \ &
$\alpha_s$ & $\Delta\alpha_{\rm had}^{(5)}$\hspace{-2mm} & 
$M_Z$\ \ & $m_t$ & $m_h$ 
\\
\hline
$\delta g_R^b$ &
$1.00$ &  &  &  &  &  &  \\
$\delta g_L^b$ &
$0.90$ & $1.00$ &  &  &  &  &  \\
$\alpha_s$ &
$0.04$ & $-0.02$ & $1.00$ &  &  &  &  \\
$\Delta\alpha_{\rm had}^{(5)}$ &
$-0.22$ & $-0.21$ & $0.00$ & $1.00$ &  &  &  \\
$M_Z$ &
$0.01$ & $0.01$ & $0.00$ & $0.08$ & $1.00$ &  &  \\
$m_t$ &
$0.01$ & $0.02$ & $0.00$ & $0.18$ & $-0.06$ & $1.00$ &  \\
$m_h$ &
$0.00$ & $0.00$ & $0.00$ & $-0.01$ & $0.00$ & $0.00$ & $1.00$ \\
\hline
\end{tabular}
\caption{Correlation matrix for the fit results in the second column
  of table~\ref{tab:Zbb}.}
\label{tab:ZbbCorr_oldRb}
\end{table}

\begin{table}[ht]
\centering
\begin{tabular}{crrrcrcc} 
\hline
& 
$\delta g_R^b$\ \ & $\delta g_L^b$\ \ &
$\alpha_s$\ \ & $\Delta\alpha_{\rm had}^{(5)}$\hspace{-2mm} & 
$M_Z$\ & $m_t$ & $m_h$ 
\\
\hline
$\delta g_R^b$ &
$1.00$ &  &  &  &  &  &  \\
$\delta g_L^b$ &
$0.82$ & $1.00$ &  &  &  &  &  \\
$\alpha_s$ &
$-0.01$ & $0.00$ & $1.00$ &  &  &  &  \\
$\Delta\alpha_{\rm had}^{(5)}$ &
$-0.19$ & $-0.22$ & $-0.01$ & $1.00$ &  &  &  \\
$M_Z$ &
$0.01$ & $0.01$ & $0.00$ & $0.08$ & $1.00$ &  &  \\
$m_t$ &
$-0.01$ & $0.02$ & $0.01$ & $0.17$ & $-0.05$ & $1.00$ &  \\
$m_h$ &
$0.00$ & $0.00$ & $0.00$ & $0.00$ & $0.00$ & $0.00$ & $1.00$ \\
\hline
\end{tabular}
\caption{Correlation matrix for the fit results in the third column
  of table~\ref{tab:Zbb}.}
\label{tab:ZbbCorr_newRb}
\end{table}

\begin{table}[ht]
\centering
\begin{tabular}{crcrccc} 
\hline
& 
$a$\ \ \ & 
$\alpha_s$ & $\Delta\alpha_{\rm had}^{(5)}$\hspace{-2mm} & 
$M_Z$ & $m_t$ & $m_h$ 
\\
\hline
$a$ &
$1.00$ &  &  &  &  &  \\
$\alpha_s$ &
$-0.02$ & $1.00$ &  &  &  &  \\
$\Delta\alpha_{\rm had}^{(5)}$ &
$0.53$ & $0.00$ & $1.00$ &  &  &  \\
$M_Z$ &
$-0.17$ & $0.00$ & $-0.02$ & $1.00$ &  &  \\
$m_t$ &
$-0.32$ & $0.01$ & $-0.02$ & $0.00$ & $1.00$ &  \\
$m_h$ &
$0.01$ & $0.00$ & $0.00$ & $0.00$ & $0.00$ & $1.00$ \\
\hline
\end{tabular}
\caption{Correlation matrix for the fit result in top left entry of
  table~\ref{tab:HVV}.} 
\label{tab:HVVCorr_oldRb}
\end{table}

\begin{table}[ht]
\centering
\begin{tabular}{crcrccc}
\hline
& $a$\ \ \ & $\alpha_s$ & 
$\Delta\alpha_{\rm had}^{(5)}$\hspace{-2mm} & 
$M_Z$ & $m_t$ & $m_h$ \\
\hline
$a$ &
$1.00$ &  &  &  &  &  \\
$\alpha_s$ &
$0.02$ & $1.00$ &  &  &  &  \\
$\Delta\alpha_{\rm had}^{(5)}$ &
$0.53$ & $0.00$ & $1.00$ &  &  &  \\
$M_Z$ &
$-0.15$ & $0.00$ & $-0.01$ & $1.00$ &  &  \\
$m_t$ &
$-0.32$ & $0.00$ & $-0.02$ & $0.00$ & $1.00$ &  \\
$m_h$ &
$0.01$ & $0.00$ & $0.00$ & $0.00$ & $0.00$ & $1.00$ \\
\hline
\end{tabular}
\caption{Same as table \ref{tab:HVVCorr_oldRb} but for the top-right
  entry.} 
\label{tab:HVVCorr_newRb}
\end{table}

\begin{table}[ht]
\centering
\begin{tabular}{crcccccc}
\hline
& $C_{WB}$\ & $C_{H}$ 
& $\alpha_s$\ \ & $\Delta\alpha_{\rm had}^{(5)}$\hspace{-1mm} &
$M_Z$\ & $m_t$ & $m_h$
\\
\hline
$C_{WB}$ &
$1.00$ &  &  &  &  &  &  \\
$C_{H}$ &
$-0.86$ & $1.00$ &  &  &  &  &  \\
$\alpha_s$ &
$-0.03$ & $0.03$ & $1.00$ &  &  &  &  \\
$\Delta\alpha_{\rm had}^{(5)}$ &
$-0.40$ & $0.12$ & $0.01$ & $1.00$ &  &  &  \\
$M_Z$ &
$-0.04$ & $0.12$ & $0.00$ & $0.00$ & $1.00$ &  &  \\
$m_t$ &
$0.00$ & $0.17$ & $0.01$ & $0.00$ & $0.00$ & $1.00$ &  \\
$m_h$ &
$0.00$ & $0.00$ & $0.00$ & $0.00$ & $0.00$ & $0.00$ & $1.00$ \\
\hline
\end{tabular}
\caption{Correlation matrix for the fit results in
  table~\ref{tab:Dim6Fit} (first set of operators, using the
  large-$m_t$ expansion).}
\label{tab:Dim6FitCorr1}
\end{table}

\begin{table}[ht]
\setlength{\tabcolsep}{0.7\tabcolsep}
\centering
\begin{tabular}{crrrrrrrccccc}
\hline
& $C_{HL}^\prime$\ & $C_{HQ}^\prime$\ & $C_{HL}$\ & $C_{HQ}$\
& $C_{HE}$\ & $C_{HU}$\ & $C_{HD}$\
& $\alpha_s$ & $\Delta\alpha_{\rm had}^{(5)}$\hspace{-1mm} &
$M_Z$ & $m_t$ & $m_h$
\\
\hline
$C_{HL}^\prime$ &
$1.00$ &  &  &  &  &  &  &  &  &  &  &  \\
$C_{HQ}^\prime$ &
$0.24$ & $1.00$ &  &  &  &  &  &  &  &  &  &  \\
$C_{HL}$ &
$0.56$ & $0.04$ & $1.00$ &  &  &  &  &  &  &  &  &  \\
$C_{HQ}$ &
$0.00$ & $-0.38$ & $-0.09$ & $1.00$ &  &  &  &  &  &  &  &  \\
$C_{HE}$ &
$0.58$ & $-0.05$ & $0.72$ & $-0.11$ & $1.00$ &  &  &  &  &  &  &  \\
$C_{HU}$ &
$-0.01$ & $-0.78$ & $0.05$ & $0.62$ & $0.04$ & $1.00$ &  &  &  &  &  &  \\
$C_{HD}$ &
$0.01$ & $0.61$ & $-0.24$ & $0.36$ & $-0.26$ & $-0.11$ & $1.00$ &  &  &  &  &  \\
$\alpha_s$ &
$-0.02$ & $-0.03$ & $-0.01$ & $0.00$ & $-0.01$ & $0.00$ & $0.00$ & $1.00$ &  &  &  &  \\
$\Delta\alpha_{\rm had}^{(5)}$ &
$-0.36$ & $-0.04$ & $0.12$ & $-0.01$ & $0.12$ & $-0.01$ & $0.01$ & $0.00$ & $1.00$ &  &  &  \\
$M_Z$ &
$0.16$ & $0.02$ & $0.06$ & $0.00$ & $0.12$ & $0.00$ & $-0.01$ & $0.00$ & $0.00$ & $1.00$ &  &  \\
$m_t$ &
$0.32$ & $0.06$ & $0.12$ & $0.02$ & $0.17$ & $0.00$ & $0.00$ & $0.00$ & $0.00$ & $0.00$ & $1.00$ &  \\
$m_h$ &
$-0.01$ & $0.00$ & $0.00$ & $0.00$ & $0.00$ & $0.00$ & $0.00$ & $0.00$ & $0.00$ & $0.00$ & $0.00$ & $1.00$ \\
\hline
\end{tabular}
\caption{Same as table \ref{tab:Dim6FitCorr1}, but for the second set
  of operators.}
\label{tab:Dim6FitCorr2}
\end{table}

\begin{table}[ht]
\small
\setlength{\tabcolsep}{0.5\tabcolsep}
\centering
\begin{tabular}{crrrrrrrrrccccc}
\hline
& $C_{HL}^\prime$ & $C_{HL}$ & $C_{HE}$ & $C_{HU_2}$ & $C_{HD_3}$ &
$C_{12}$ & $C_{2}$\ \ & $C_{3}$\ \ & $C[\Gamma_{uds}]$\hspace{-2mm} &
$\alpha_s$ & $\Delta\alpha_{\rm had}^{(5)}$\hspace{-1mm} & 
$M_Z$ & $m_t$ & $m_h$ \\
\hline
$C_{HL}^\prime$ &
$1.00$ &  &  &  &  &  &  &  &  &  &  &  &  &  \\
$C_{HL}$ &
$0.56$ & $1.00$ &  &  &  &  &  &  &  &  &  &  &  &  \\
$C_{HE}$ &
$0.58$ & $0.72$ & $1.00$ &  &  &  &  &  &  &  &  &  &  &  \\
$C_{HU_2}$ &
$-0.03$ & $0.04$ & $0.04$ & $1.00$ &  &  &  &  &  &  &  &  &  &  \\
$C_{HD_3}$ &
$0.03$ & $-0.23$ & $-0.26$ & $-0.13$ & $1.00$ &  &  &  &  &  &  &  &  &  \\
$C_{12}$ &
$0.02$ & $0.02$ & $0.01$ & $-0.02$ & $0.00$ & $1.00$ &  &  &  &  &  &  &  &  \\
$C_{2}$ &
$0.11$ & $0.05$ & $0.02$ & $-0.30$ & $0.14$ & $0.02$ & $1.00$ &  &  &  &  &  &  &  \\
$C_{3}$ &
$0.23$ & $-0.04$ & $-0.14$ & $-0.17$ & $0.87$ & $0.01$ & $0.11$ & $1.00$ &  &  &  &  &  &  \\
$C[\Gamma_{uds}]$ &
$-0.01$ & $0.08$ & $-0.03$ & $-0.30$ & $-0.03$ & $0.00$ & $-0.75$ & $-0.01$ & $1.00$ &  &  &  &  &  \\
$\alpha_s$ &
$-0.02$ & $-0.01$ & $-0.01$ & $0.00$ & $0.00$ & $-0.01$ & $-0.01$ & $-0.03$ & $-0.03$ & $1.00$ &  &  &  &  \\
$\Delta\alpha_{\rm had}^{(5)}$ &
$-0.36$ & $0.12$ & $0.12$ & $-0.01$ & $0.00$ & $0.00$ & $-0.02$ & $-0.05$ & $0.04$ & $0.00$ & $1.00$ &  &  &  \\
$M_Z$ &
$0.16$ & $0.06$ & $0.12$ & $0.00$ & $-0.01$ & $0.00$ & $0.01$ & $0.02$ & $-0.02$ & $0.00$ & $0.00$ & $1.00$ &  &  \\
$m_t$ &
$0.31$ & $0.11$ & $0.16$ & $-0.01$ & $0.01$ & $0.00$ & $0.03$ & $0.08$ & $-0.02$ & $0.00$ & $0.00$ & $0.00$ & $1.00$ &  \\
$m_h$ &
$-0.01$ & $0.00$ & $0.00$ & $0.00$ & $0.00$ & $0.00$ & $0.00$ & $0.00$ & $0.00$ & $0.00$ & $0.00$ & $0.00$ & $0.00$ & $1.00$ \\
\hline
\end{tabular}
\caption{Same as table \ref{tab:Dim6FitCorr1}, but for the third set
  of operators, where $C_{12}$, $C_2$ and $C_3$ denote 
$C_{HQ_1}^\prime + C_{HQ_2}^\prime$, 
$C_{HQ_2}^\prime - C_{HQ_2}$ and 
$C_{HQ_3}^\prime + C_{HQ_3}$, respectively.}
\label{tab:Dim6FitCorr3}
\end{table}

\begin{table}[ht]
\centering
\begin{tabular}{crrccccc}
\hline
& $C_{WB}$\ & $C_{H}$\ & $\alpha_s$ & $\Delta\alpha_{\rm had}^{(5)}$\hspace{-1mm} & $M_Z$ & $m_t$ & $m_h$ \\
\hline
$C_{WB}$ &
$1.00$ &  &  &  &  &  &  \\
$C_{H}$ &
$-0.89$ & $1.00$ &  &  &  &  &  \\
$\alpha_s$ &
$0.00$ & $-0.01$ & $1.00$ &  &  &  &  \\
$\Delta\alpha_{\rm had}^{(5)}$ &
$-0.40$ & $0.15$ & $0.00$ & $1.00$ &  &  &  \\
$M_Z$ &
$-0.02$ & $0.09$ & $0.00$ & $0.00$ & $1.00$ &  &  \\
$m_t$ &
$-0.02$ & $0.16$ & $0.00$ & $0.00$ & $0.00$ & $1.00$ &  \\
$m_h$ &
$0.00$ & $0.00$ & $0.00$ & $0.00$ & $0.00$ & $0.00$ & $1.00$ \\
\hline
\end{tabular}
\caption{Same as table \ref{tab:Dim6FitCorr1}, using the results from
  ref.~\cite{Freitas:2012sy,freitasprivate}.}
\label{tab:Dim6FitCorr1newRb}
\end{table}

\begin{table}[ht]
\setlength{\tabcolsep}{0.8\tabcolsep}
\centering
\begin{tabular}{crrrrrrccccc}
\hline
& $C_{HL}^\prime$ & $C_{HQ}^\prime$ & 
$C_{HQ}$ & $C_{HU}$ & $C_{HD}$ & $C[\mathcal{A}_\ell]$\hspace{-1mm} &
$\alpha_s$ & $\Delta\alpha_{\rm had}^{(5)}$\hspace{-1mm} & $M_Z$ & $m_t$ & $m_h$ \\
\hline
$C_{HL}^\prime$ &
$1.00$ &  &  &  &  &  &  &  &  &  &  \\
$C_{HQ}^\prime$ &
$0.00$ & $1.00$ &  &  &  &  &  &  &  &  &  \\
$C_{HQ}$ &
$-0.03$ & $0.31$ & $1.00$ &  &  &  &  &  &  &  &  \\
$C_{HU}$ &
$-0.04$ & $0.35$ & $0.70$ & $1.00$ &  &  &  &  &  &  &  \\
$C_{HD}$ &
$0.02$ & $-0.12$ & $0.18$ & $-0.25$ & $1.00$ &  &  &  &  &  &  \\
$C[\mathcal{A}_\ell]$ &
$-0.23$ & $0.00$ & $-0.09$ & $0.10$ & $-0.36$ & $1.00$ &  &  &  &  &  \\
$\alpha_s$ &
$-0.02$ & $-0.01$ & $0.01$ & $0.00$ & $0.00$ & $0.01$ & $1.00$ &  &  &  &  \\
$\Delta\alpha_{\rm had}^{(5)}$ &
$-0.35$ & $0.01$ & $0.00$ & $0.00$ & $0.00$ & $0.52$ & $0.00$ & $1.00$ &  &  &  \\
$M_Z$ &
$0.15$ & $0.00$ & $-0.01$ & $0.00$ & $0.00$ & $-0.05$ & $0.00$ & $0.00$ & $1.00$ &  &  \\
$m_t$ &
$0.31$ & $-0.01$ & $0.02$ & $-0.02$ & $0.00$ & $-0.13$ & $0.00$ & $0.00$ & $0.00$ & $1.00$ &  \\
$m_h$ &
$-0.01$ & $0.00$ & $0.00$ & $0.00$ & $0.00$ & $0.01$ & $0.00$ & $0.00$ & $0.00$ & $0.00$ & $1.00$ \\
\hline
\end{tabular}
\caption{Same as table \ref{tab:Dim6FitCorr1newRb}, but for the second set
  of operators.}
\label{tab:Dim6FitCorr2newRb}
\end{table}

\begin{table}[ht]
\small
\setlength{\tabcolsep}{0.5\tabcolsep}
\centering
\begin{tabular}{crrrrrrrrccccc}
\hline
& $C_{HL}^\prime$ & $C_{HU_2}$ & $C_{HD_3}$ & 
$C_{12}$ & $C_{2}$\ \ & $C_{3}$\ \ &
$C[\mathcal{A}_\ell]$\hspace{-1mm} & $C[\Gamma_{uds}]$\hspace{-2mm} &
$\alpha_s$ & $\Delta\alpha_{\rm had}^{(5)}$\hspace{-1mm} & $M_Z$ & $m_t$ & $m_h$ \\
\hline
$C_{HL}^\prime$ &
$1.00$ &  &  &  &  &  &  &  &  &  &  &  &  \\
$C_{HU_2}$ &
$-0.04$ & $1.00$ &  &  &  &  &  &  &  &  &  &  &  \\
$C_{HD_3}$ &
$0.02$ & $-0.26$ & $1.00$ &  &  &  &  &  &  &  &  &  &  \\
$C_{12}$ &
$0.00$ & $0.28$ & $-0.11$ & $1.00$ &  &  &  &  &  &  &  &  &  \\
$C_{2}$ &
$0.01$ & $0.50$ & $-0.19$ & $0.43$ & $1.00$ &  &  &  &  &  &  &  &  \\
$C_{3}$ &
$0.00$ & $0.61$ & $-0.14$ & $0.46$ & $0.92$ & $1.00$ &  &  &  &  &  &  &  \\
$C[\mathcal{A}_\ell]$ &
$-0.23$ & $0.07$ & $-0.35$ & $0.00$ & $-0.01$ & $-0.04$ & $1.00$ &  &  &  &  &  &  \\
$C[\Gamma_{uds}]$ &
$-0.03$ & $0.61$ & $-0.26$ & $0.46$ & $0.89$ & $0.99$ & $0.01$ & $1.00$ &  &  &  &  &  \\
$\alpha_s$ &
$-0.02$ & $0.00$ & $0.00$ & $0.00$ & $0.00$ & $0.00$ & $0.01$ & $0.00$ & $1.00$ &  &  &  &  \\
$\Delta\alpha_{\rm had}^{(5)}$ &
$-0.35$ & $0.00$ & $0.00$ & $0.01$ & $0.00$ & $0.00$ & $0.52$ & $0.01$ & $0.00$ & $1.00$ &  &  &  \\
$M_Z$ &
$0.15$ & $0.00$ & $0.00$ & $0.00$ & $0.00$ & $0.00$ & $-0.05$ & $-0.01$ & $0.00$ & $0.00$ & $1.00$ &  &  \\
$m_t$ &
$0.31$ & $-0.01$ & $0.00$ & $-0.01$ & $0.00$ & $0.00$ & $-0.13$ & $-0.01$ & $0.00$ & $0.00$ & $0.00$ & $1.00$ &  \\
$m_h$ &
$-0.01$ & $0.00$ & $0.00$ & $0.00$ & $0.00$ & $0.00$ & $0.01$ & $0.00$ & $0.00$ & $0.00$ & $0.00$ & $0.00$ & $1.00$ \\
\hline
\end{tabular}
\caption{Same as table \ref{tab:Dim6FitCorr1newRb}, but for the third set
  of operators, where $C_{12}$, $C_2$ and $C_3$ denote 
$C_{HQ_1}^\prime + C_{HQ_2}^\prime$, 
$C_{HQ_2}^\prime - C_{HQ_2}$ and 
$C_{HQ_3}^\prime + C_{HQ_3}$, respectively.}
\label{tab:Dim6FitCorr3newRb}
\end{table}

\clearpage
\section{Fit results for the observables with the large-$m_t$
  expansion for two-loop fermionic EW corrections to $\rho_Z^f$}
\label{app:obs_oldRb}

In this Appendix we present a graphical summary of the fit results for
all observables obtained within the various scenarios considered in
this work, obtained using the large-$m_t$ expansion for the two-loop
fermionic EW corrections to $\rho_Z^f$. The labels in the figures
refer to the various fits performed with a self-explanatory
notation. The blue band corresponds to the direct measurement, also
reported with the ``Data'' label. 

\begin{figure}[htbp]
\centering
\includegraphics[width=.49\textwidth]{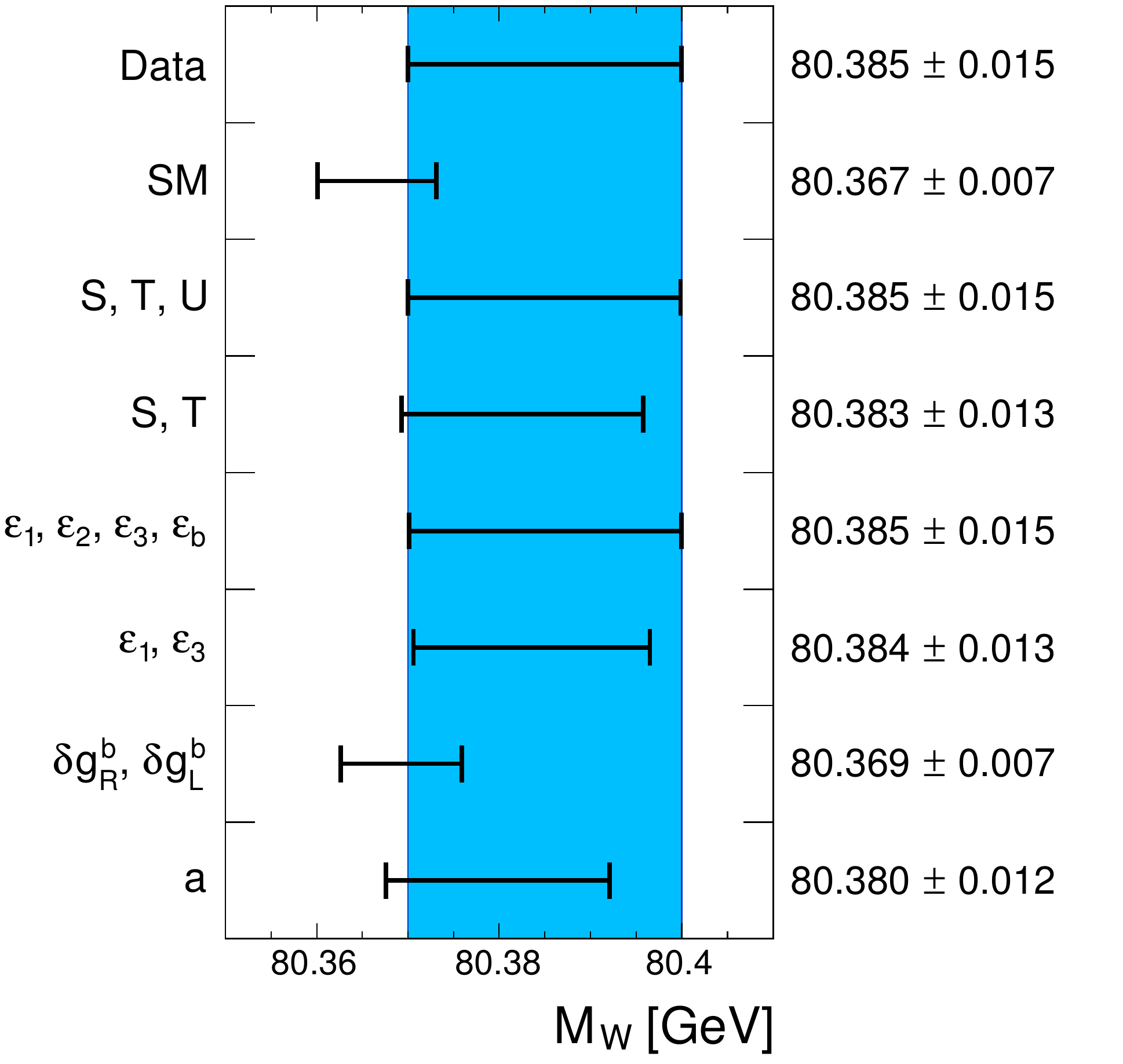}
\hfill
\includegraphics[width=.49\textwidth]{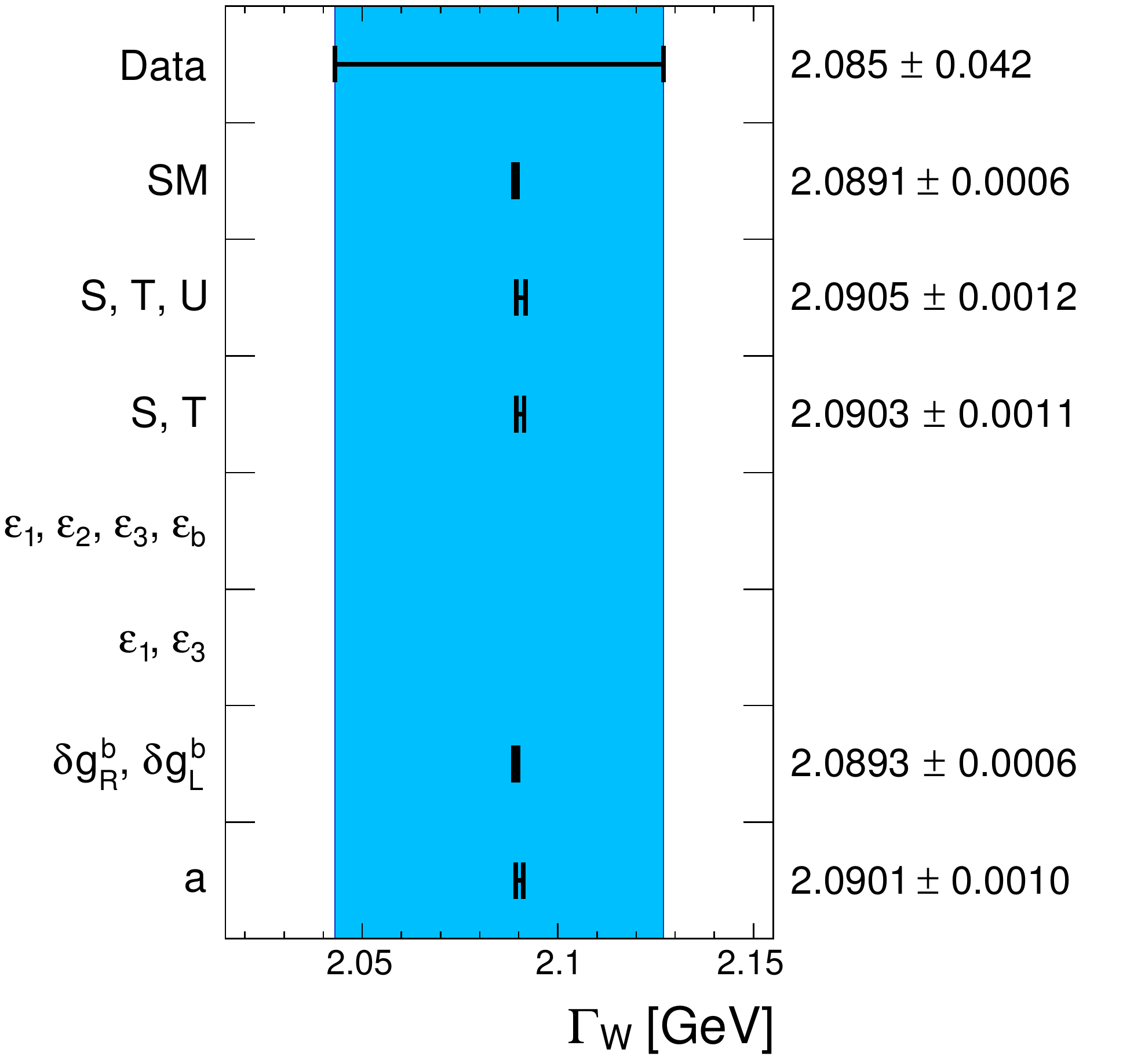}
\\[2mm]
\includegraphics[width=.49\textwidth]{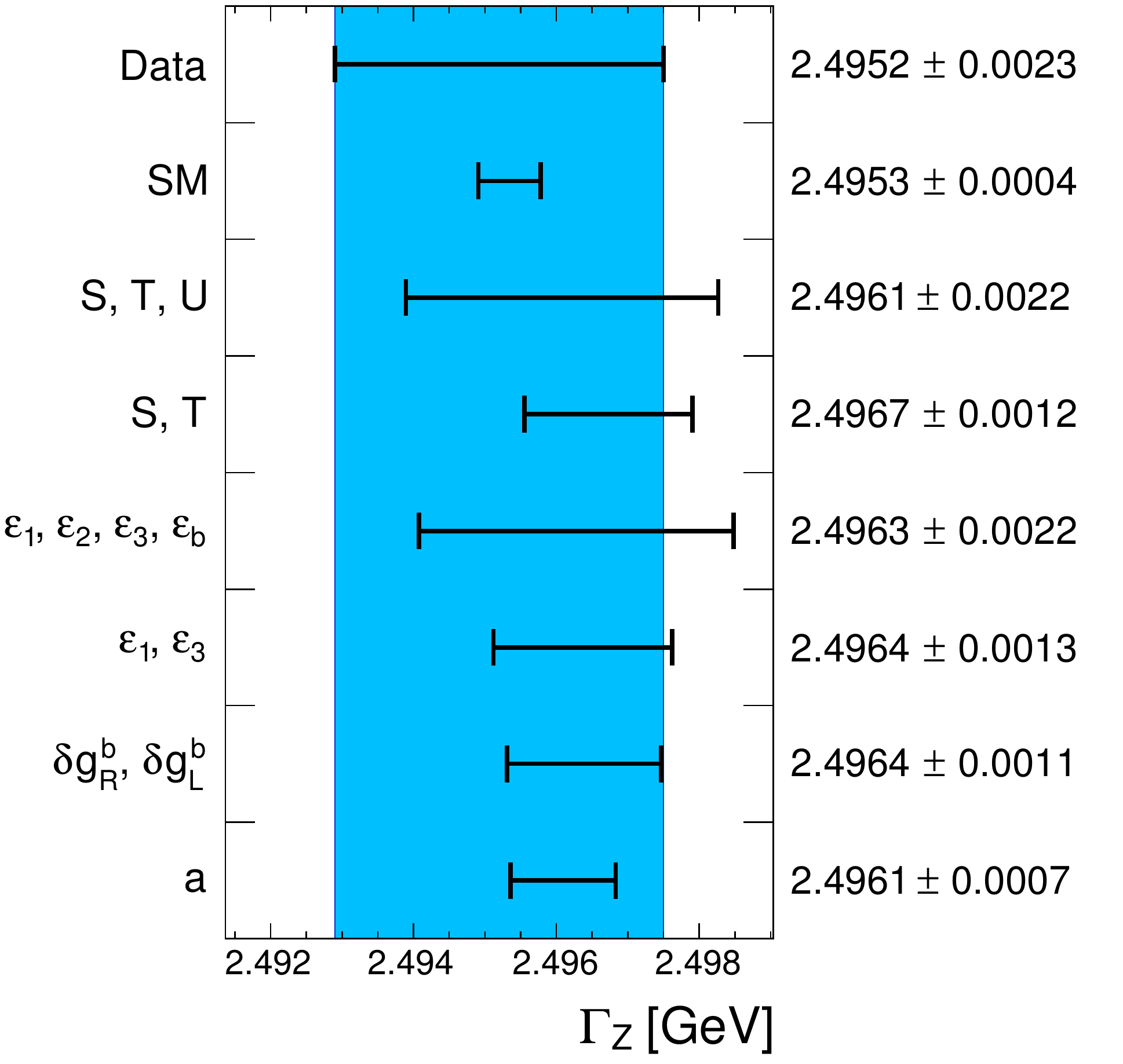}
\hfill
\includegraphics[width=.49\textwidth]{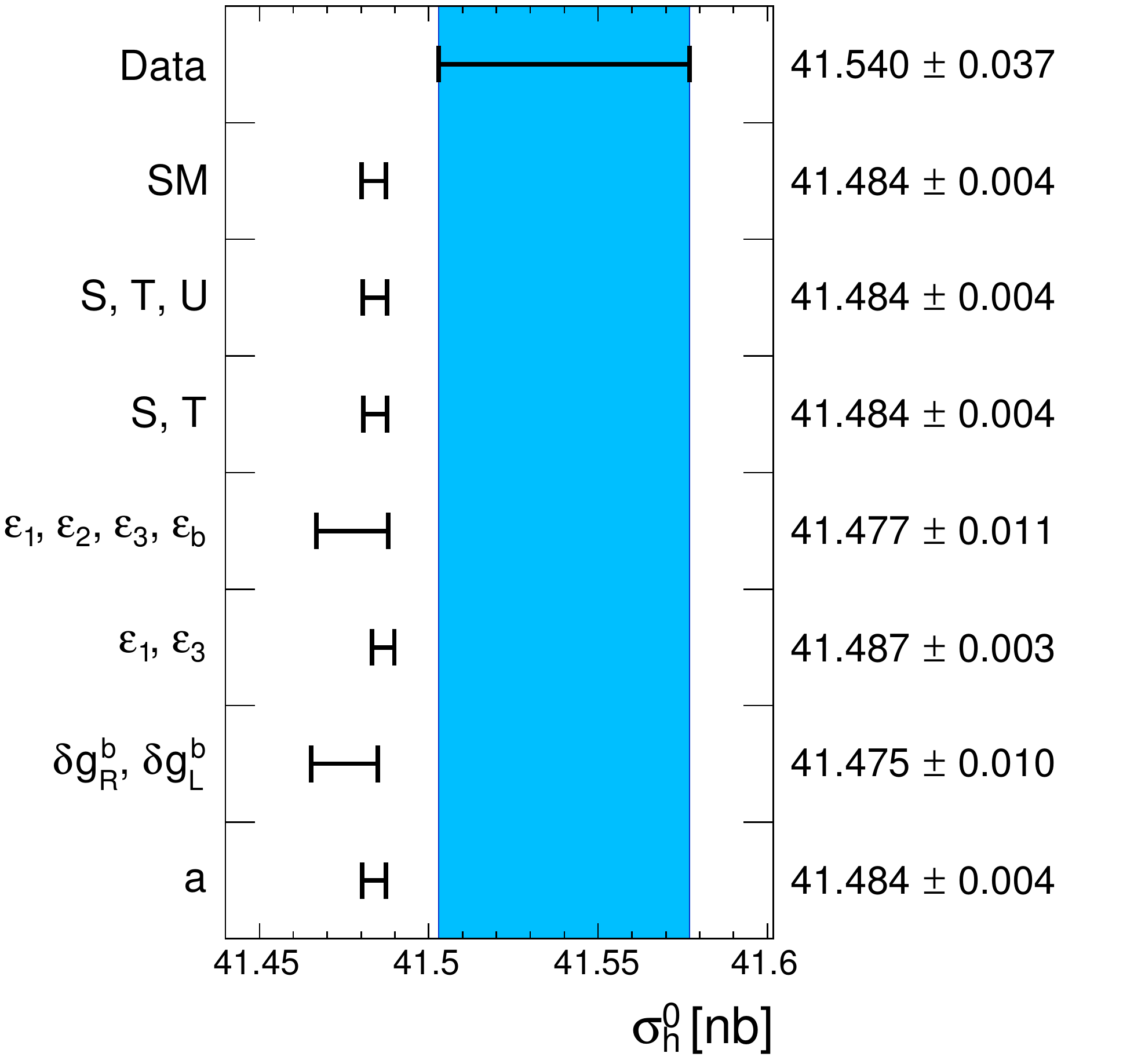}
\caption{Fit results, with the large-$m_t$ expansion for the two-loop
  fermionic EW corrections to the coupling $\rho_Z^f$.}
\label{fig:oldRb_summary1}
\end{figure}

\begin{figure}[htbp]
\centering
\includegraphics[width=.49\textwidth]{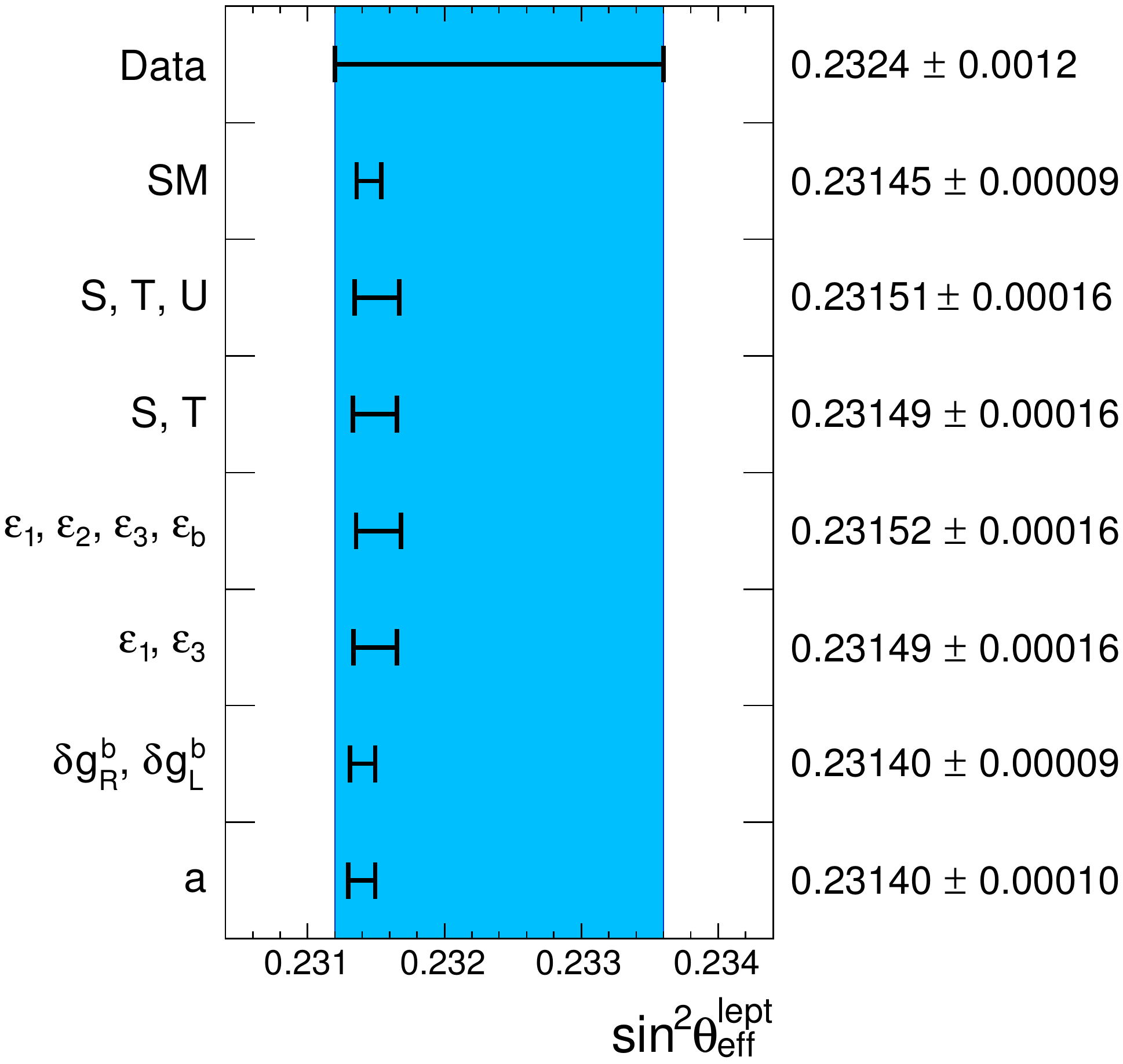}
\hfill
\includegraphics[width=.49\textwidth]{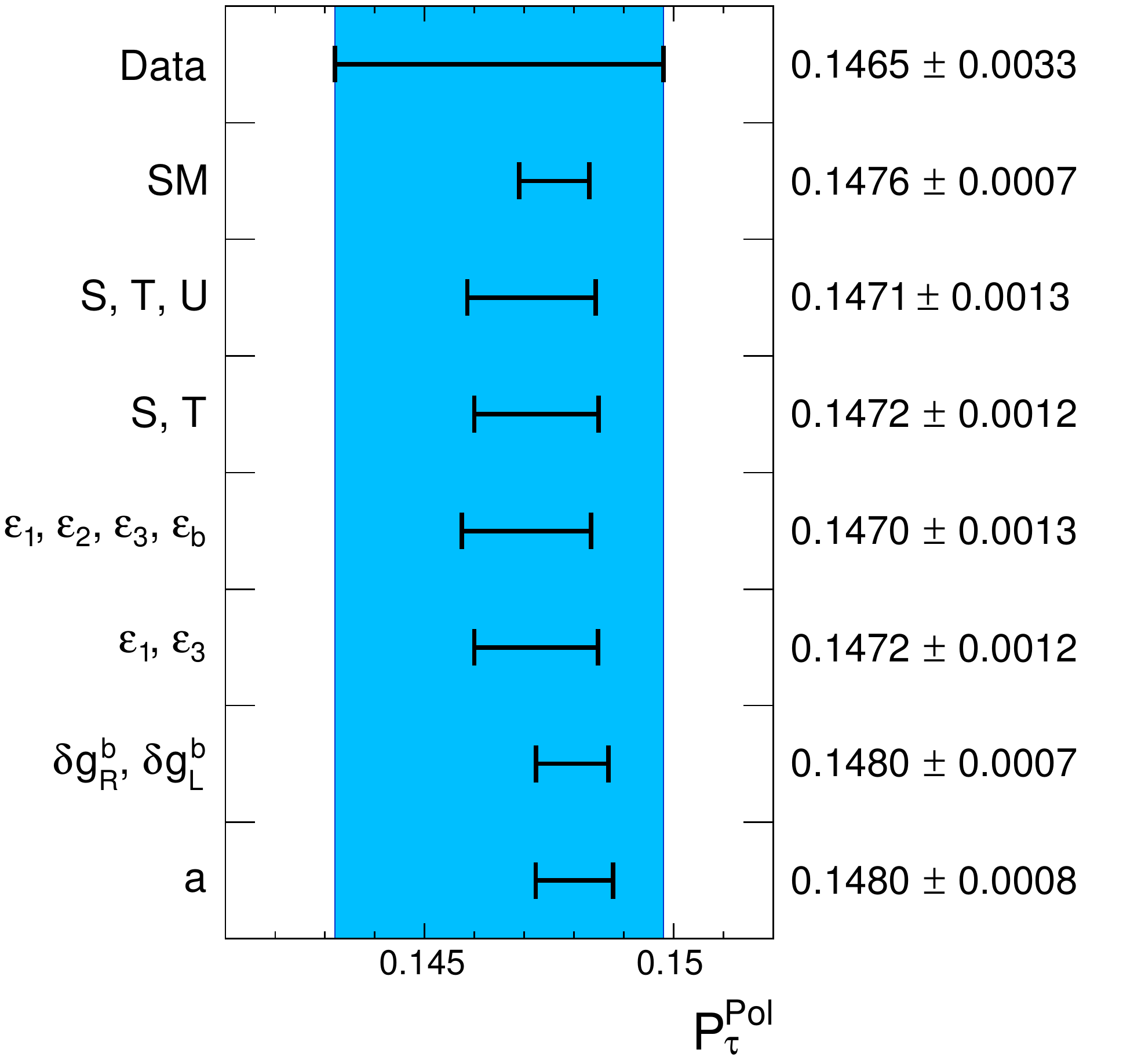}
\\[2mm]
\includegraphics[width=.49\textwidth]{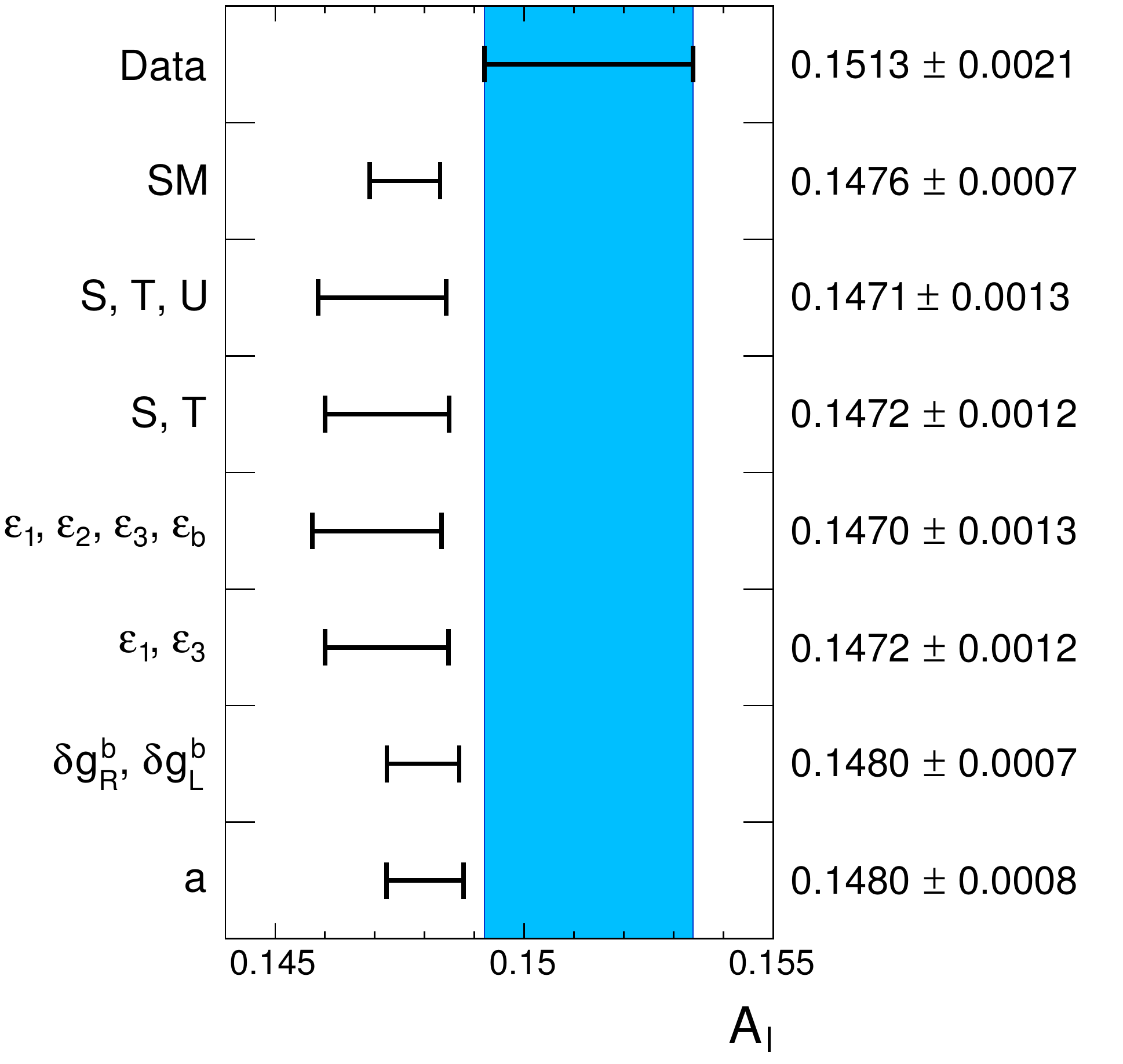}
\hfill
\includegraphics[width=.49\textwidth]{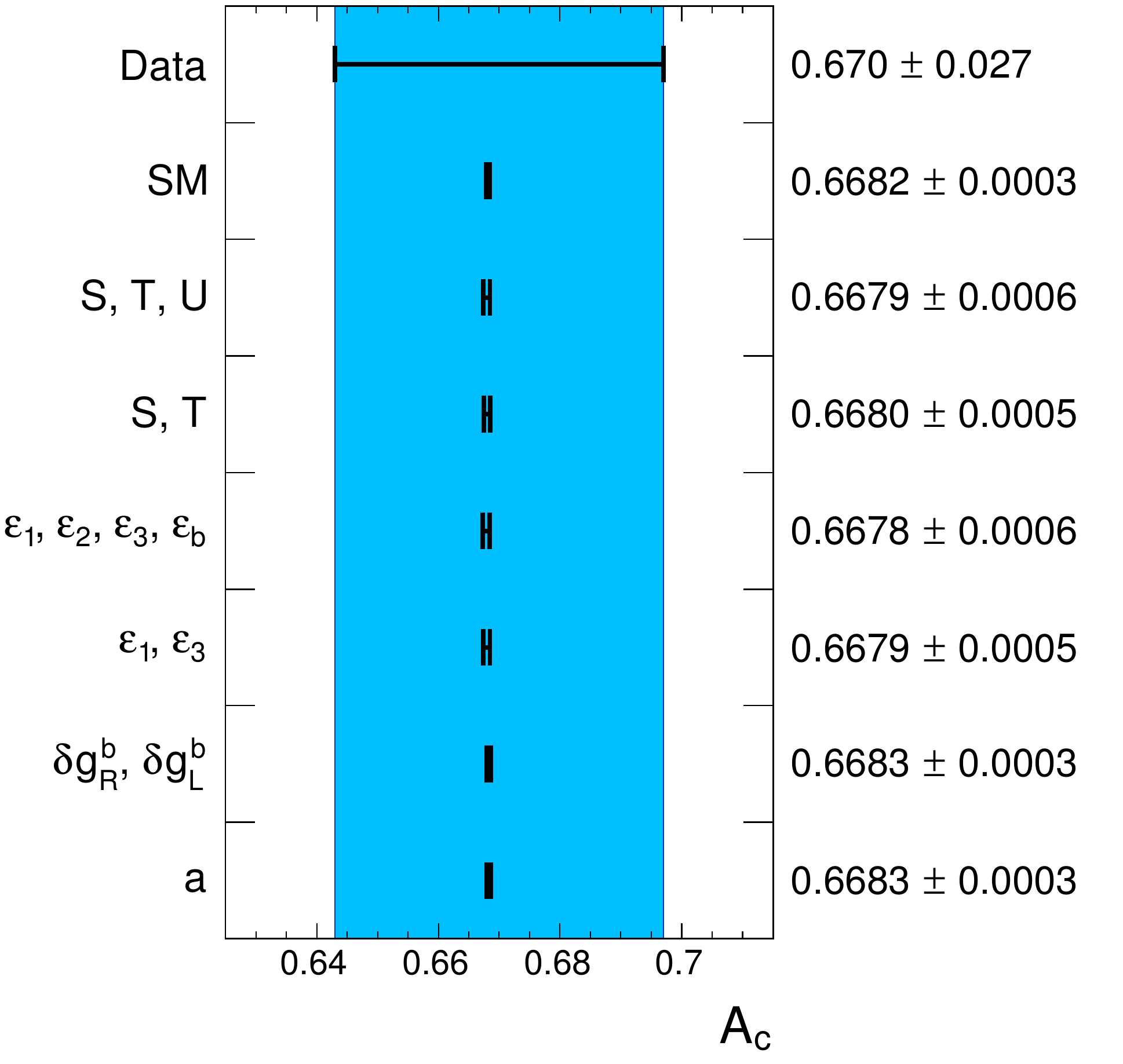}
\\[2mm]
\includegraphics[width=.49\textwidth]{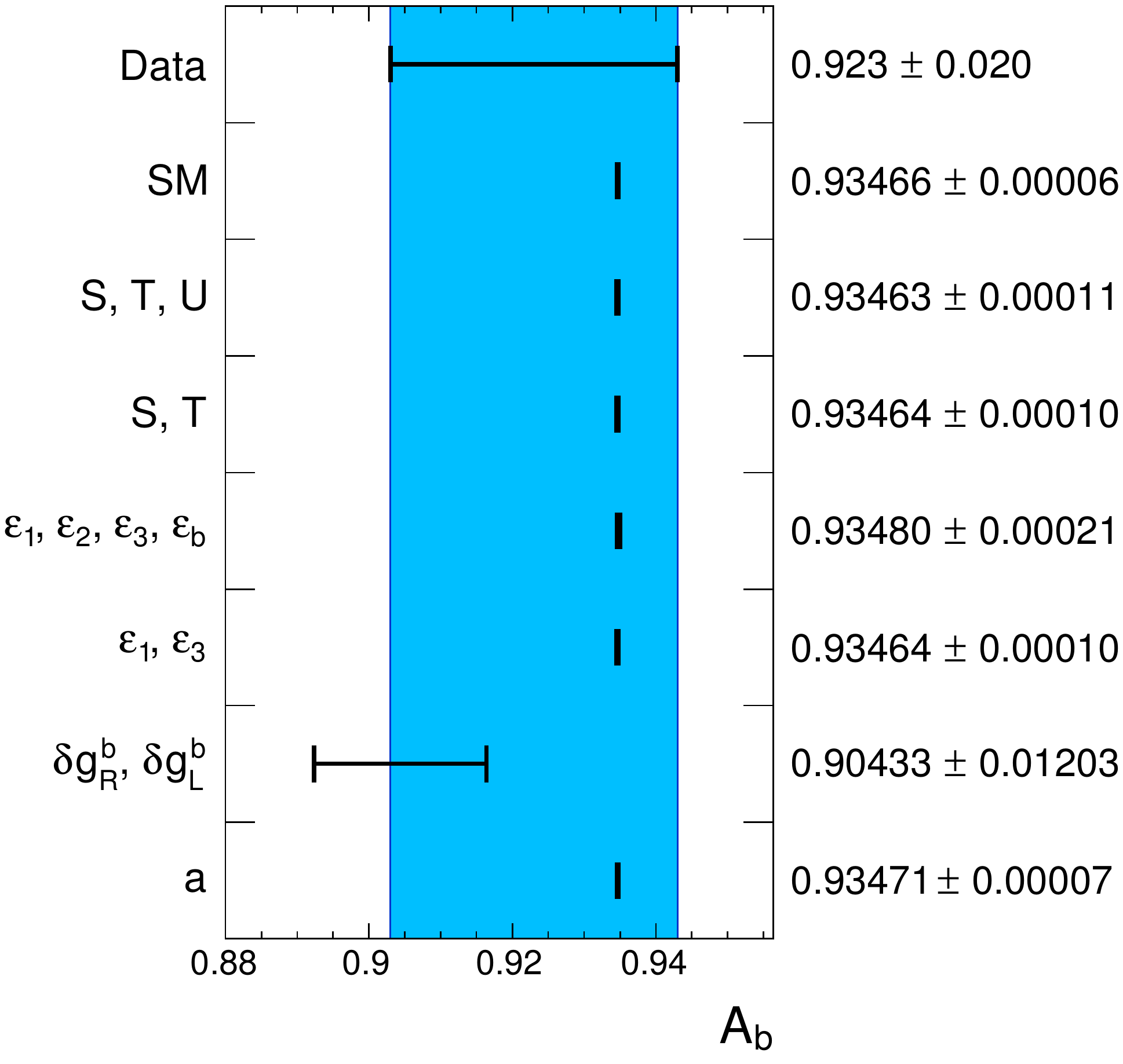}
\hfill
\includegraphics[width=.49\textwidth]{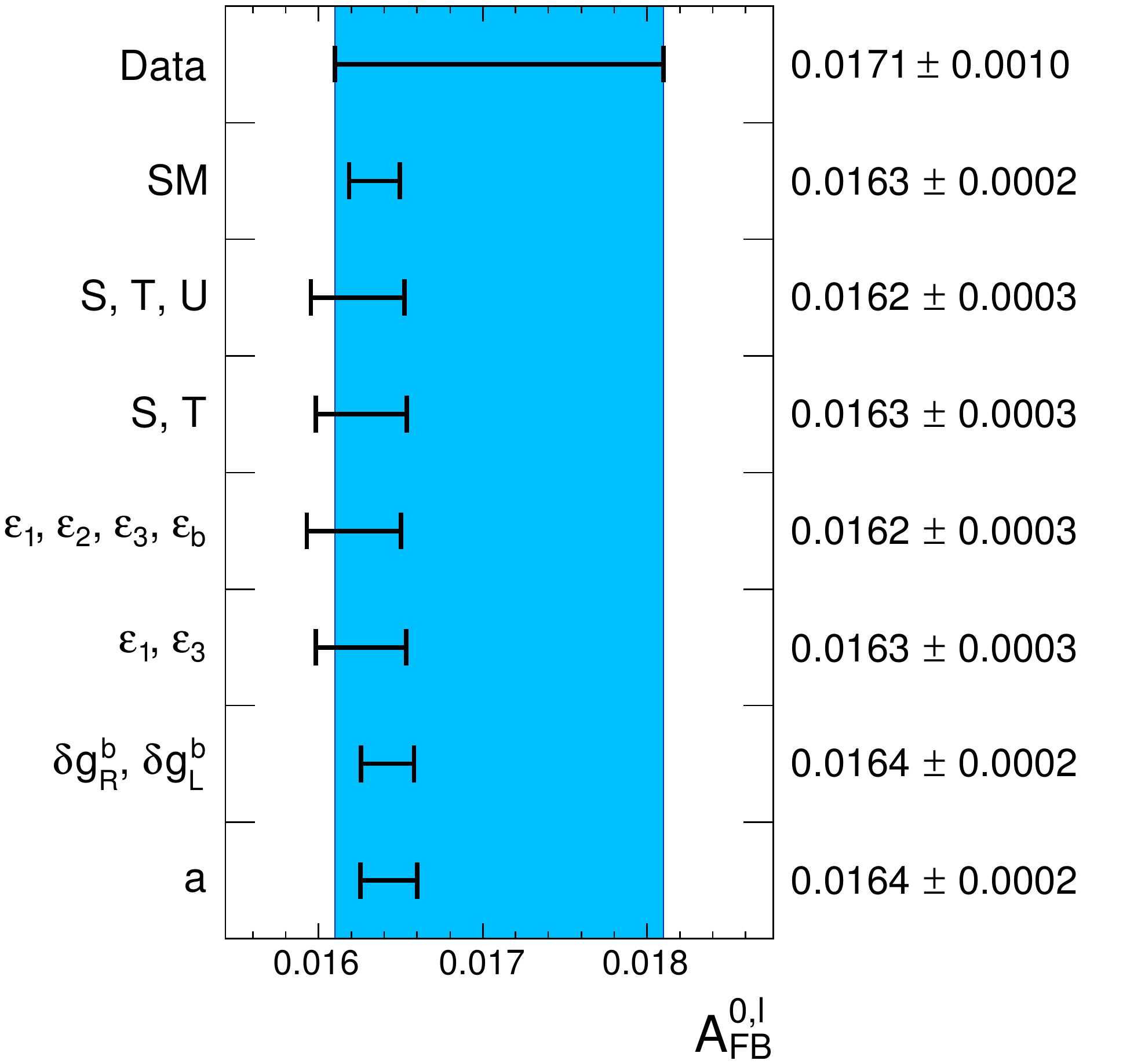}
\caption{Same as figure~\ref{fig:oldRb_summary1}.}
\label{fig:oldRb_summary2}
\end{figure}

\begin{figure}[htbp]
\centering
\includegraphics[width=.49\textwidth]{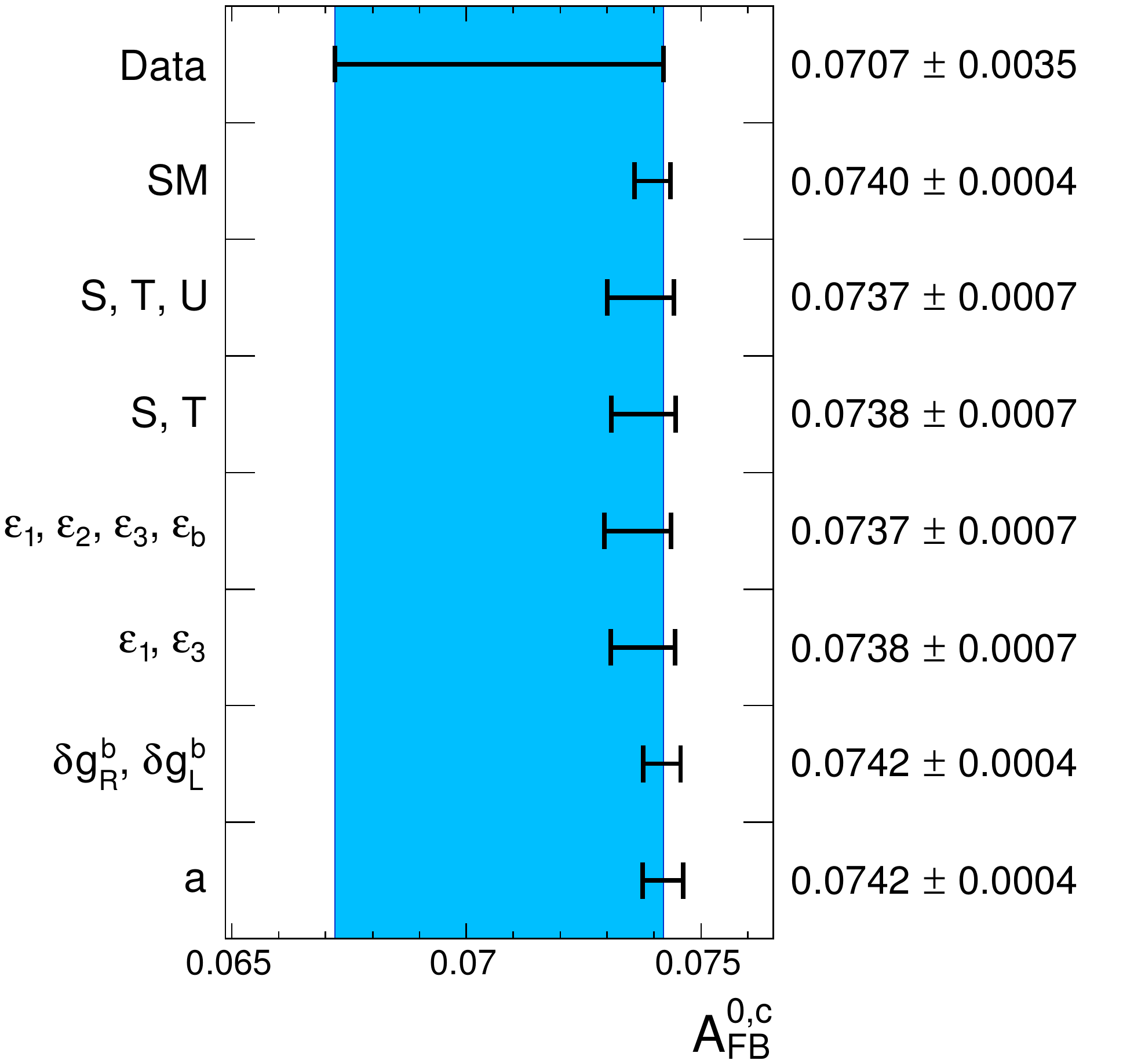}
\hfill
\includegraphics[width=.49\textwidth]{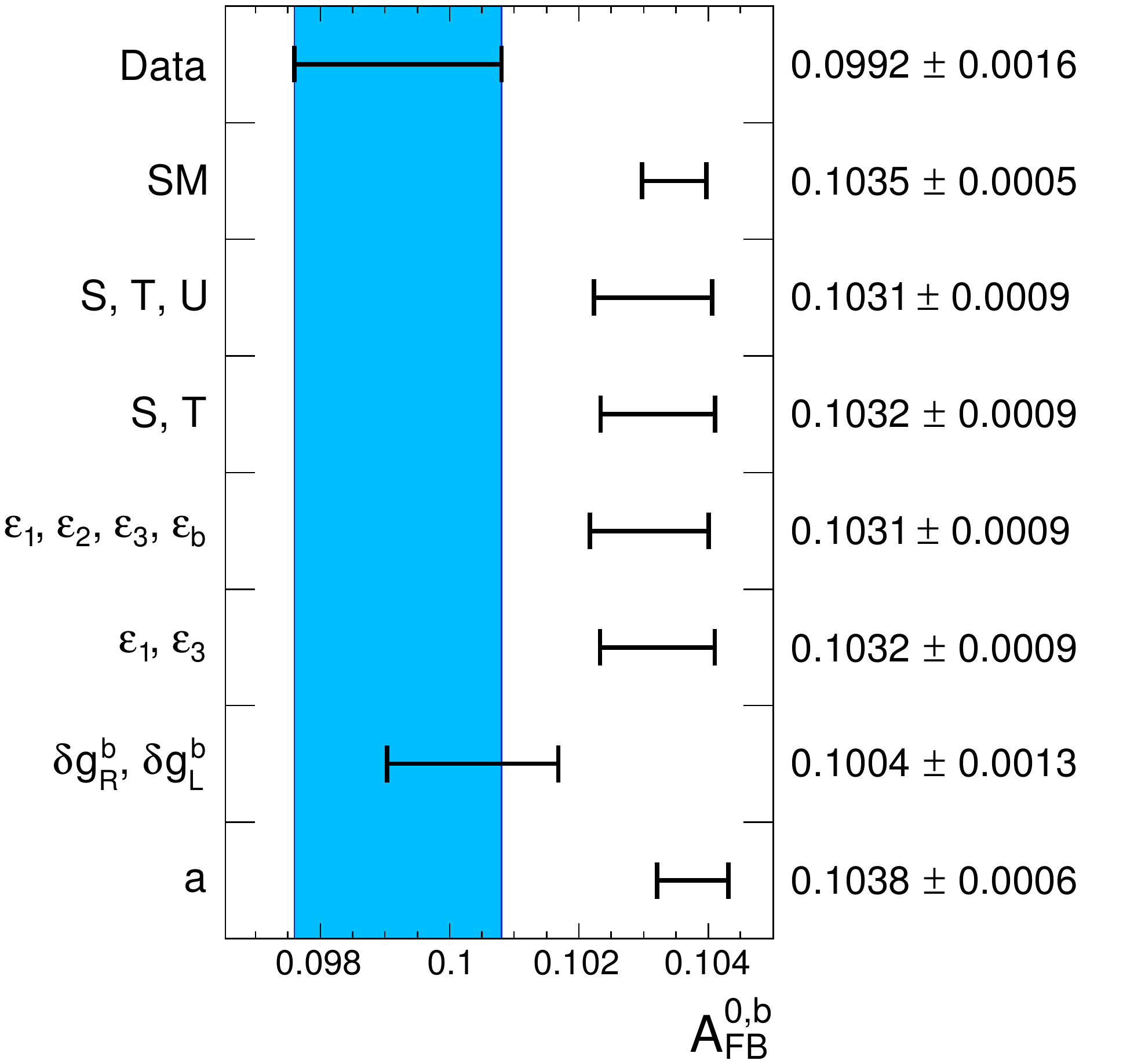}
\\[2mm]
\includegraphics[width=.49\textwidth]{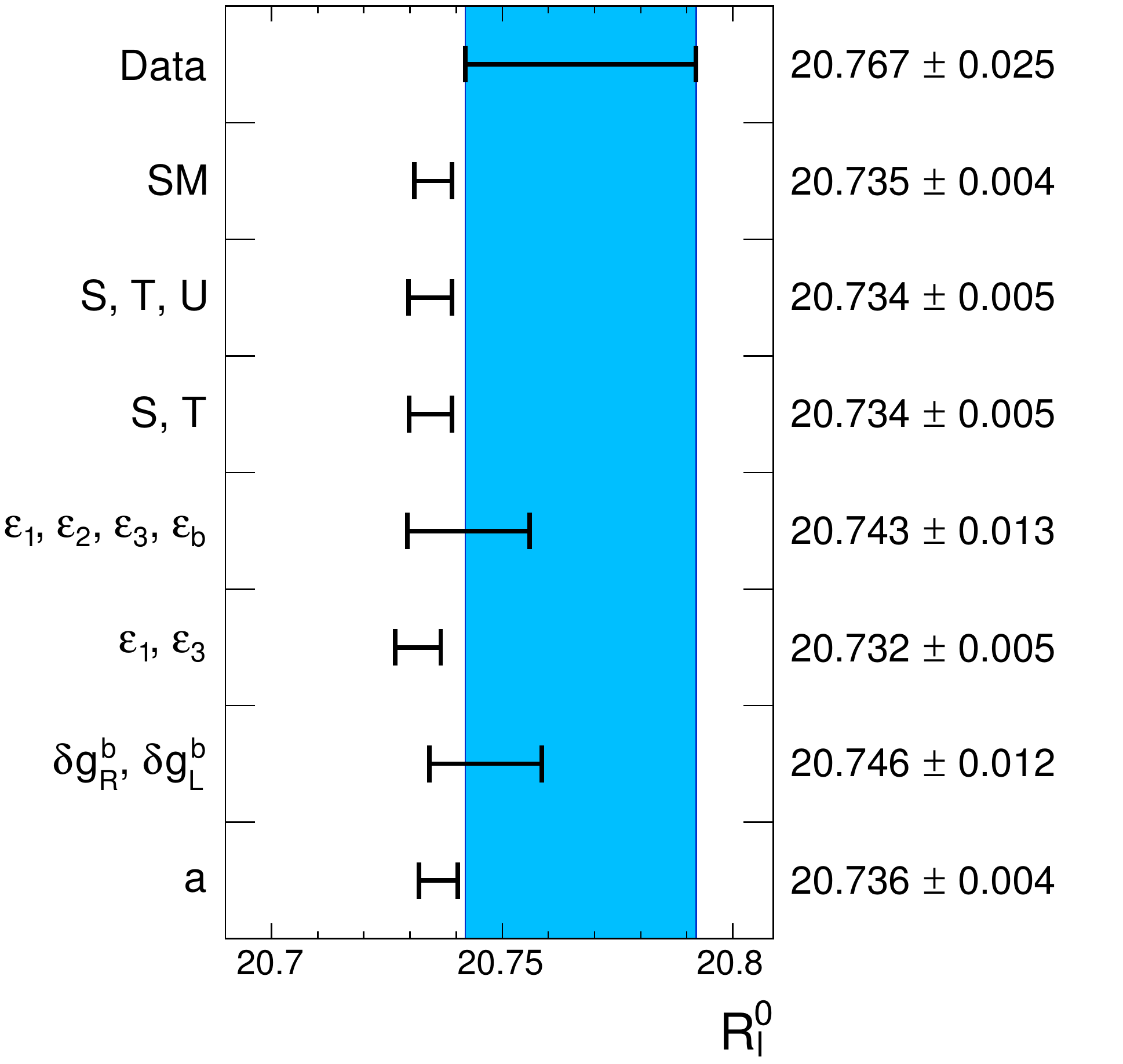}
\hfill
\includegraphics[width=.49\textwidth]{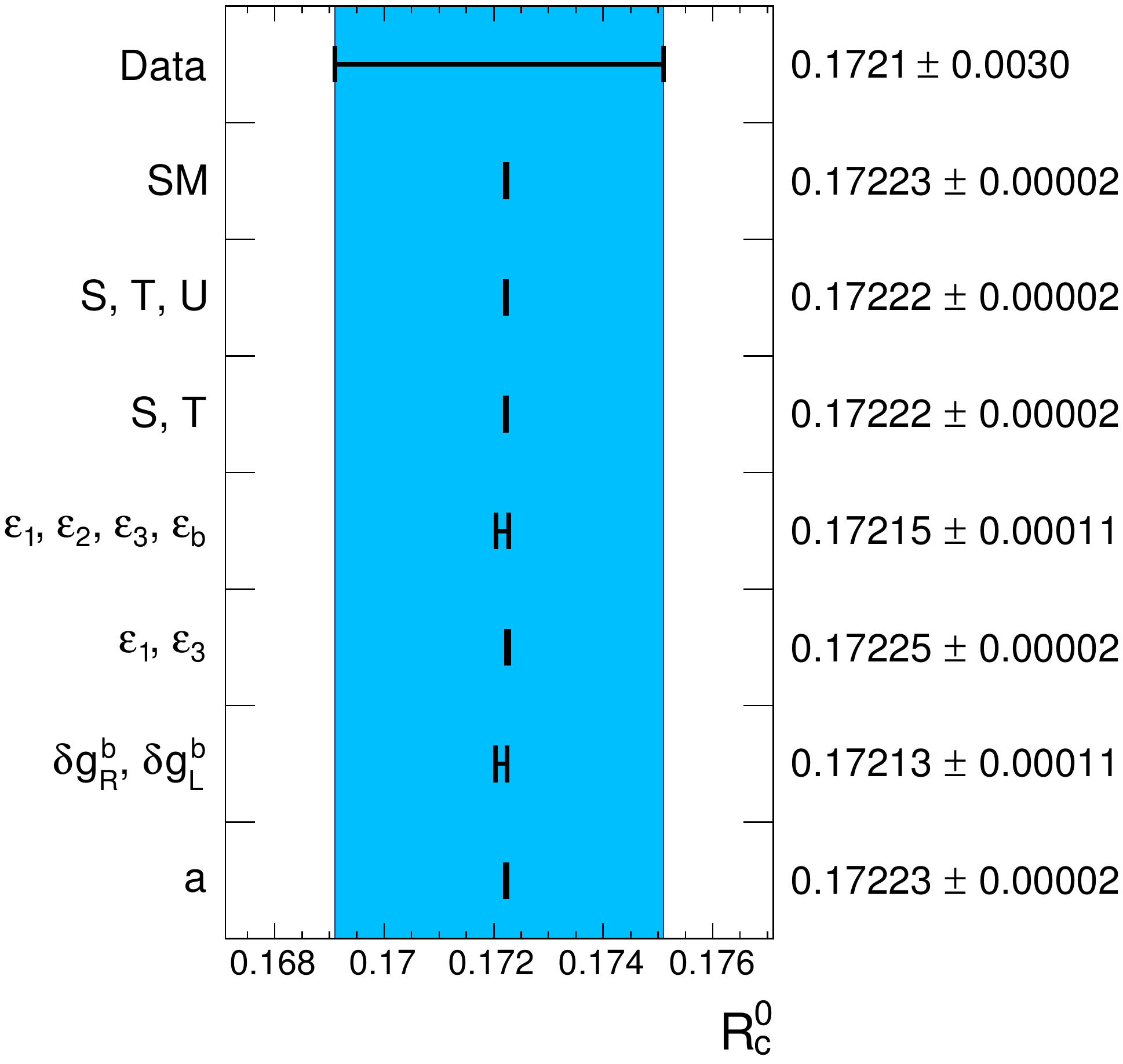}
\\[2mm]
\includegraphics[width=.49\textwidth]{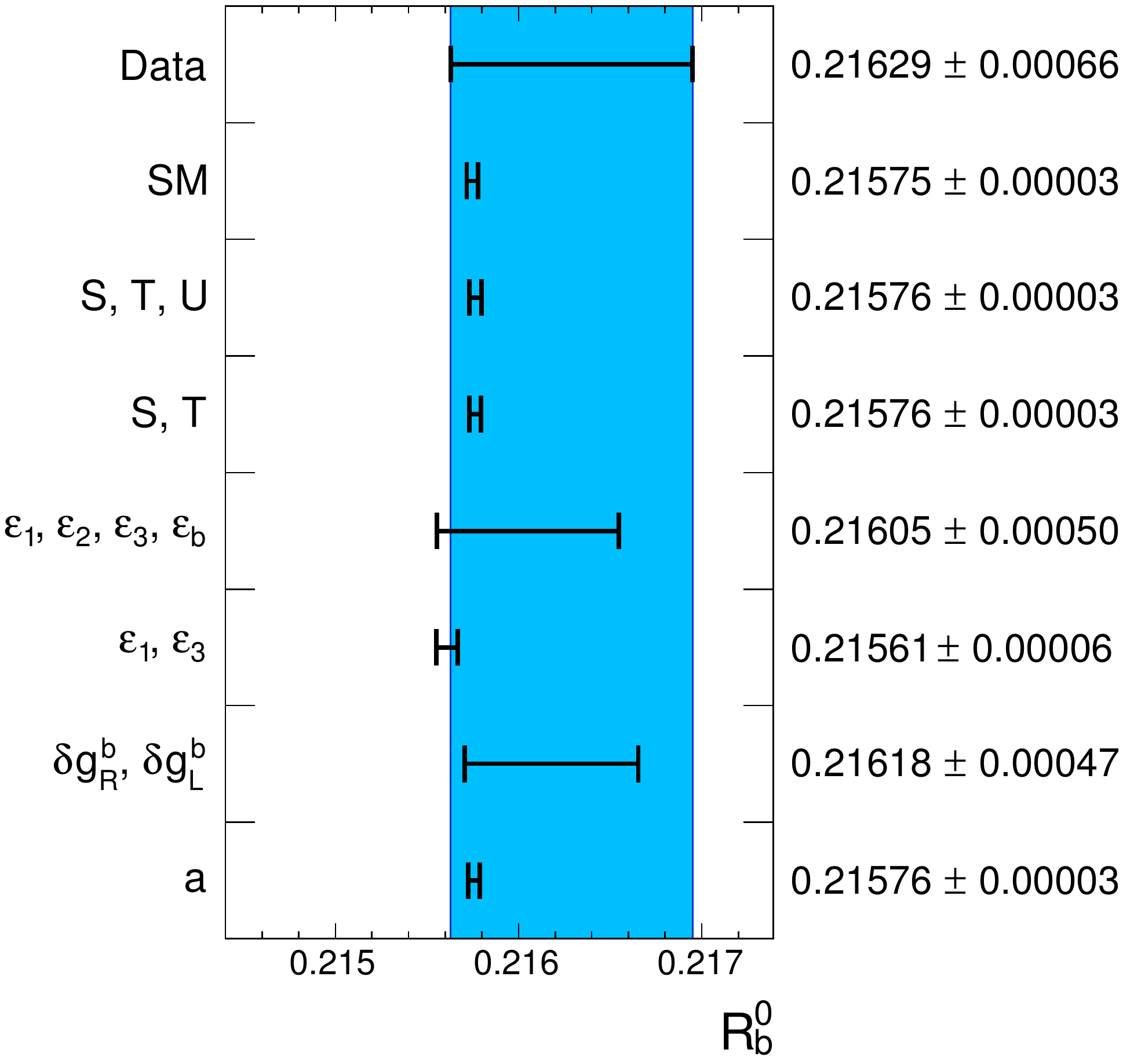}
\caption{Same as figure~\ref{fig:oldRb_summary1}.}
\label{fig:oldRb_summary3}
\vspace{-10mm}
\end{figure}

\clearpage
\section{Fit results for the observables using the full two-loop
  fermionic EW corrections to $\rho_Z^f$}
\label{app:obs_newRb}

In this Appendix we present a graphical summary of the fit results for
all observables obtained within the various scenarios considered in
this work, obtained using the results from
ref.~\cite{Freitas:2012sy,freitasprivate} for the two-loop fermionic
EW corrections to $\rho_Z^f$. In the NP fits, we neglect the
observables $\Gamma_Z$, $\sigma^0_h$ and $R_\ell^0$. 
The labels in the figures refer to the
various fits performed with a self-explanatory notation. The orange band
corresponds to the direct measurement, also reported with the ``Data''
label. 

\begin{figure}[!htbp!]
\centering
\includegraphics[width=.49\textwidth]{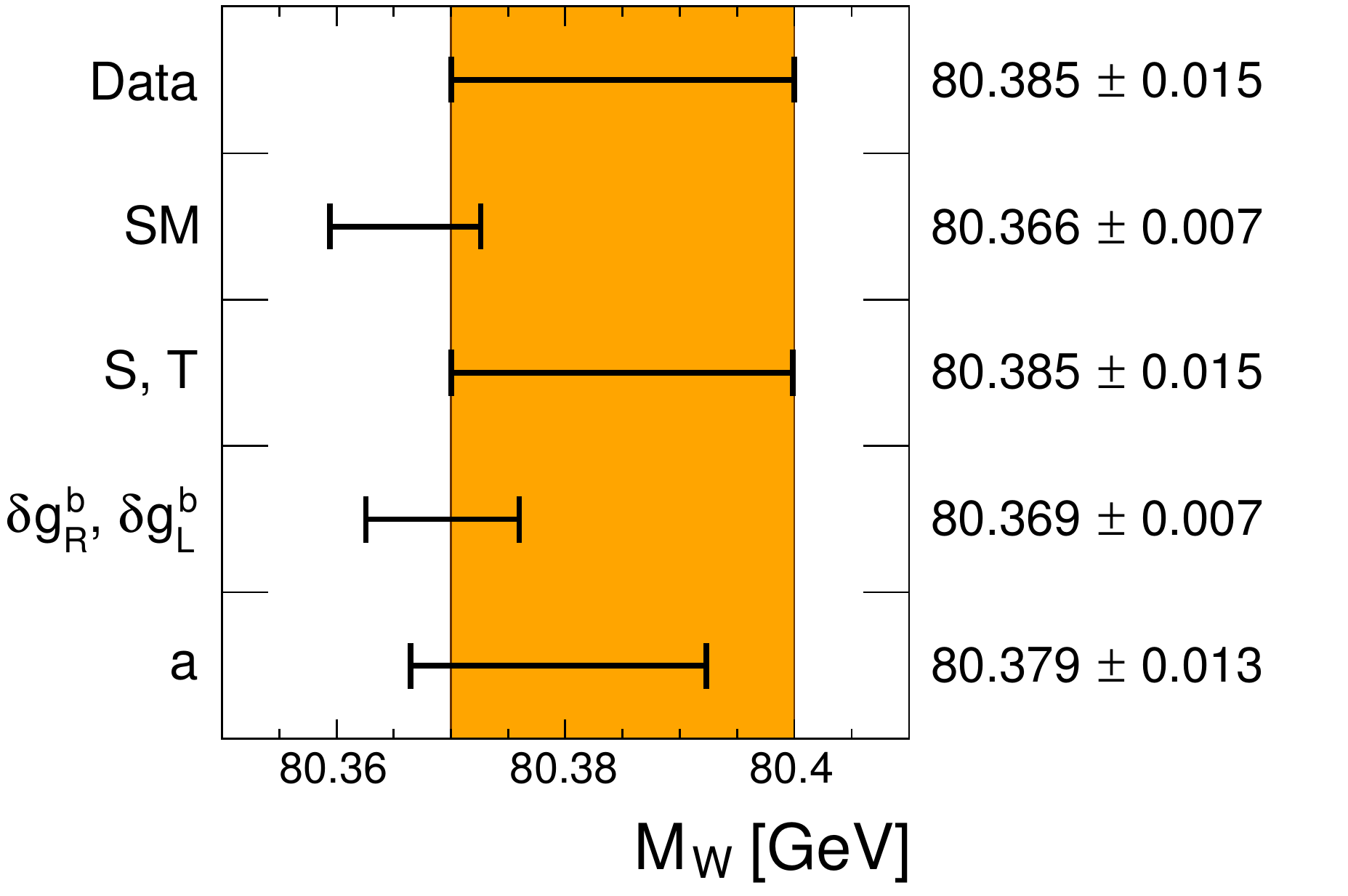}
\hfill
\includegraphics[width=.49\textwidth]{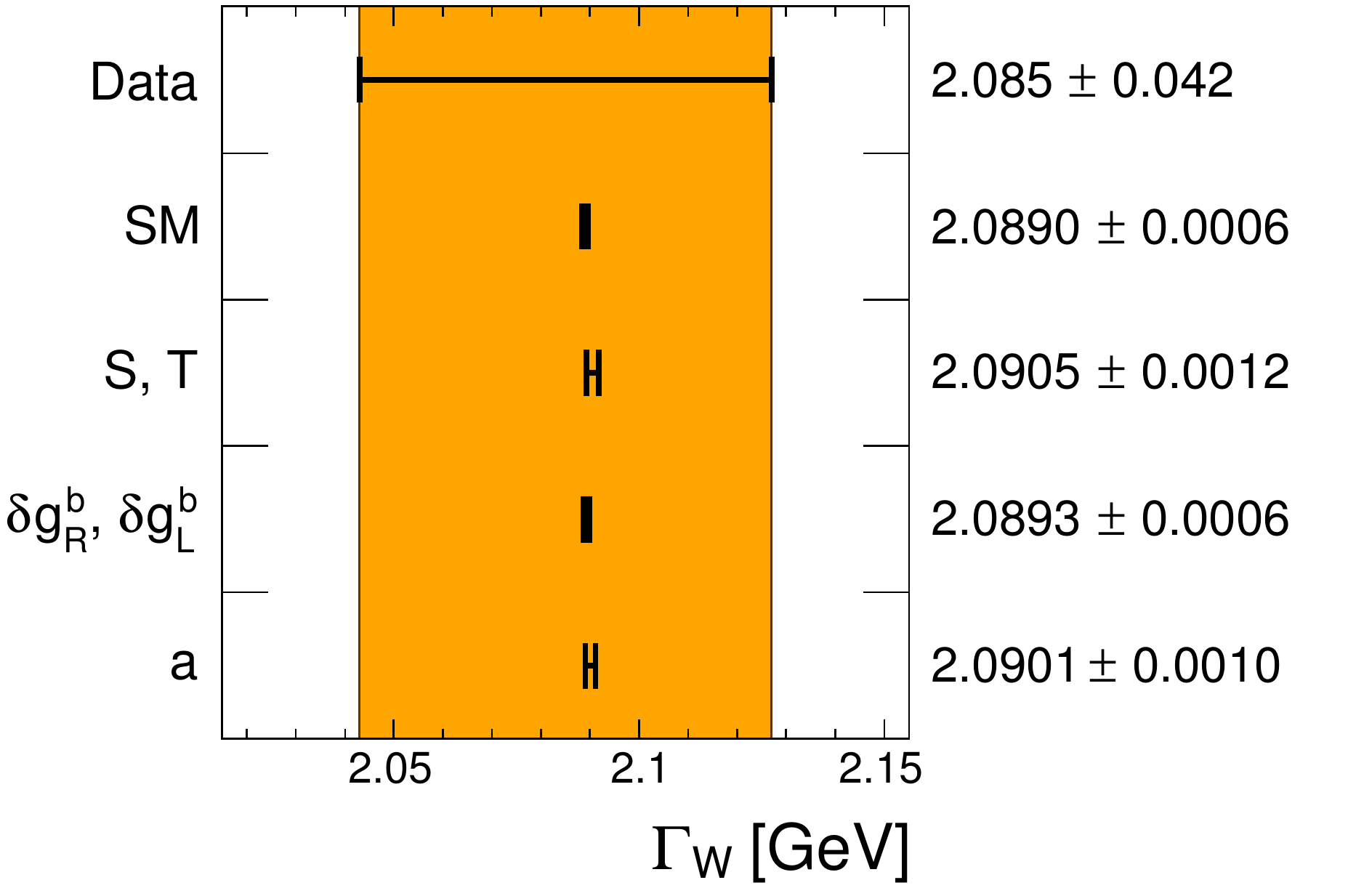}
\\[2mm]
\includegraphics[width=.49\textwidth]{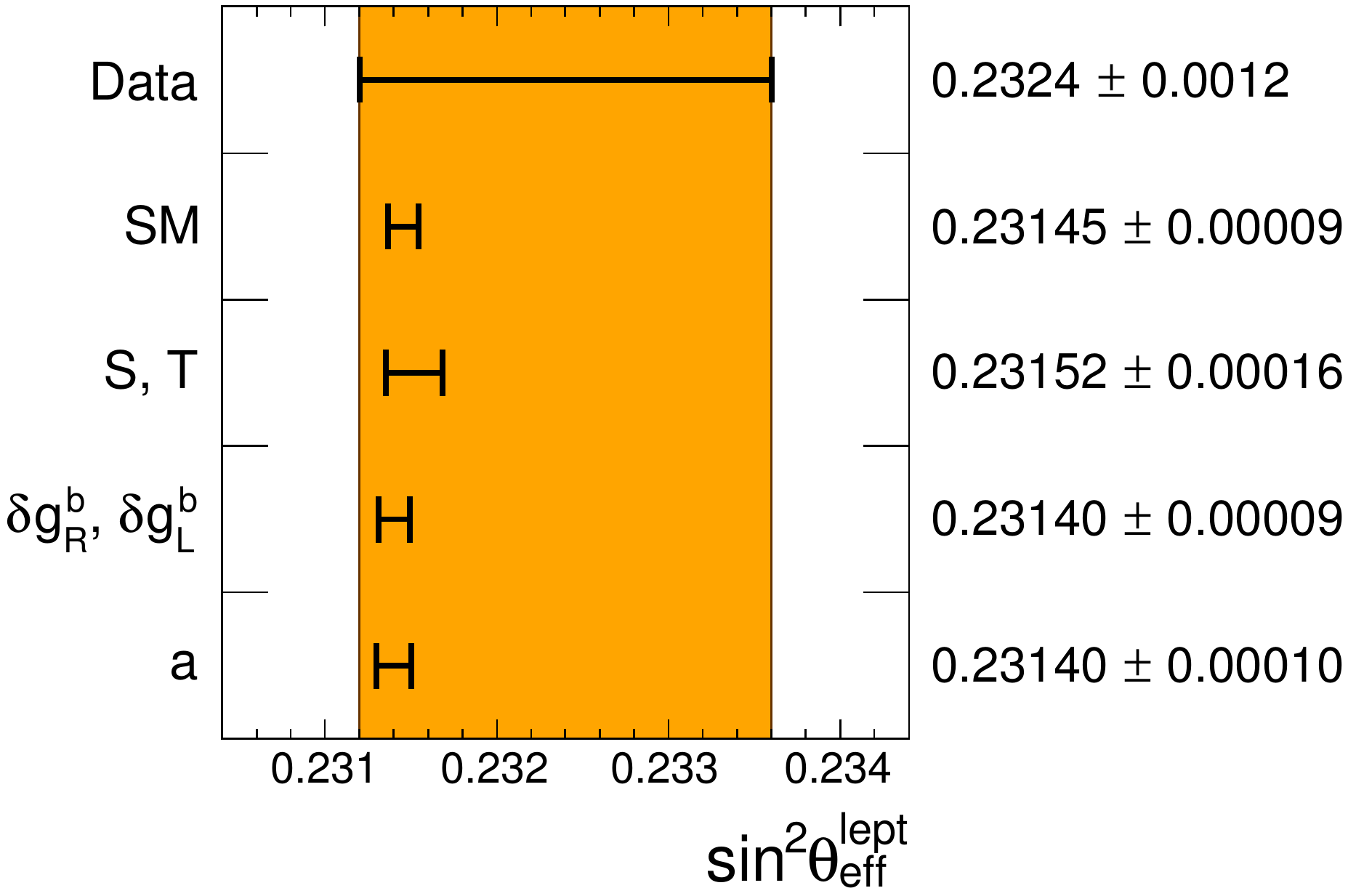}
\hfill
\includegraphics[width=.49\textwidth]{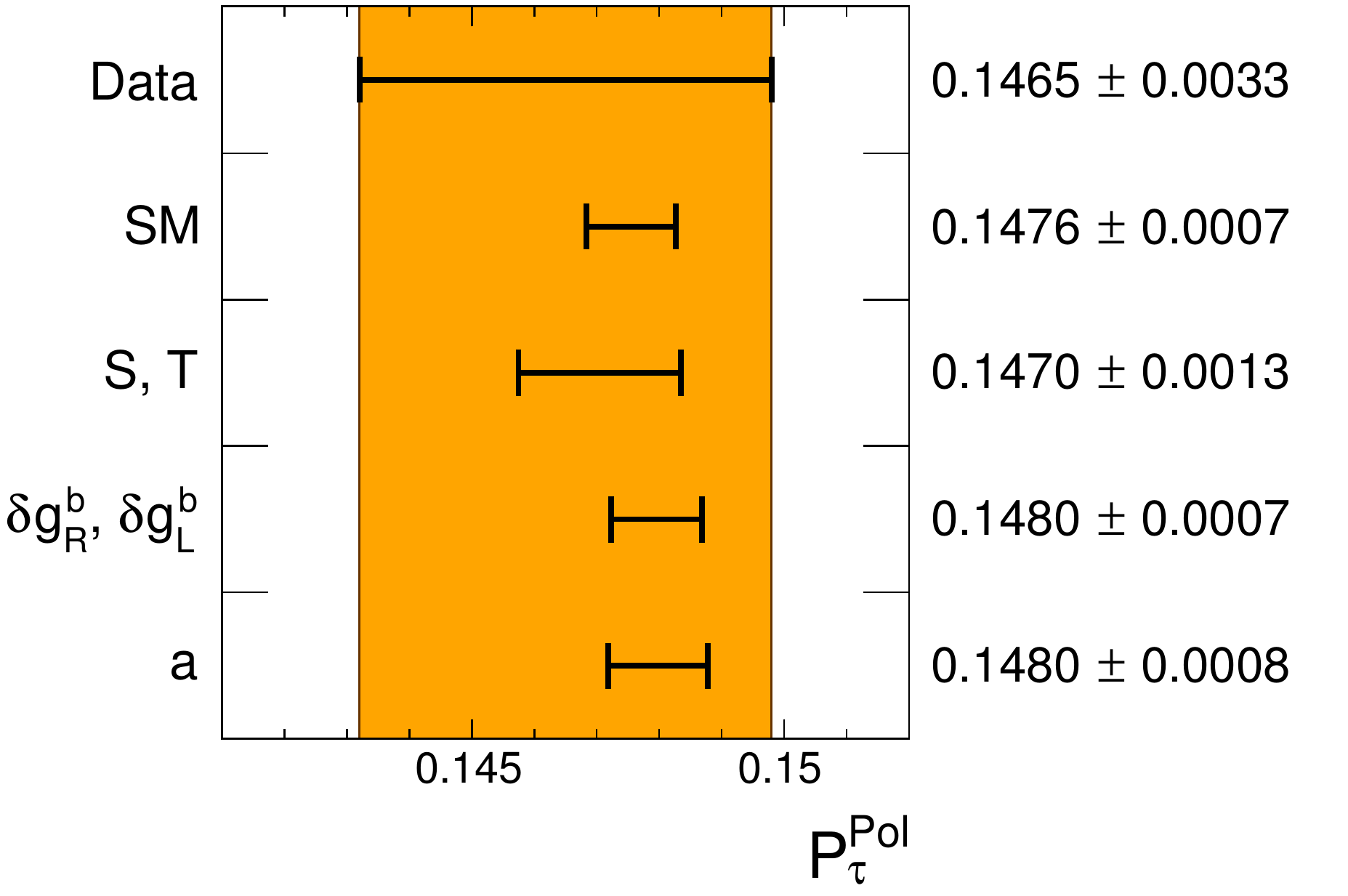}
\\[2mm]
\includegraphics[width=.49\textwidth]{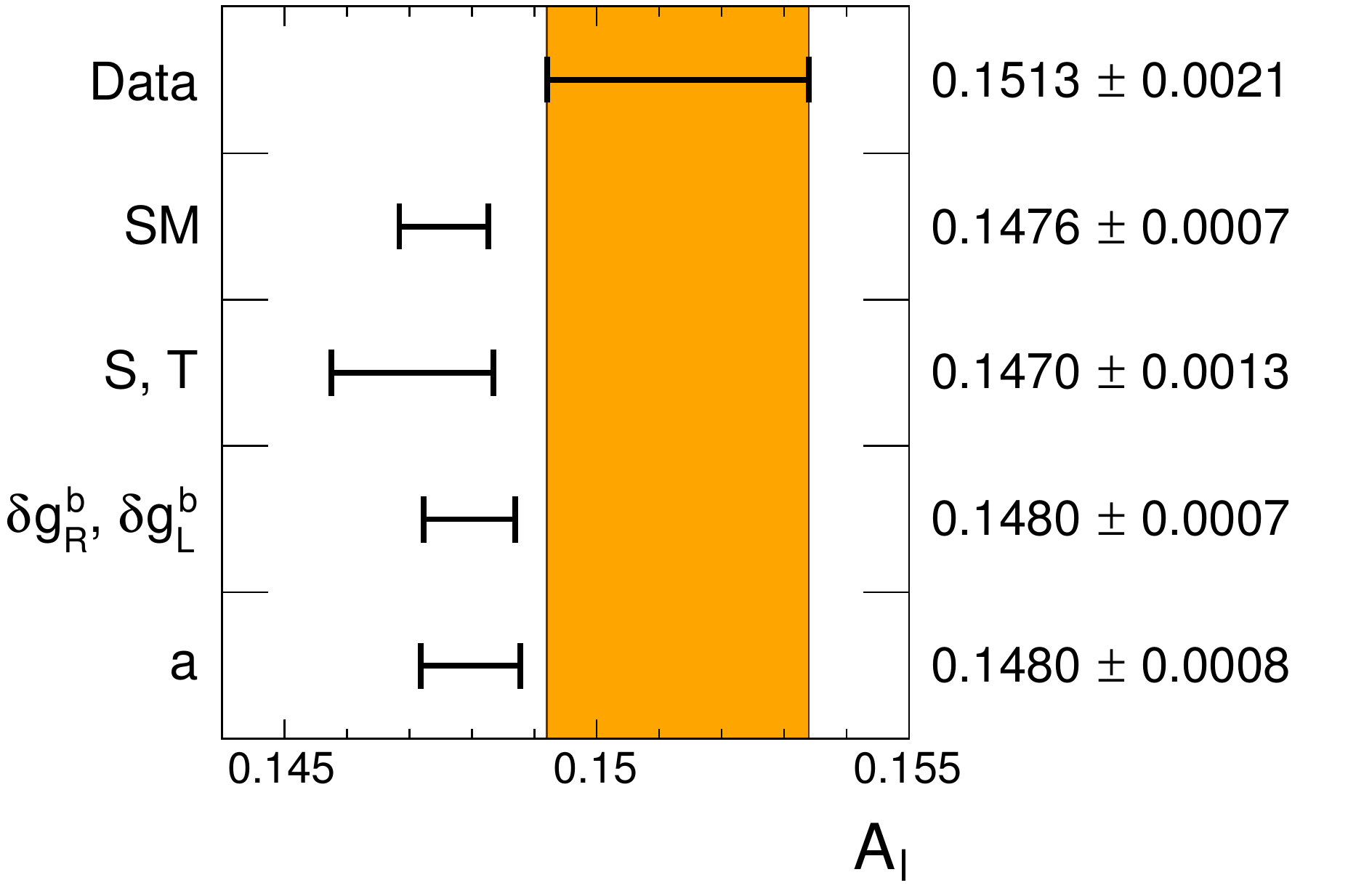}
\hfill
\includegraphics[width=.49\textwidth]{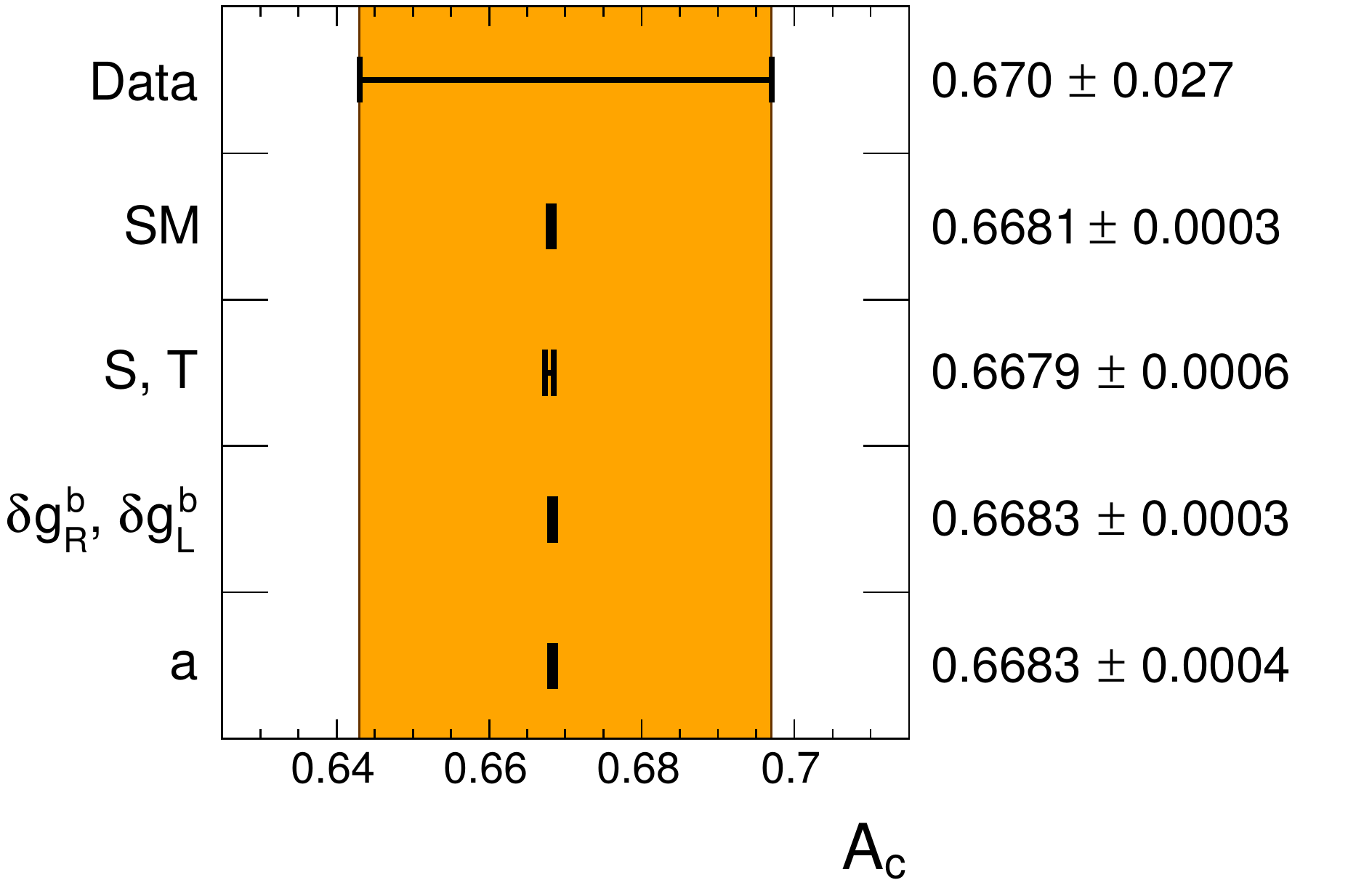}
\caption{Fit results, including the full two-loop fermionic EW 
  corrections to the coupling $\rho_Z^f$ with the results of
  ref.~\cite{Freitas:2012sy,freitasprivate} and the parameters $\delta 
  \rho_Z^{\nu,\ell,b}$.}
\label{fig:newRb_summary1}
\end{figure}

\begin{figure}[!htbp!]
\centering
\includegraphics[width=.49\textwidth]{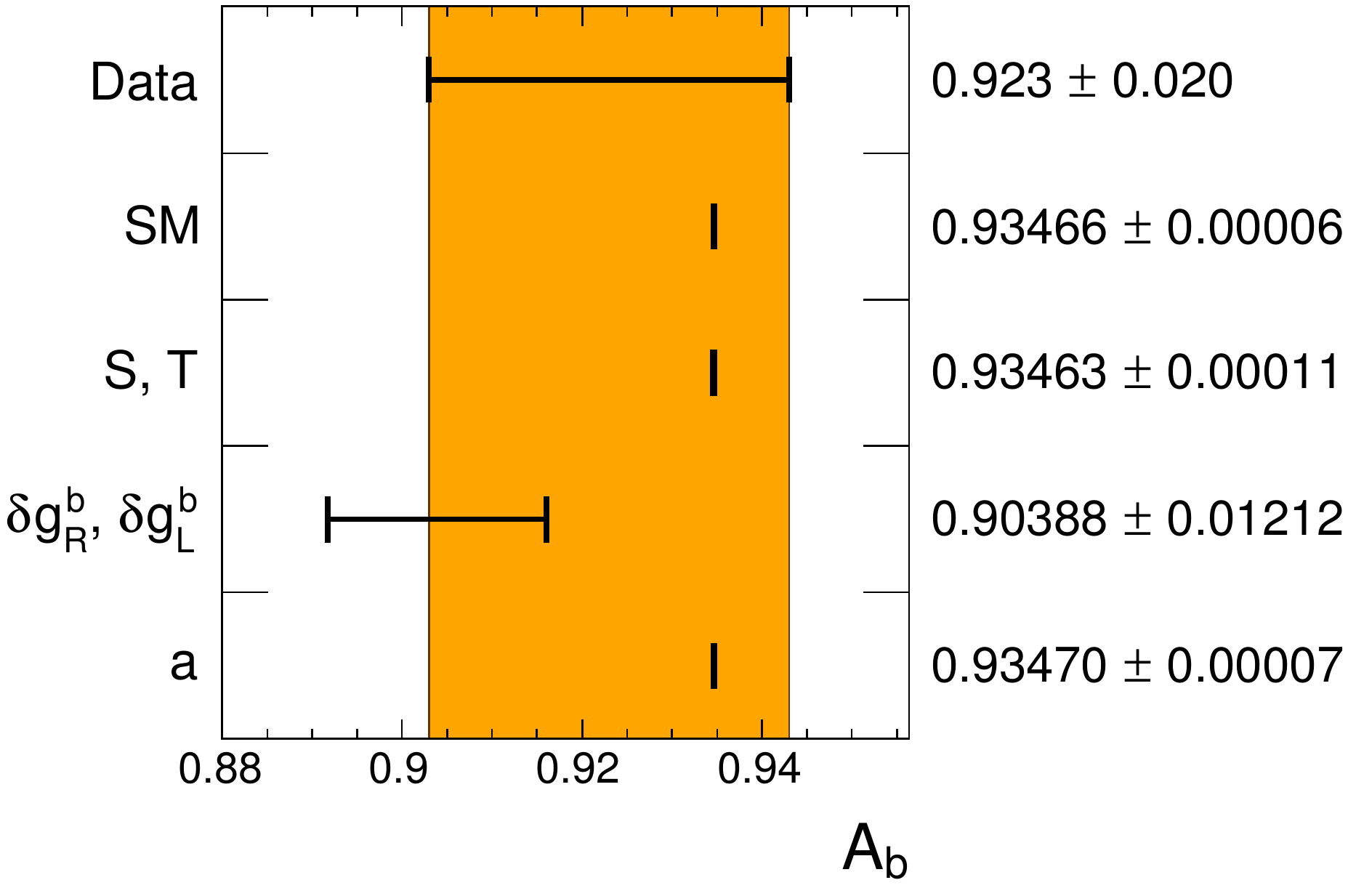}
\hfill
\includegraphics[width=.49\textwidth]{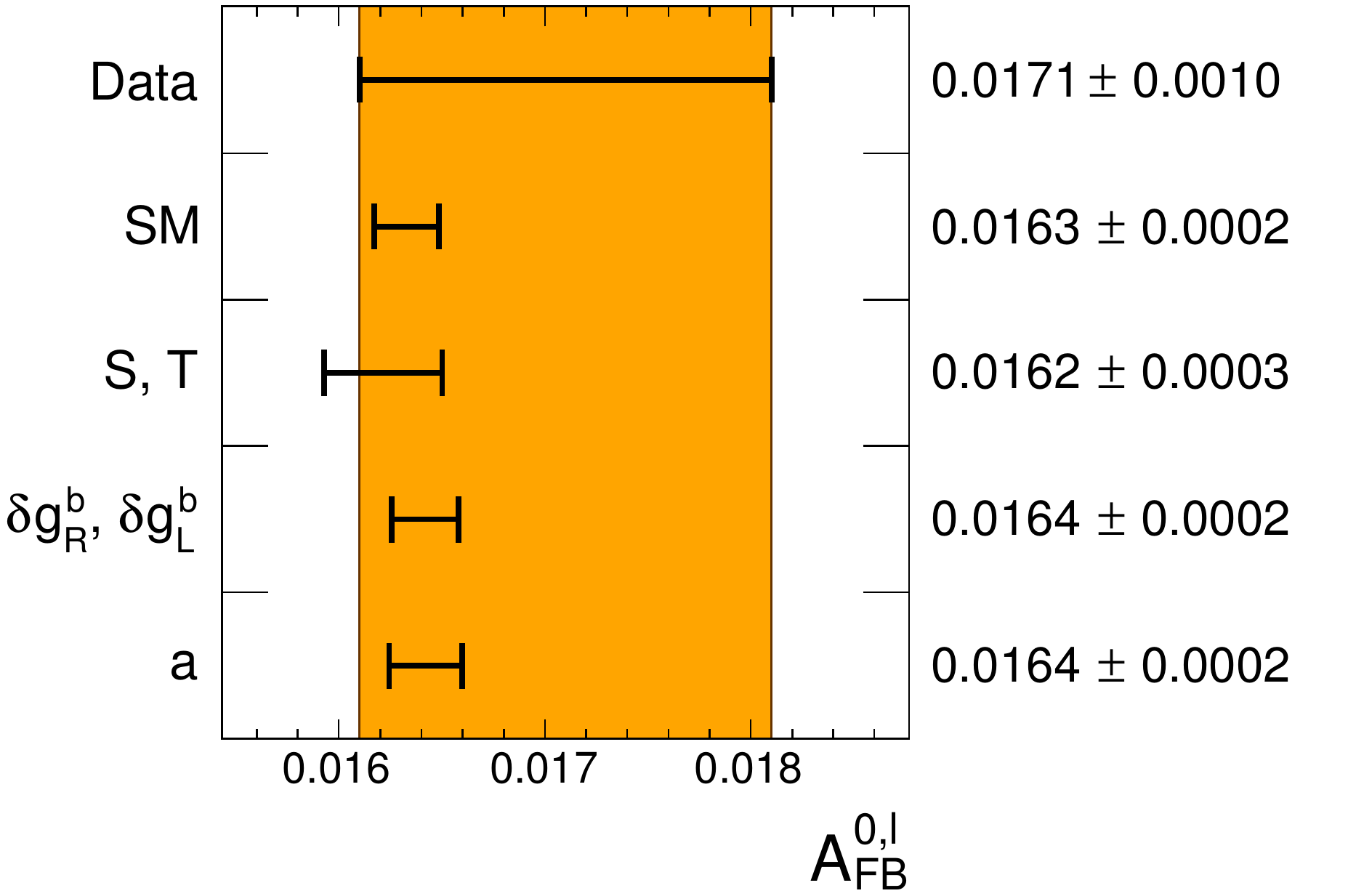}
\\[2mm]
\includegraphics[width=.49\textwidth]{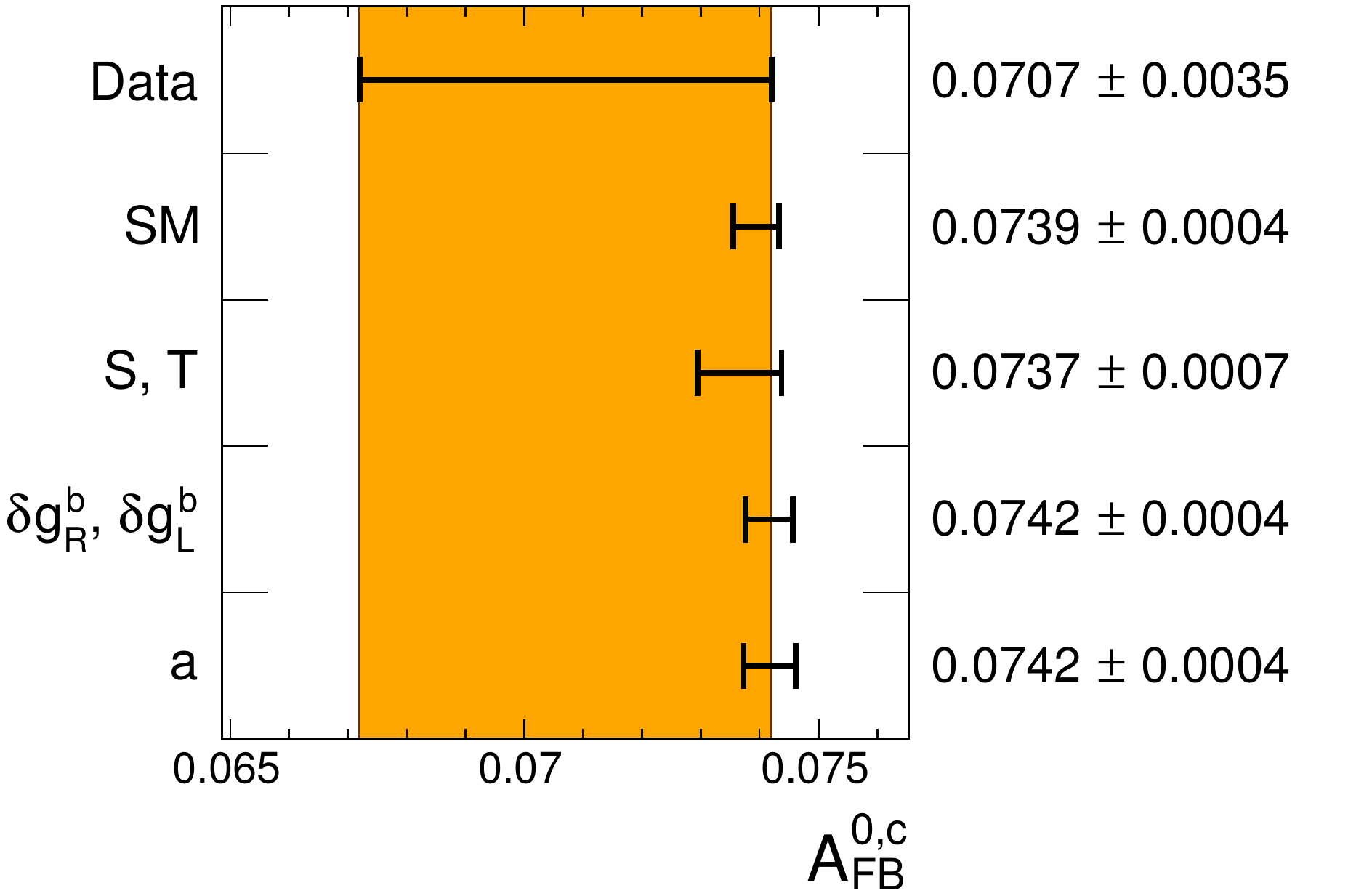}
\hfill
\includegraphics[width=.49\textwidth]{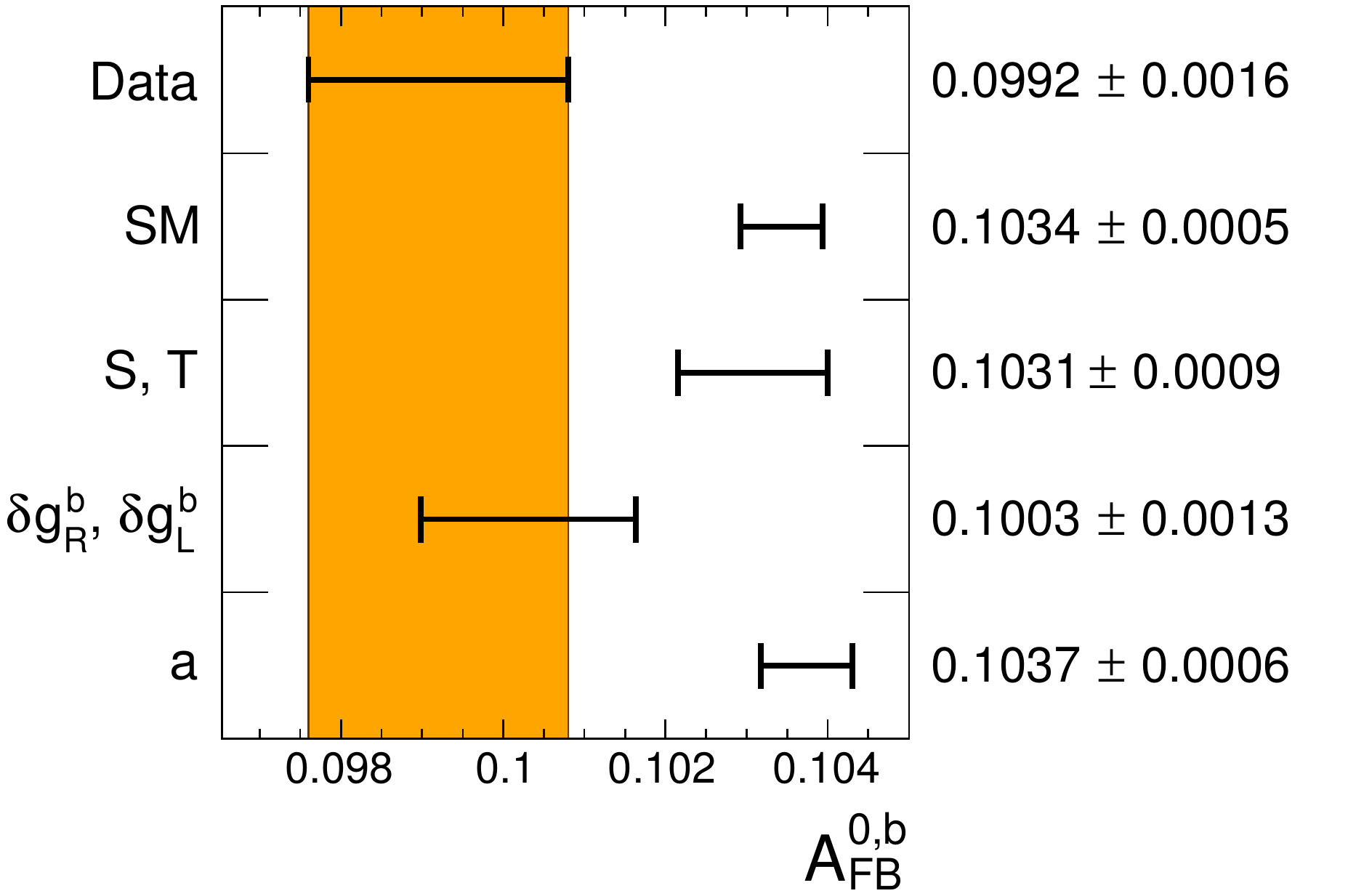}
\\[2mm]
\includegraphics[width=.49\textwidth]{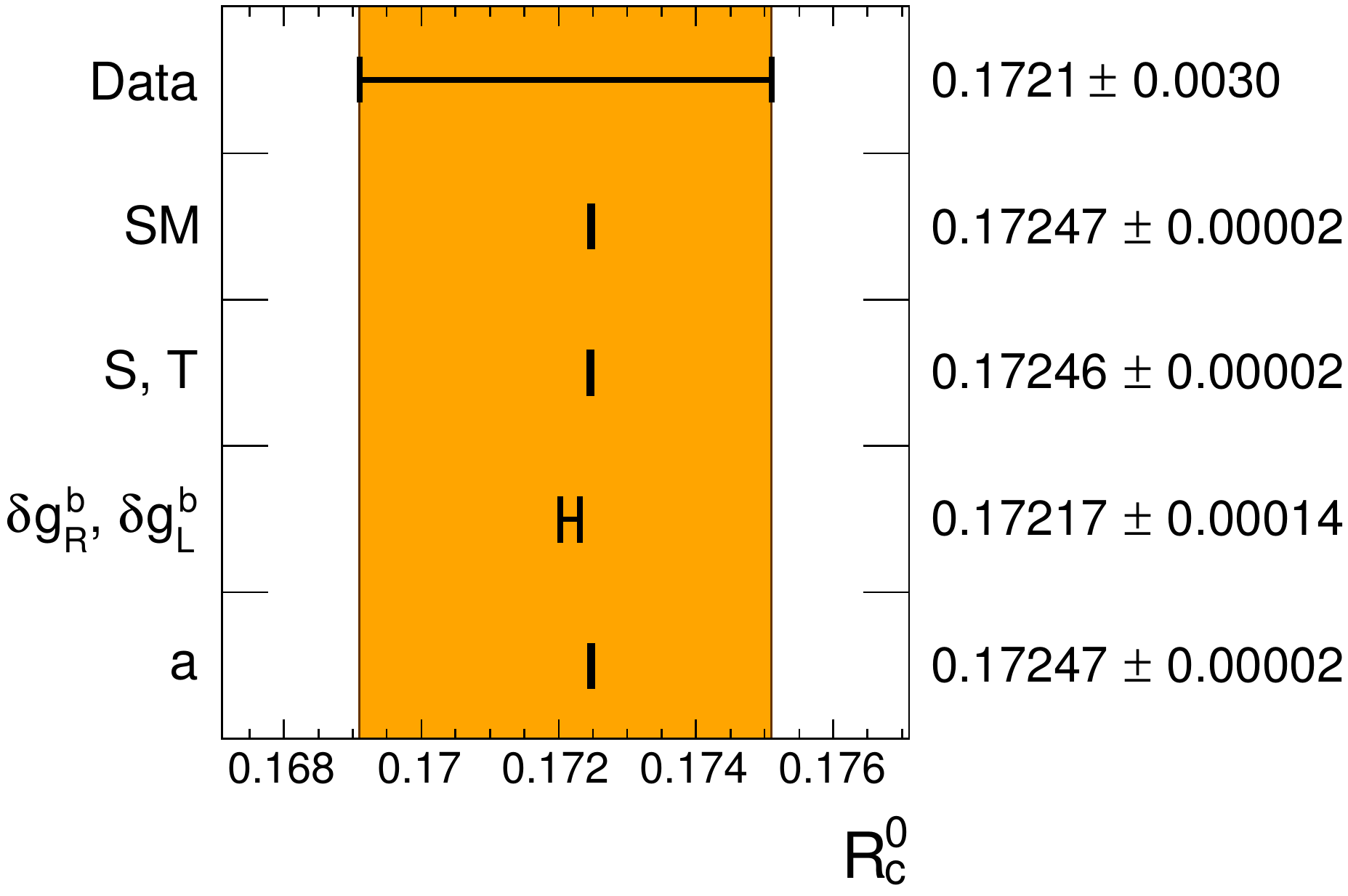}
\hfill
\includegraphics[width=.49\textwidth]{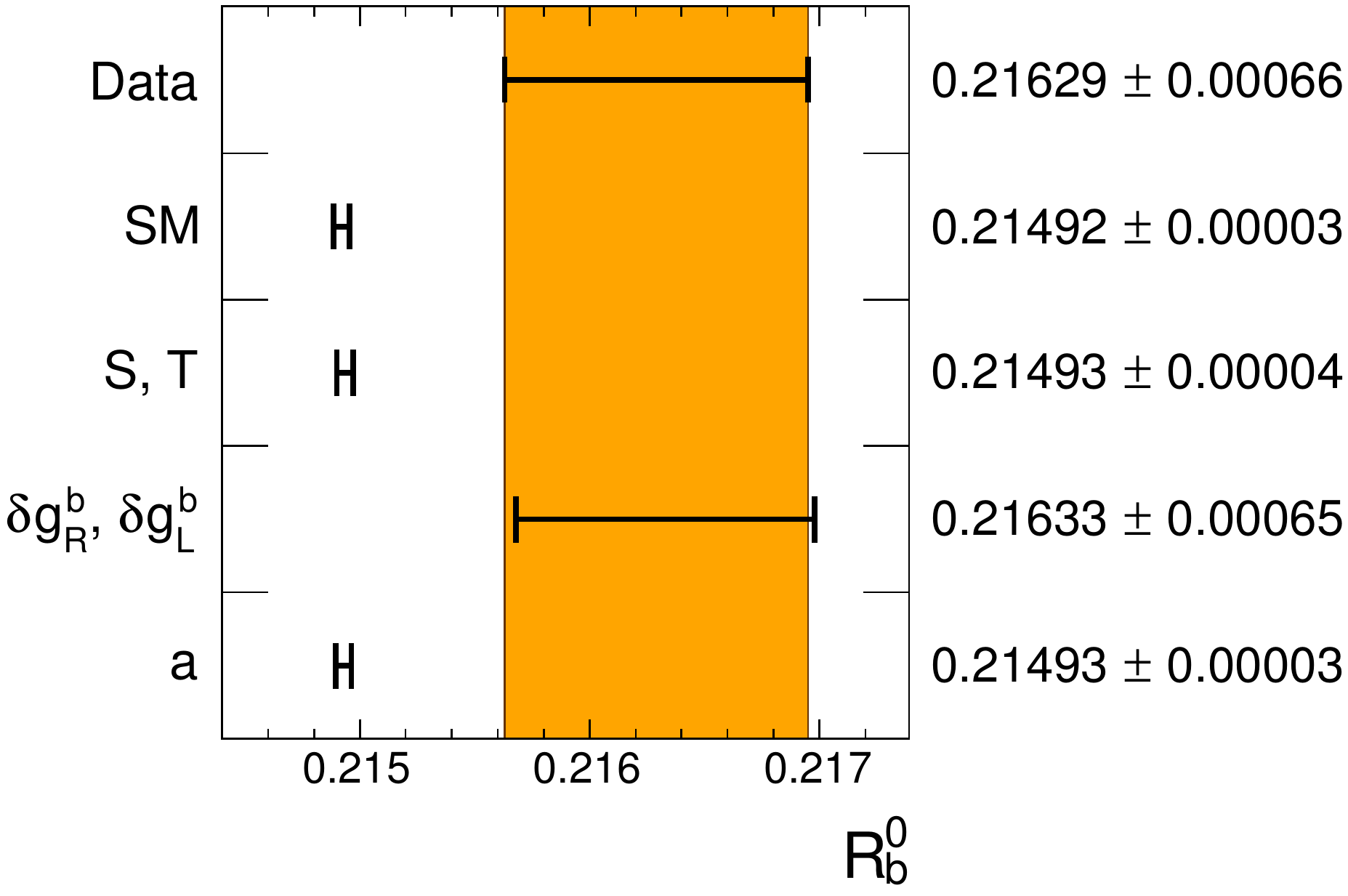}
\caption{Same as figure~\ref{fig:newRb_summary1}.}
\label{fig:newRb_summary2}
\end{figure}

\newpage
\bibliographystyle{JHEP}
\bibliography{ewfit}

\end{document}